\documentclass[preprint]{aastex631}

\newcommand{\fwidth}{4.4cm}
\usepackage{color}
\usepackage{natbib}
\usepackage{amsmath}
\usepackage{nicefrac}
\tightenlines

\newcommand \AU	        {\,{\rm au}}
\newcommand \bahat      {\boldsymbol{\hat{a}}}
\newcommand \bB         {\boldsymbol{B}}

\newcommand \bbhat      {\boldsymbol{\hat{b}}}

\newcommand \bE         {\boldsymbol{E}}
\newcommand \bEinc      {\boldsymbol{E}_{\rm inc}}

\newcommand \bJ         {\boldsymbol{J}}

\newcommand \bkhat      {\hat{\bf k}}

\newcommand \bP         {{\bf P}}

\newcommand \bxhat      {\boldsymbol{\hat{x}}}
\newcommand \byhat      {\boldsymbol{\hat{y}}}
\newcommand \bzhat      {\boldsymbol{\hat{z}}}

\newcommand \beq        {\begin{equation}}
\newcommand \beqa	{\begin{eqnarray}}

\newcommand \cm         {\,{\rm cm}}

\newcommand \eeq	{\end{equation}}
\newcommand \eeqa	{\end{eqnarray}}

\newcommand \gtsim	{\gtrsim}		 

\newcommand \Ha 	{{\rm H}}

\newcommand \kms	{\,{\rm km~s}^{-1}}
\newcommand \kpc	{\,{\rm kpc}}

\newcommand \Lsol	{L_{\odot}}
\newcommand \ltsim	{\lesssim}		 

\newcommand \Ndip       {N}              
\newcommand \NH         {N_{\rm H}}
\newcommand \obs        {{\rm obs}}

\newcommand \pmax       {p_{\rm max}}

\newcommand \poromicro  {{\cal P}_{\rm micro}}

\newcommand \Rone       {{\cal A}}
\newcommand \Rtwo       {{\cal S}}

\newcommand \xtimes     {{\!\,\times\!\,}}

\newcommand \yr  	{\,{\rm yr}}

\newcommand \mm         {\,{\rm mm}}

\newcommand \aeff       {a_{\rm eff}}
\newcommand \aeffp      {a_{{\rm eff},p}}

\newcommand \Cran       {C_{\rm ran}}
\newcommand \CextPSA    {C_{\rm ext,PSA}}
\newcommand \Cpol       {C_{\rm pol}}
\newcommand \CpolPSA    {C_{\rm pol,PSA}}

\newcommand \falign     {f_{\rm align}}
\newcommand \lambdap    {\lambda_{p}}
\newcommand \lambdamax  {\lambda_{\rm max}}

\newcommand \lambdapobs {\lambda_{p,{\rm obs}}}
\newcommand \PhiPSA     {\Phi_{\rm PSA}}
\newcommand \Qabs       {Q_{\rm abs}}
\newcommand \Qext       {Q_{\rm ext}}
\newcommand \Qextran    {Q_{\rm ext,ran}}
\newcommand \QextPSA    {Q_{\rm ext,PSA}}
\newcommand \Qpol       {Q_{\rm pol}}
\newcommand \QpolPSA    {Q_{\rm pol,PSA}}
\newcommand \Qran       {Q_{\rm ext,ran}}
\newcommand \shape      {{\rm shape}}
\newcommand \sigmaobs   {\sigma_{\rm obs}}
\newcommand \sigmap     {\sigma_p}
\newcommand \Vgr        {V_{\rm gr}}

\newcommand{\btdnote}[1]{{\color{red}\it [btd note: #1]}}

\countdef\decade=200
\advance\decade by \year
\countdef\hours=201
\advance\hours by \time
\divide\hours by 60
\countdef\mins=202
\advance\mins by \hours
\multiply\mins by 60
\multiply\hours by 100
\countdef\miltime=203
\advance\miltime by \hours
\advance\miltime by \time
\advance\miltime by -\mins

\begin{document}

\title{%
        \vspace*{-2.0em}
        {\normalsize\rm {\it The Astrophysical Journal}, accepted}\\ 
        \vspace*{1.0em}
        {\bf Sensitivity of Polarization to Grain Shape: I. Convex Shapes}
	}

\author[0000-0002-0846-936X]{B.~T.~Draine}
\affiliation{Dept.\ of Astrophysical Sciences,
  Princeton University, Princeton, NJ 08544, USA}

\email{draine@astro.princeton.edu}

\begin{abstract}

Aligned interstellar grains produce polarized extinction (observed at
wavelengths from the far-ultraviolet to the mid-infrared), and
polarized thermal emission (observed at far-infrared and submm
wavelengths).  The grains must be quite nonspherical, but the actual
shapes are unknown.  The \emph{relative} efficacy for aligned grains
to produce polarization at optical vs.\ infrared wavelengths depends
on particle shape.  The discrete dipole approximation is used to
calculate polarization cross sections for 20 different convex shapes,
for wavelengths from $0.1\micron$ to $100\micron$, and grain sizes
$\aeff$ from $0.05\micron$ to $0.3\micron$.  Spheroids, cylinders,
square prisms, and triaxial ellipsoids are considered.  Minimum aspect
ratios required by the observed starlight polarization are determined.
Some shapes can also be ruled out because they provide too little or
too much polarization at far-infrared and sub-mm wavelengths.  The
ratio of $10\micron$ polarization to integrated optical polarization
is almost independent of grain shape, varying by only $\pm8\%$ among
the viable convex shapes; thus, at least for convex grains,
uncertainties in grain shape cannot account for the discrepancy
between predicted and observed 10$\micron$ polarization toward Cyg
OB2-12.

\end{abstract}


\section{Introduction
         \label{sec:intro}}

Since the discovery of starlight polarization over 70 years ago
\citep{Hiltner_1949a,Hall_1949}, polarization has become a valuable
tool for studying both the physical properties of interstellar dust
and the structure of the Galactic magnetic field.  Initially
unpolarized starlight propagating through the interstellar medium
(ISM) becomes linearly polarized as a result of polarization-dependent
extinction (linear dichroism) by the aligned dust grains in the ISM.
While the physics of dust grain alignment is not yet fully understood,
early investigations \citep{Davis+Greenstein_1951} showed how spinning
dust grains could become aligned with shortest axes tending to be
parallel to the local magnetic field $\bB_0$.  Subsequent studies
identified a number of important physical processes that were
initially overlooked \citep[see the review
  by][]{Andersson+Lazarian+Vaillancourt_2015}, but it remains clear
that in the diffuse ISM the magnetic field establishes the direction
of grain alignment, with dust grains tending to align with their short
axes parallel to $\bB_0$.

The aligned grains responsible for starlight polarization also produce
polarized thermal emission at far-infrared and submillimeter
wavelengths, now mapped over the full sky \citep{Planck_2015_I_2016}.
The fractional polarization of the thermal emission measured by Planck
\citep{Planck_2018_XII} and Blastpol
\citep{Ashton+Ade+Angile+etal_2018} is nearly independent of
wavelength from $250\micron$ to $3\mm$ \citep[see Figure 13
  in][]{Hensley+Draine_2021}, motivating the ``astrodust'' model
\citep{Hensley+Draine_2023}, where the extinction and emission are
dominated by a single composite grain material, incorporating both
silicate and nonsilicate constituents.  In the astrodust model, a
single grain type accounts for the observed polarization across the
entire wavelength range, from far-UV (FUV) to far-infrared (FIR).

The present paper is a study of the optics of the small particles
responsible for starlight polarization and polarized FIR
emission.  For reasons of analytic and computational convenience,
spheroids -- the simplest nonspherical shape -- are often assumed when
modeling polarization by interstellar grains
\citep[e.g.][]{Kim+Martin_1995b, Draine+Fraisse_2009,
  Siebenmorgen+Voshchinnikov+Bagnulo_2014,
  Guillet+Fanciullo+Verstraete+etal_2018, Draine+Hensley_2021a,
  Hensley+Draine_2023}.
However, interstellar grains are not ideal spheroids. Here we use the
discrete dipole approximation (DDA) to calculate accurate extinction
and polarization cross sections for a variety of grain shapes, to
compare with spheroids, and to find shapes consistent with
observational constraints.

Starlight polarization in the ISM is due to aligned nonspherical
grains with sizes comparable to the vacuum wavelength $\lambda$;
scattering is important.  At mid-IR and submm wavelengths, however,
the grain optics enters a different regime: the grains are small
compared to $\lambda$, absorption dominates, and scattering is
negligible.  We study the grain optics from the FUV ($0.1\micron$) to
the FIR ($100\micron$).  The following questions are investigated:
\begin{enumerate}
\item What shapes are consistent with the observed \emph{strength} of
  starlight polarization at optical wavelengths?
\item What shapes are compatible with the observed \emph{width} of the
  starlight polarization as a function of $\lambda$?
\item For a given strength of starlight polarization, how does the
  fractional polarization of the submm emission depend on grain shape?
  Do small-scale features such as sharp edges or corners affect the
  polarizing ability at optical wavelengths (grain size
  $\sim$$\lambda$) relative to very long wavelengths (size
  $\ll\lambda$)?
\item Can the extinction and polarization properties of nonspheroidal
  shapes be adequately approximated by spheroids with suitable axial
  ratio?
\item For a given strength of starlight polarization, how does
  $10\micron$ polarization depend on the grain shape?  A recent
  measurement by \citet{Telesco+Varosi+Wright+etal_2022} of starlight
  polarization at $10\micron$ found weaker polarization than had been
  predicted by grain models that assumed spheroidal grains.  Can
  nonspheroidal shapes explain this?
\end{enumerate}

The present paper examines a number of convex shapes.  Included are
spheroidal and ellipsoidal grains with smoothly rounded surfaces, and
also shapes (cylinders and square prisms) with sharp edges and
corners.  We consider both elongated and flattened shapes.  Some of
the studied shapes are found to be compatible with observations of
interstellar polarization, but others are not.

\begin{enumerate}
\item For spheroids, cylinders and square prisms we determine the
  minimum aspect ratios that are compatible with the observed strength
  of starlight polarization.
\item Some of the considered shapes could reproduce the integrated
  strength of starlight polarization, but are ruled out because the
  resulting polarization profile would be too broad.
\item Some of the considered geometries (e.g., elongated cylinders)
  would overproduce the observed fractional polarization at FIR
  wavelengths if required to reproduce the observed starlight
  polarization, while other geometries (e.g., the flattened cylinder)
  would fall short of the observed FIR polarization.
\item We show that grain shape does have systematic effects on the
  relative amounts of polarization at optical and infrared
  wavelengths; however, the effects are relatively modest, so that
  spheroids remain a useful approximation.
\item We find that the predicted ratio of $10\micron$ polarization to
  optical starlight polarization is only weakly dependent on grain
  shape.
\end{enumerate}

The target shapes considered are presented in section
\ref{sec:targets}.  The dielectric functions used here are discussed
in section \ref{sec:dielectric}.  The treatment of absorption and
scattering by both axisymmetric and non-axisymmetric grains is
outlined in section \ref{sec:axisymm}.  The accuracy of the DDA is
examined in section \ref{sec:DDA}.  Extinction and polarization cross
sections are presented in section \ref{sec:Q}.  Observational
constraints provided by the starlight polarization are summarized in
section \ref{sec:starlight_pol}.  The characteristic wavelength
$\lambdap$ and profile width $\sigmap$ for starlight polarization
are evaluated in section \ref{sec:single-size}, and the polarization
efficiency integral $\Phi$ is calculated for the considered shapes in
section \ref{sec:SPEI}.

The results are discussed in section \ref{sec:discuss}, and summarized
in section \ref{sec:summary}.  Certain technical matters pertaining to
application of the DDA are discussed in the Appendices.

\section{\label{sec:targets}
         Target Geometry}

\newcommand{\figwidthb}{4.2cm}
\newcommand{\figheight}{4.0cm}
\newcommand{\fclipl}{2.8cm}
\newcommand{\fclipb}{7.0cm}
\newcommand{\fclipr}{1.5cm}
\newcommand{\fclipt}{4.4cm}
\begin{figure}[b]
\begin{center}
{\bf Flattened Shapes}\hspace*{5.0cm}{\bf Elongated Shapes}\\

\includegraphics[angle=0,height=\figheight,
                 clip=true,trim=2.9cm 7.0cm 1.5cm 4.5cm]
{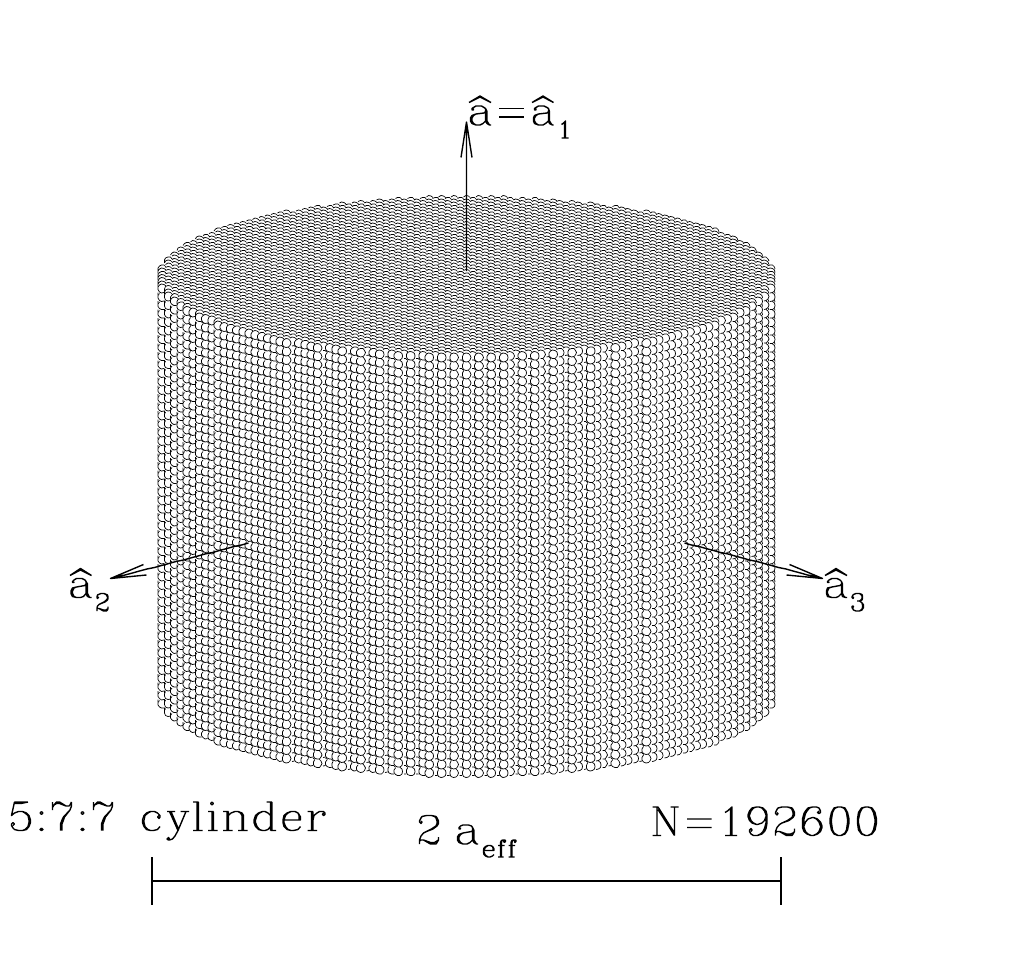}
\includegraphics[angle=0,height=\figheight,
                 clip=true,trim=2.9cm 7.0cm 1.5cm 4.5cm]
{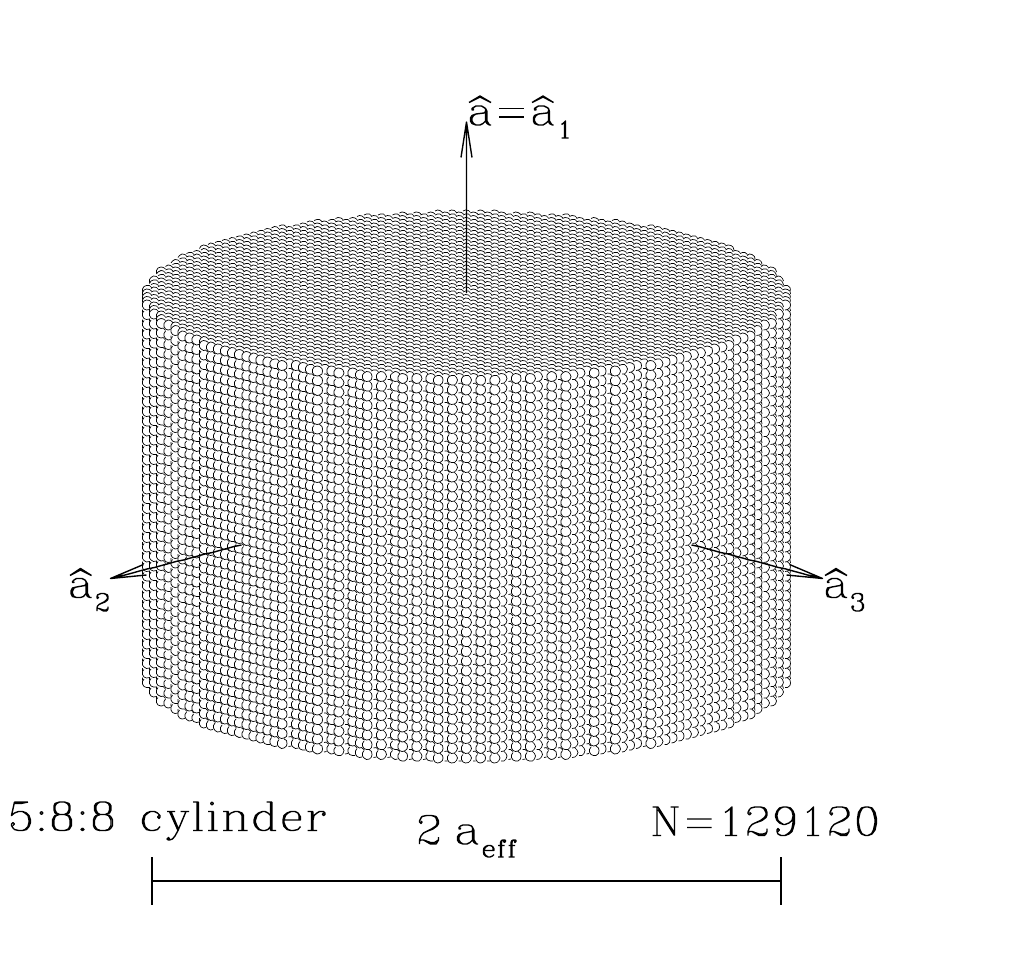}
\includegraphics[angle=0,height=\figheight,
                 clip=true,trim=2.5cm 7.0cm 1.5cm 4.5cm]
{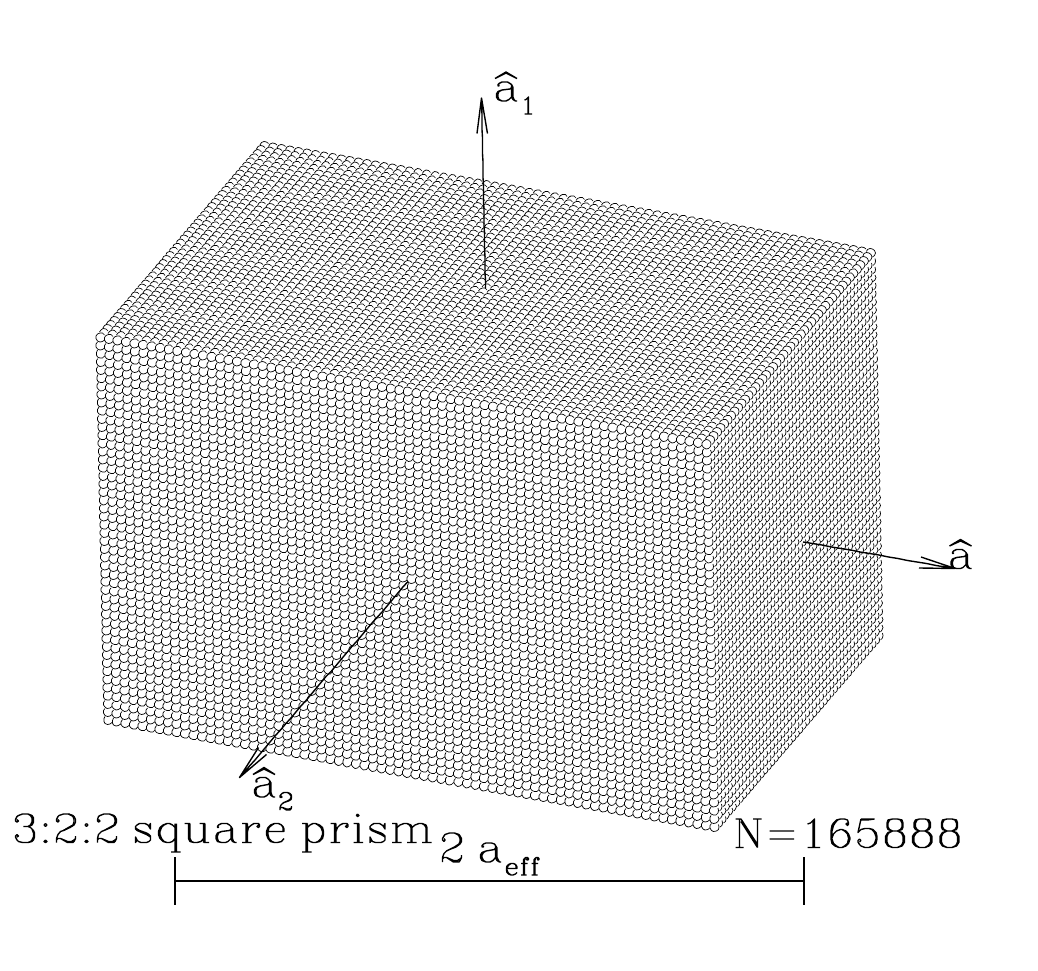}
\includegraphics[angle=0,height=\figheight,
                 clip=true,trim=2.5cm 7.0cm 1.5cm 4.5cm]
{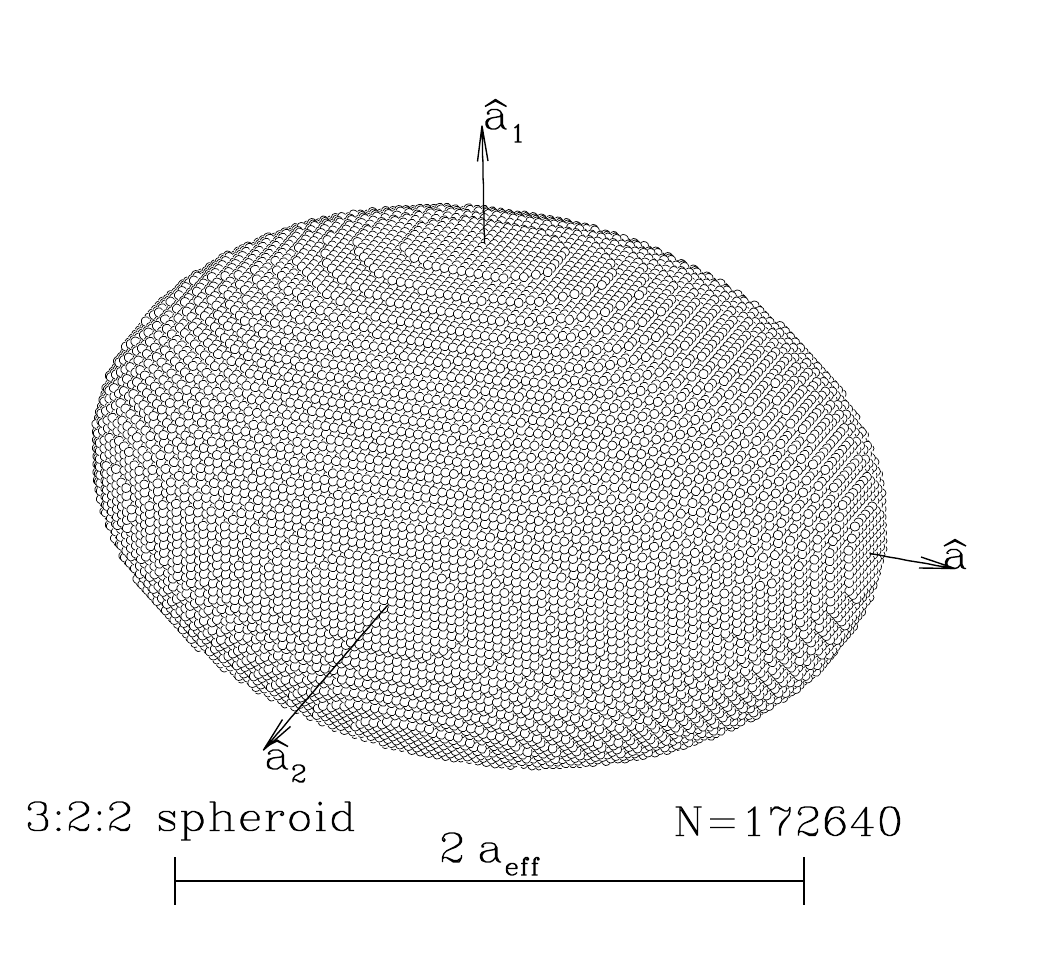}\\


\includegraphics[angle=0,height=\figheight,
                 clip=true,trim=2.9cm 7.0cm 1.5cm 4.5cm]
{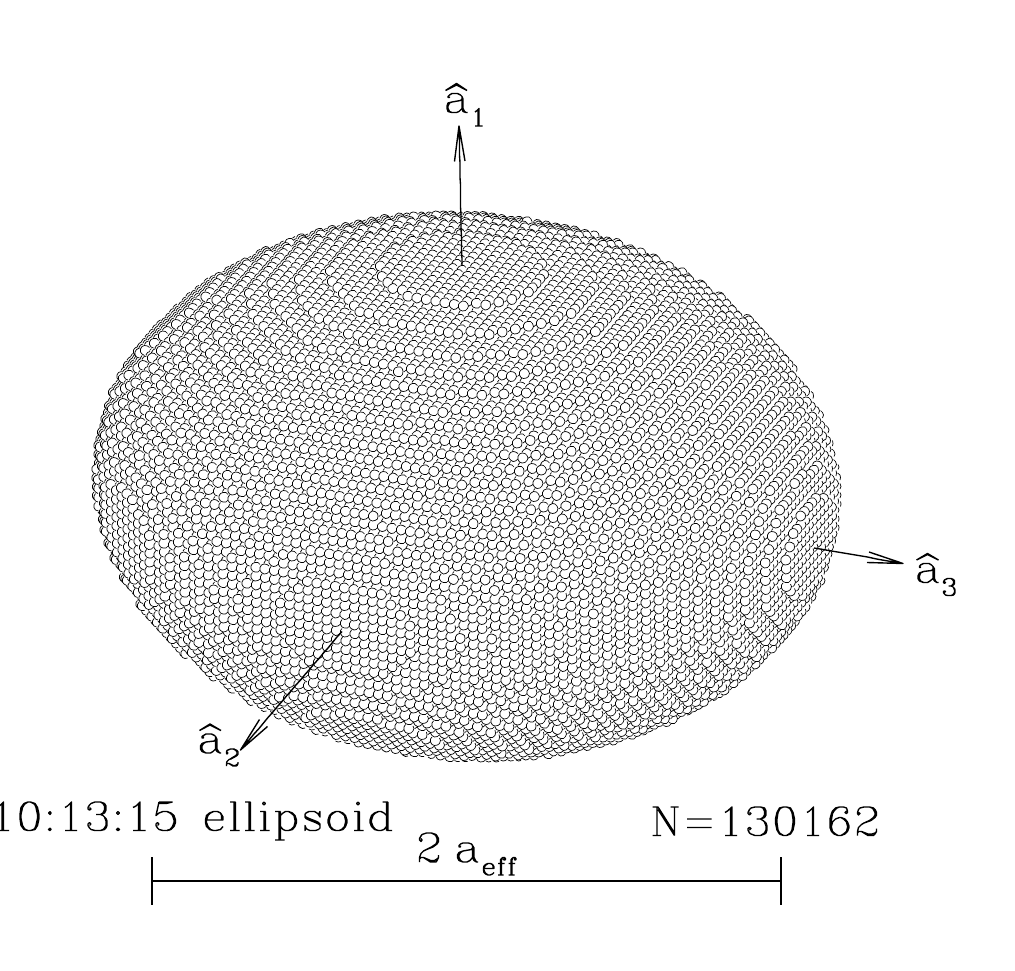}
\includegraphics[angle=0,height=\figheight,
                 clip=true,trim=2.5cm 7.0cm 1.5cm 4.5cm]
{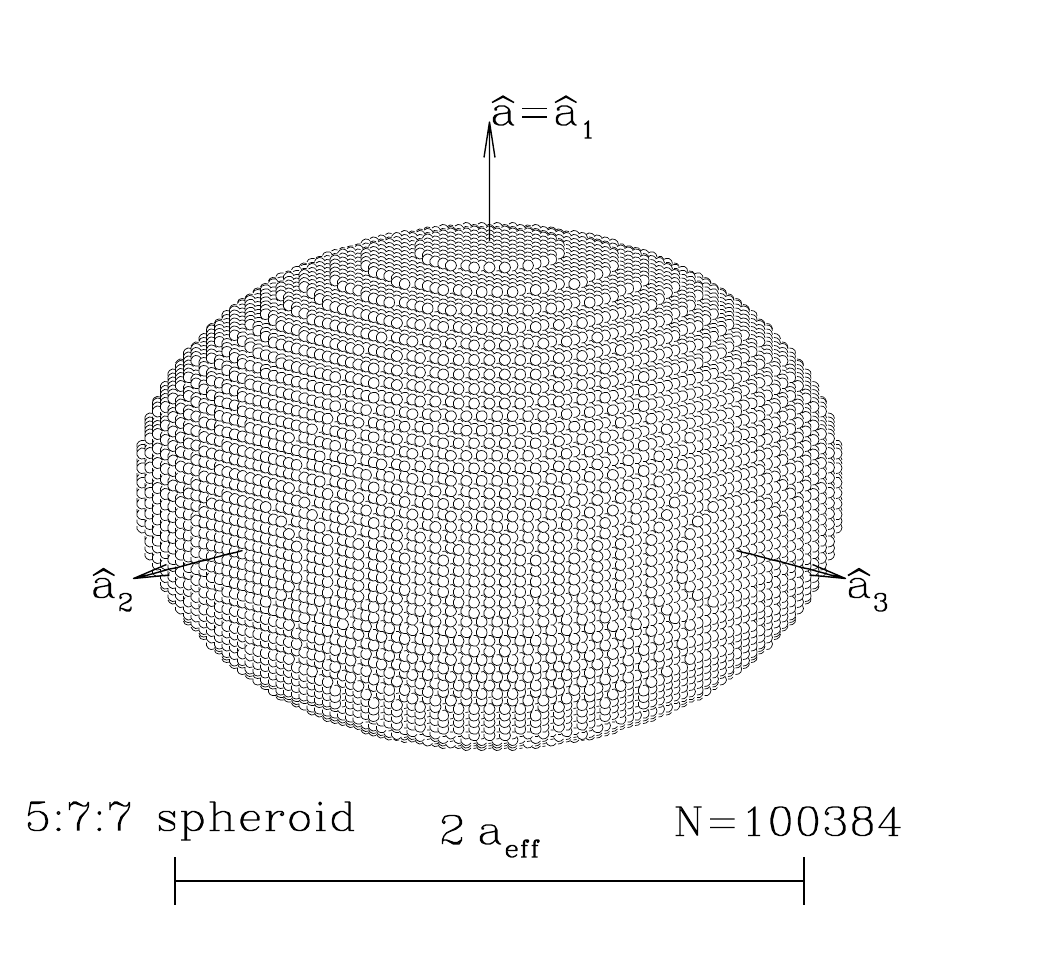}
\includegraphics[angle=0,height=\figheight,
                 clip=true,trim=2.9cm 7.0cm 1.5cm 4.5cm]
{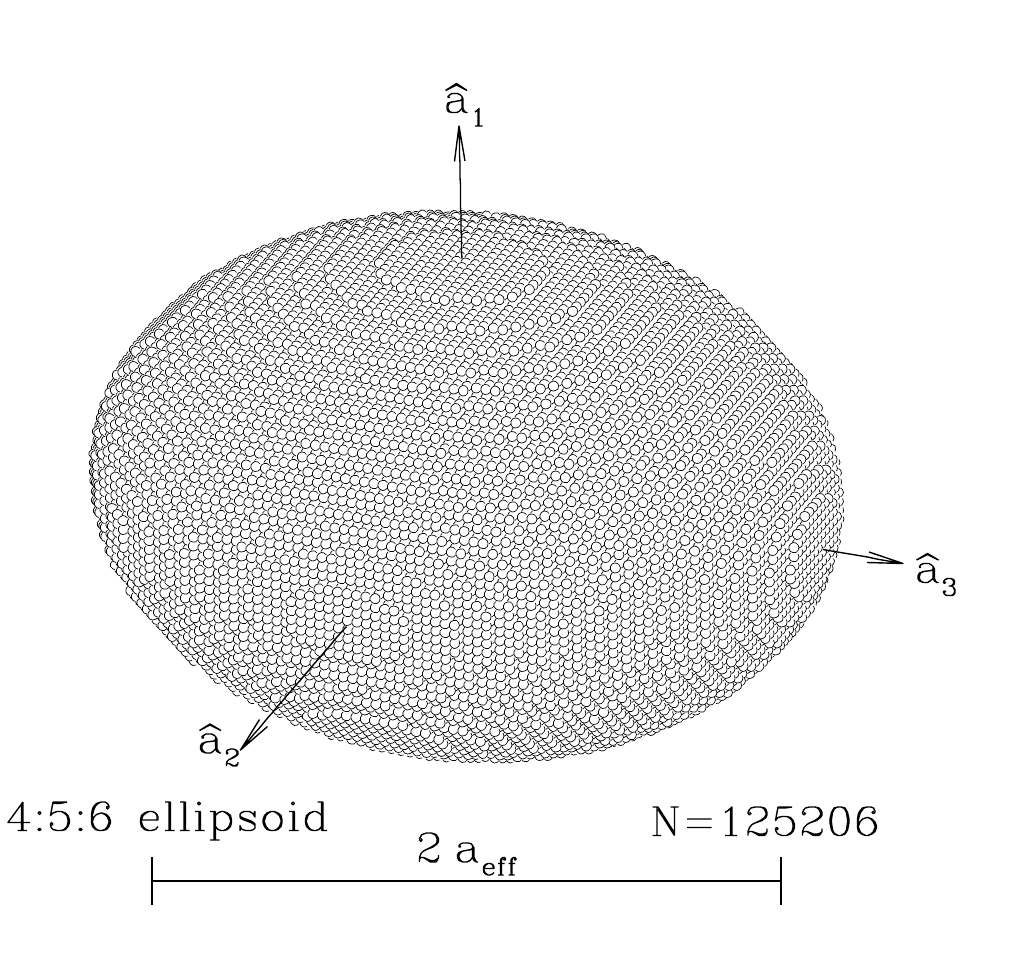}
\includegraphics[angle=0,height=\figheight,
                 clip=true,trim=2.9cm 7.0cm 1.5cm 4.5cm]
{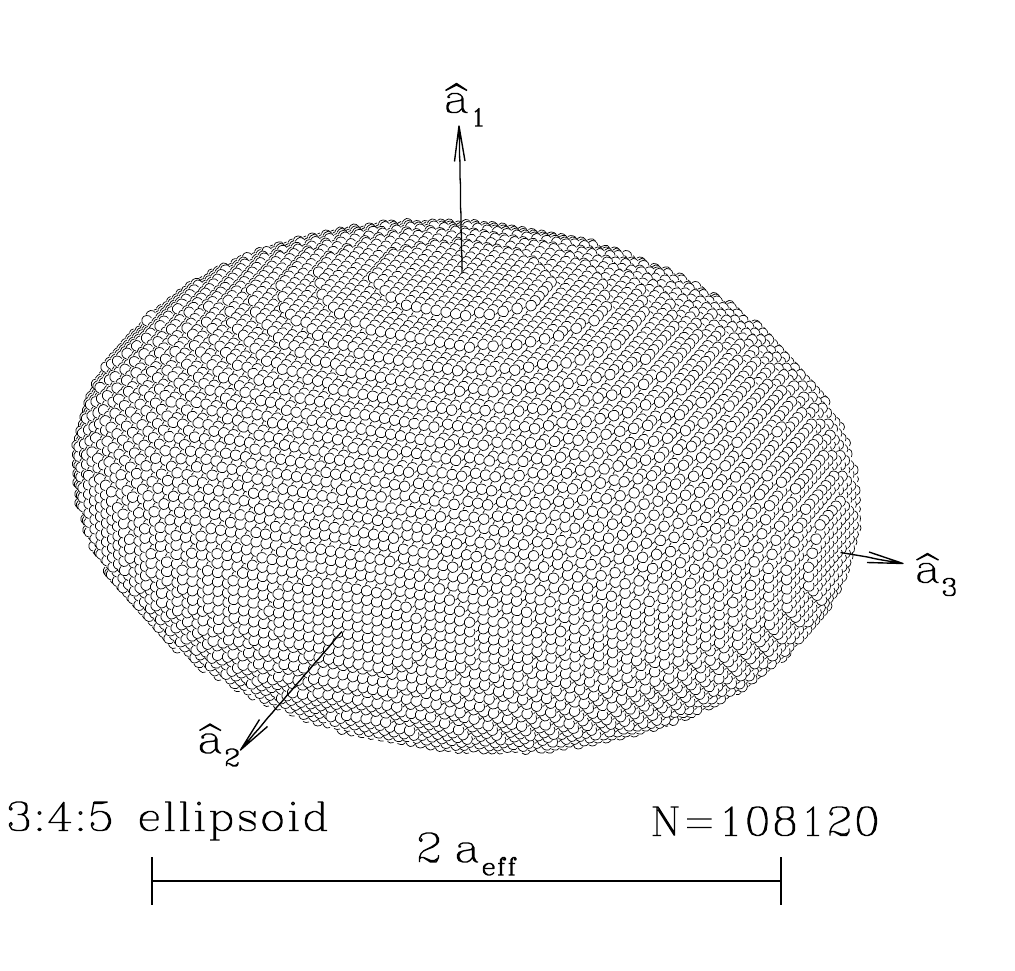} \\


\includegraphics[angle=0,height=\figheight,
                 clip=true,trim=2.5cm 7.0cm 1.5cm 4.5cm]
{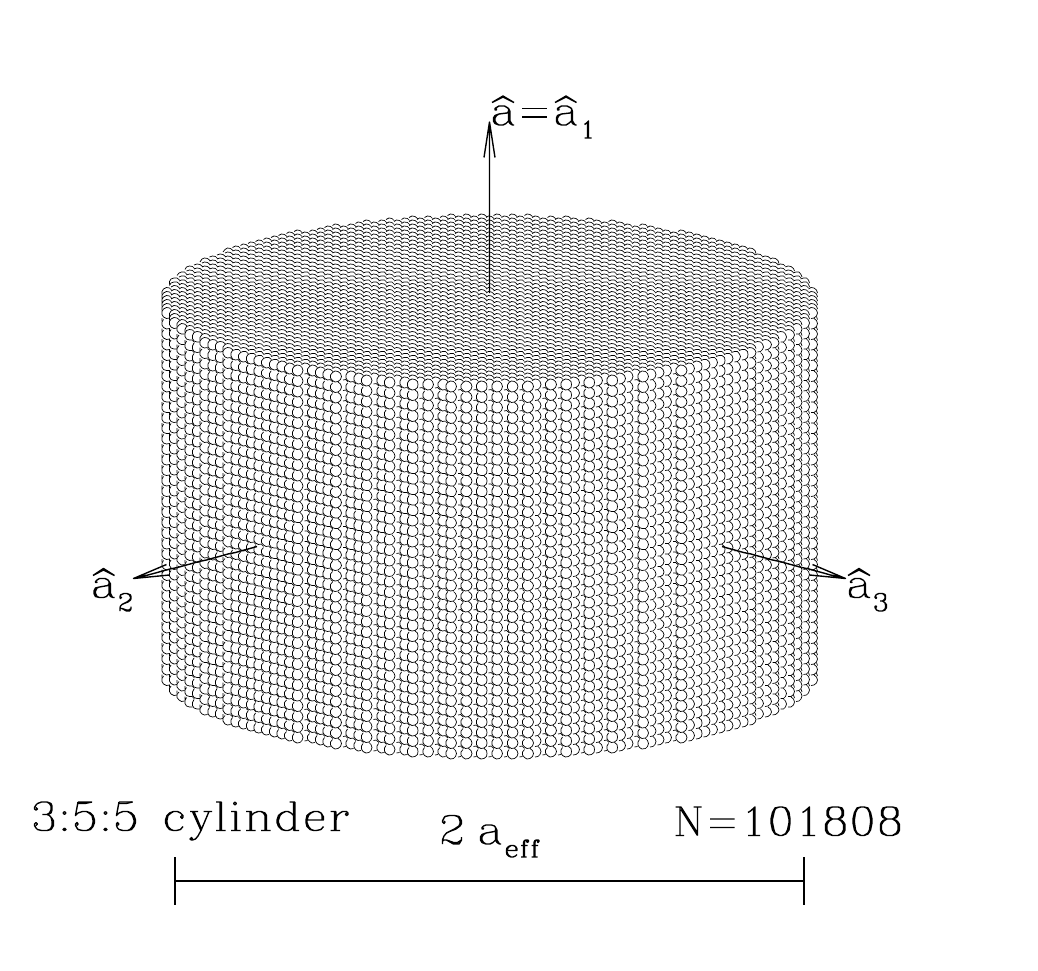}
\includegraphics[angle=0,height=\figheight,
                 clip=true,trim=2.5cm 7.0cm 1.5cm 4.5cm]
{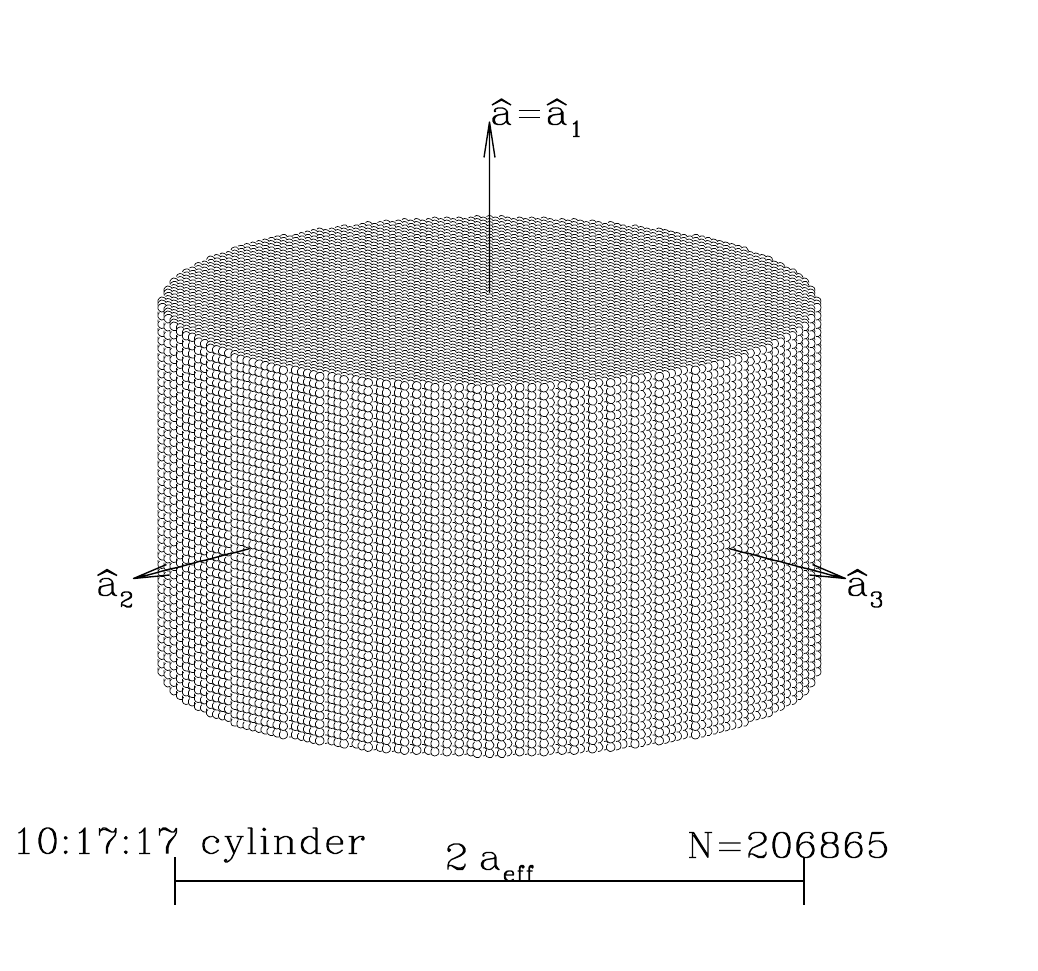}
\includegraphics[angle=0,height=\figheight,
                 clip=true,trim=2.5cm 7.0cm 1.5cm 4.5cm]
{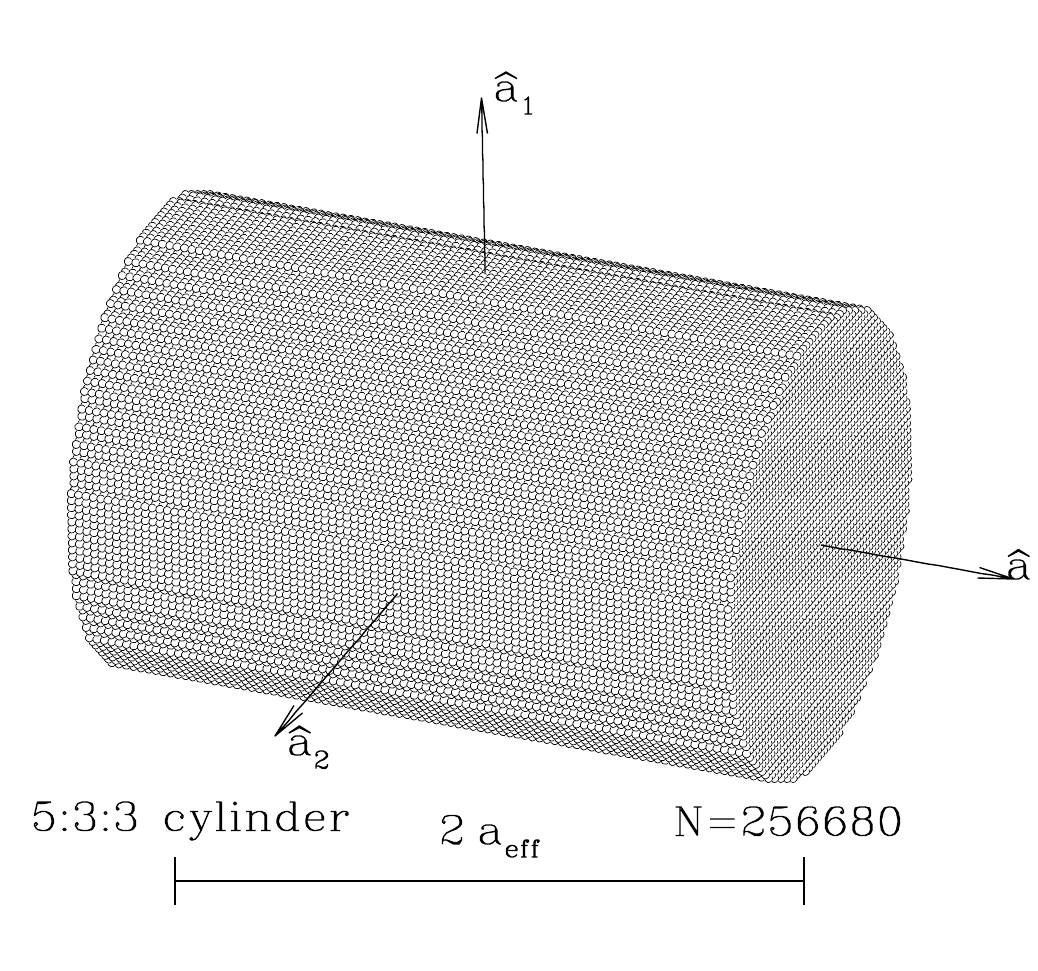}
\includegraphics[angle=0,height=\figheight,
                 clip=true,trim=2.7cm 7.0cm 1.5cm 4.5cm]
{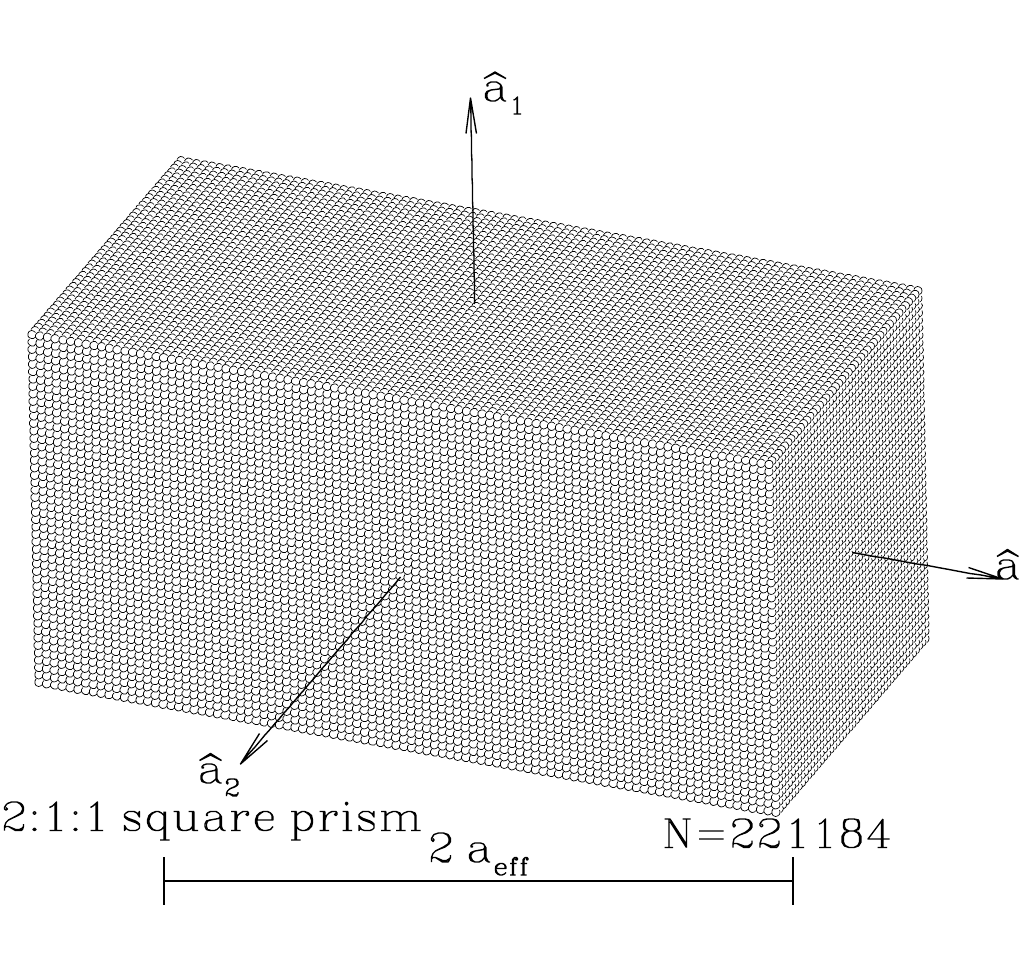}\\
\includegraphics[angle=0,height=\figheight,
                 clip=true,trim=2.5cm 7.0cm 1.5cm 4.5cm]
{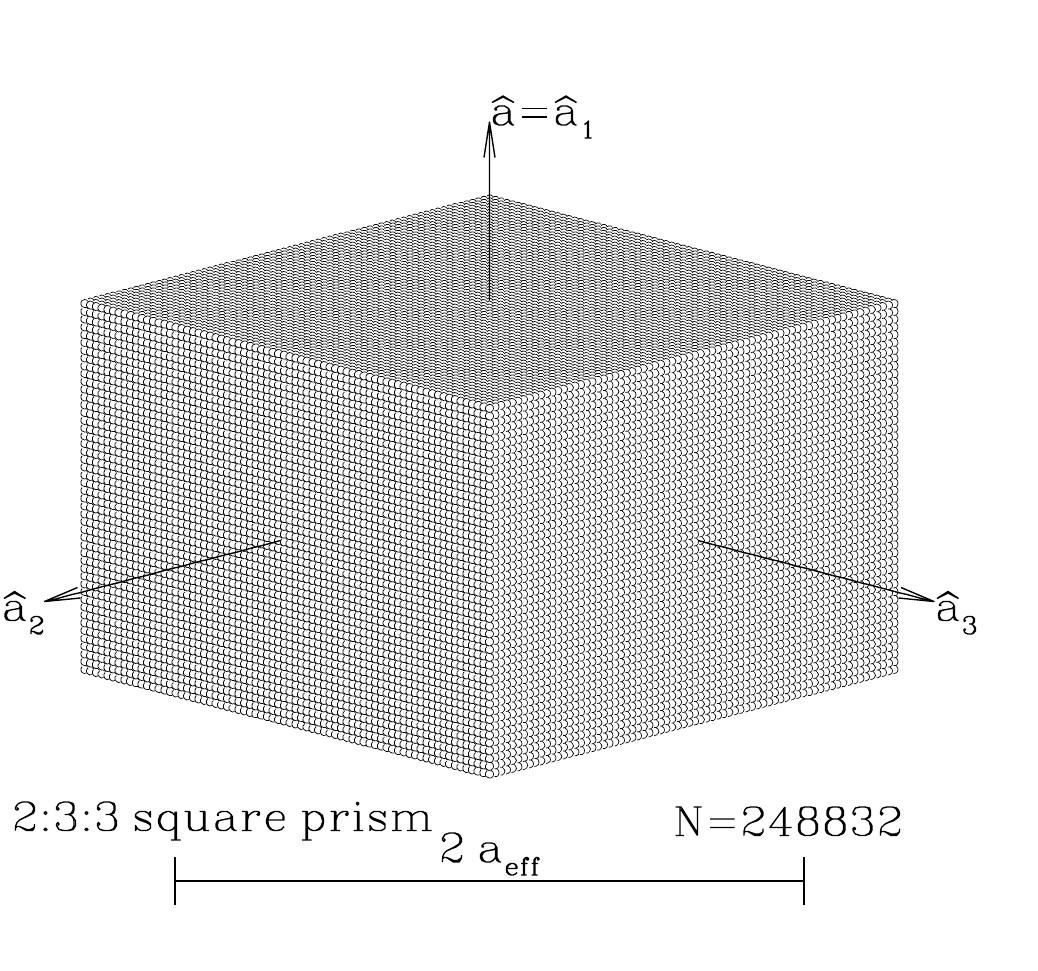}
\includegraphics[angle=0,height=\figheight,
                 clip=true,trim=2.5cm 7.0cm 1.5cm 4.5cm]
{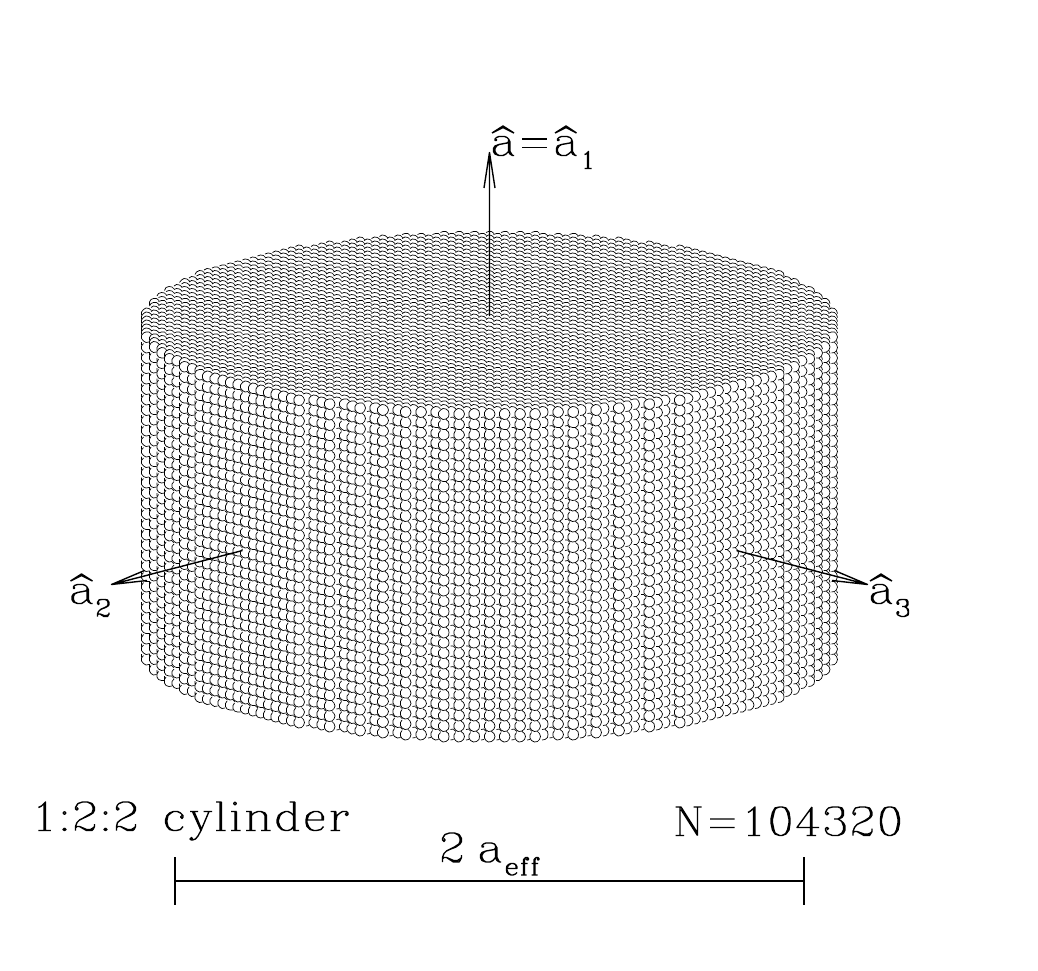}
\includegraphics[angle=0,height=\figheight,
                 clip=true,trim=2.5cm 7.0cm 1.5cm 4.5cm]
{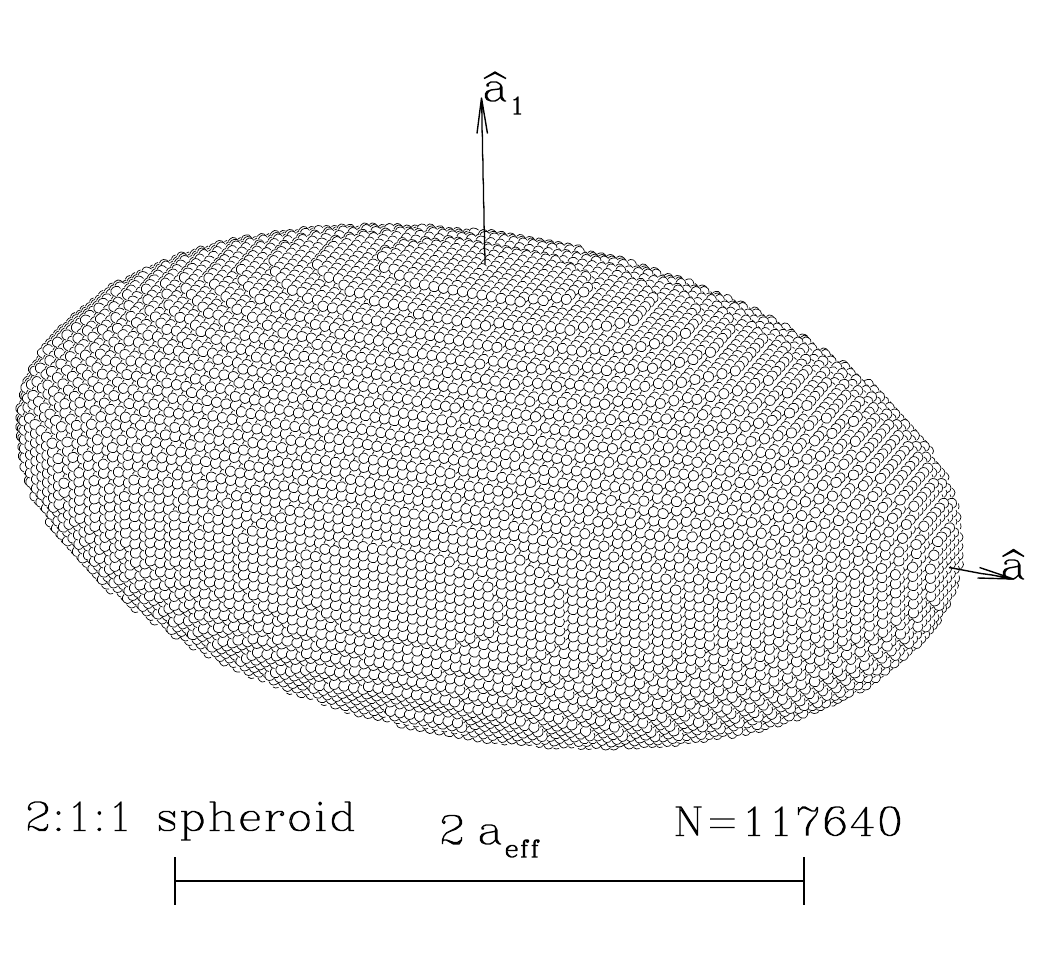}
\includegraphics[angle=0,height=\figheight,
                 clip=true,trim=2.5cm 7.0cm 1.5cm 4.5cm]
{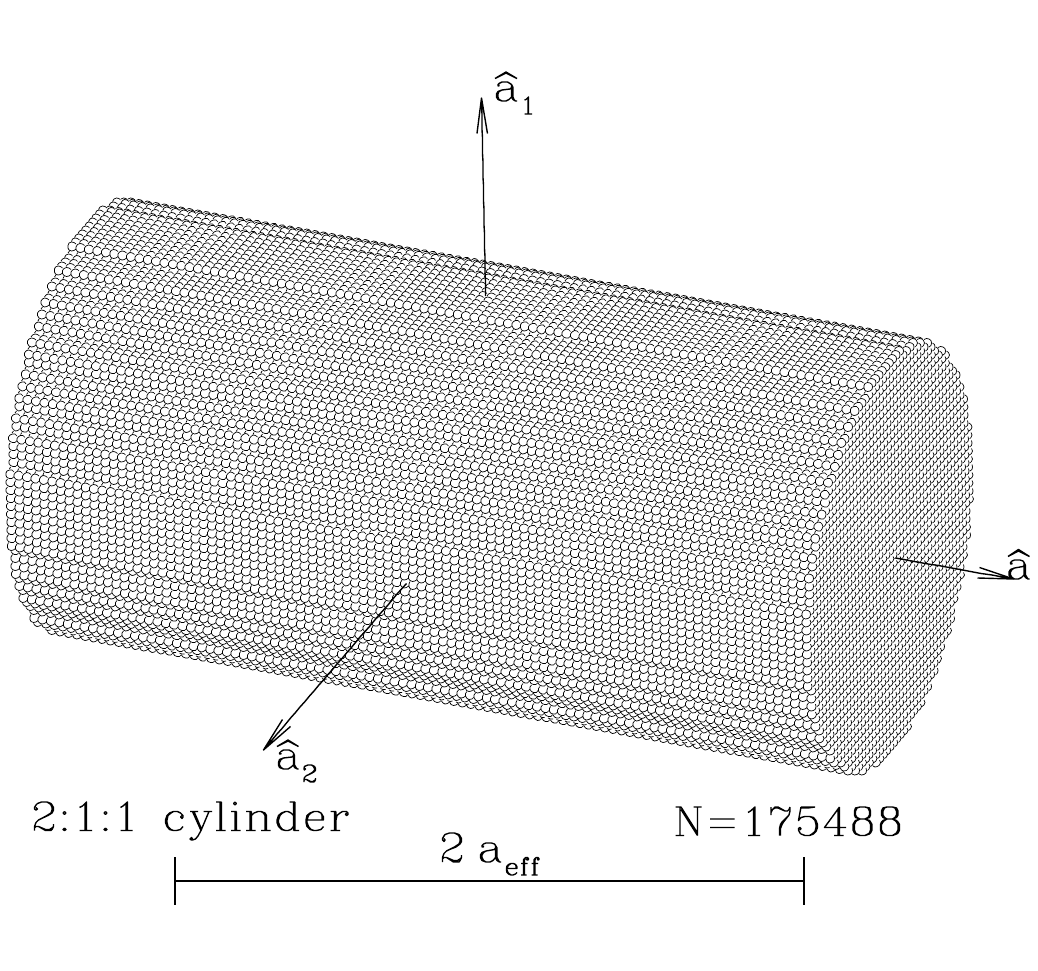}\\ 
\includegraphics[angle=0,height=\figheight,
                 clip=true,trim=2.5cm 7.0cm 1.6cm 4.5cm]
{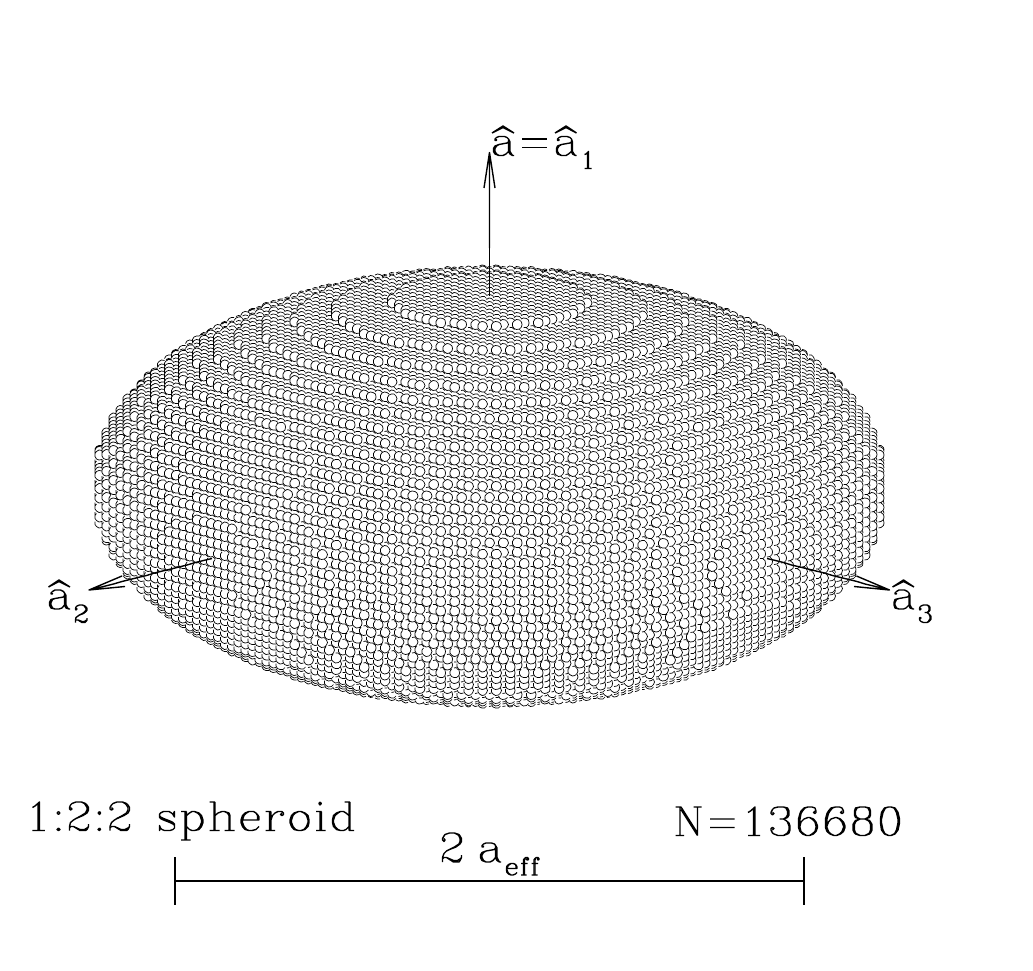}
\includegraphics[angle=0,height=\figheight,
                 clip=true,trim=2.0cm 7.0cm 1.5cm 4.5cm]
{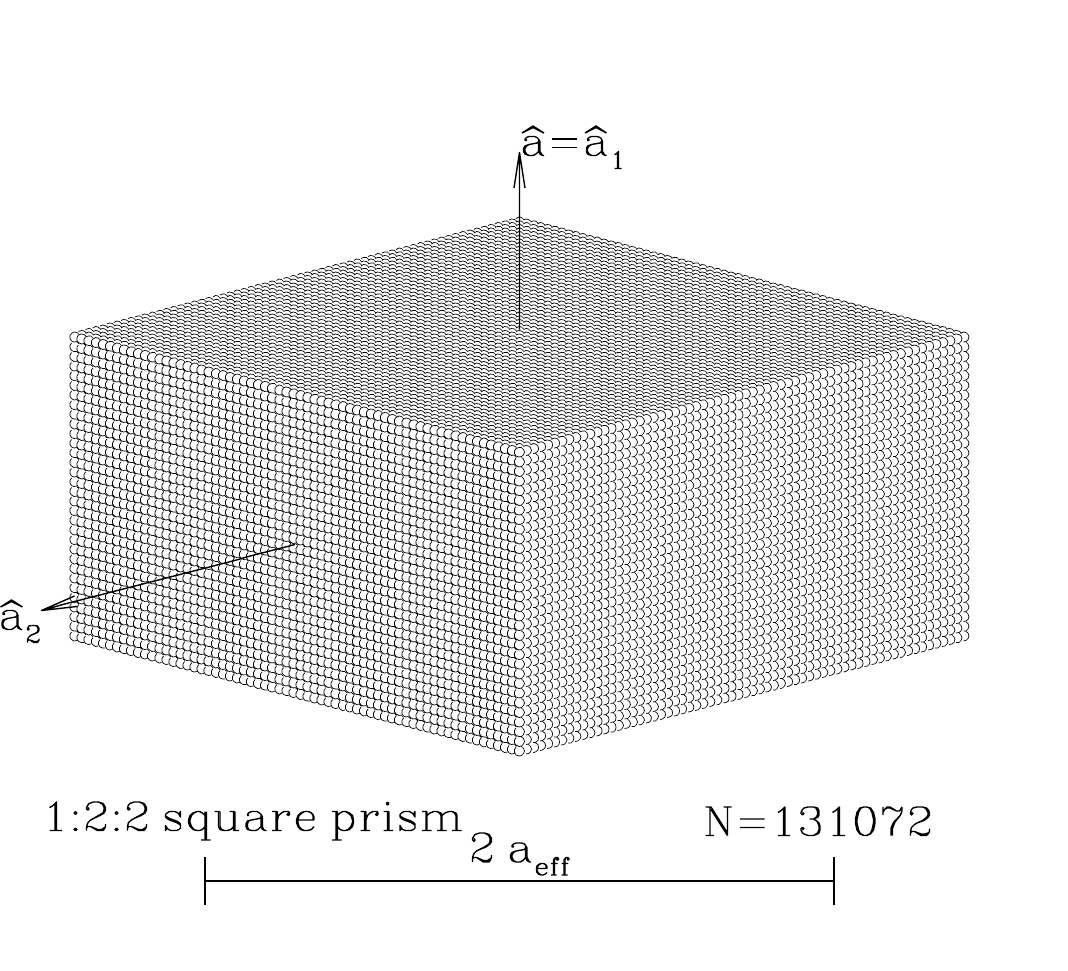}
\includegraphics[angle=0,height=\figheight,
                 clip=true,trim=2.3cm 7.0cm 1.1cm 4.5cm]
{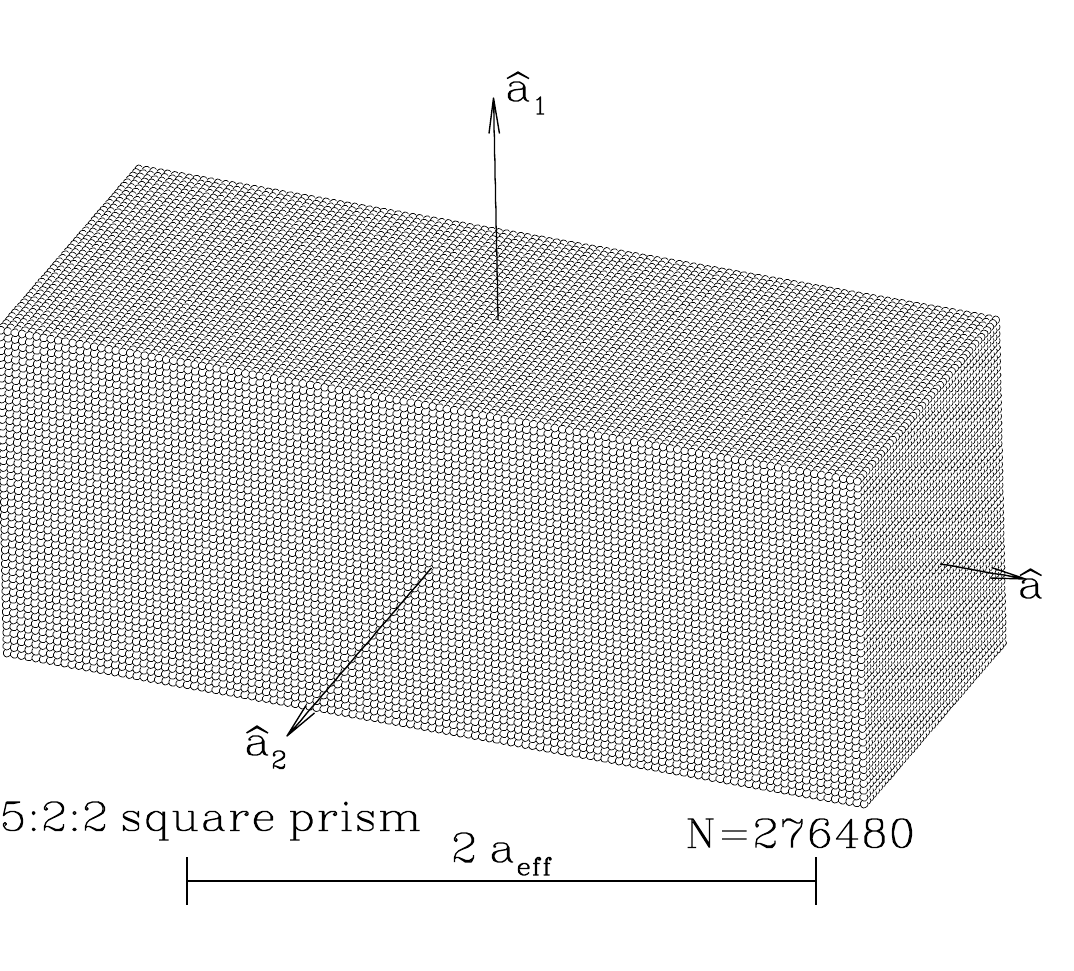}
\includegraphics[angle=0,height=\figheight,
                 clip=true,trim=1.6cm 7.0cm 1.1cm 4.5cm]
{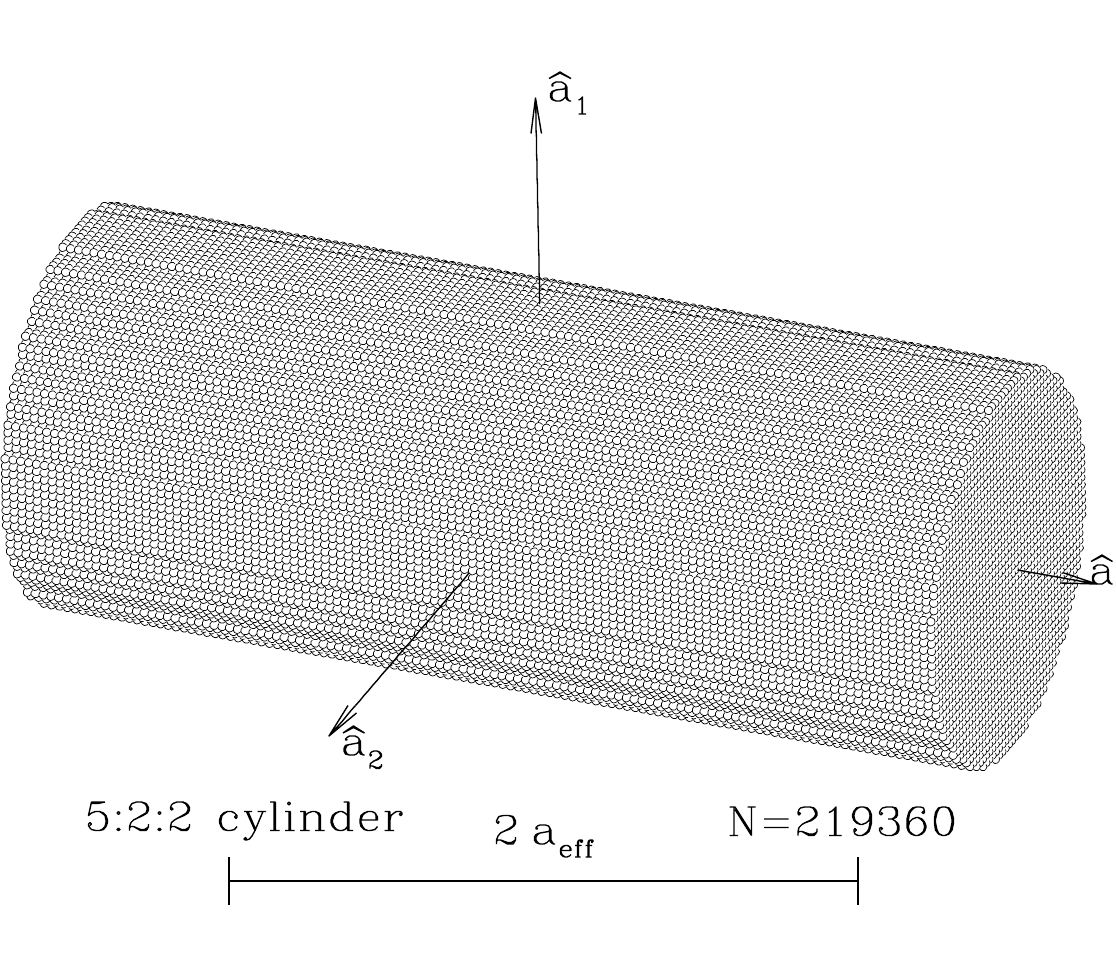}\\

\caption{\label{fig:shapes}\footnotesize 
  20 convex shapes in this study.  Scale bars show $2\aeff$
  $\bahat_1$ is the principal
  axis of largest moment of inertia; grains are asssumed to spin
  around this axis.  For grains with rotational symmetry,
  $\bahat$ is the symmetry axis: 12 of the examples
  are axisymmetric; the 5 square prisms have 4-fold rotational
  symmetry.
   }
\end{center}
\end{figure}

\citet{Draine+Hensley_2021a} found that prolate spheroids with 2:1:1
axial ratios, and oblate spheroids with 5:7:7 axial ratios, were
allowed by observations of starlight polarization, whereas spheroids
with less extreme axial ratios could not provide sufficient
polarization.  The present study includes these two spheroidal shapes,
plus other shapes with various degrees of elongation or flattening.

The target size is specified by the radius of an equal-volume sphere:
$\aeff\equiv (3V/4\pi)^{1/3}$, where $V$ is the solid volume.  Optical
cross sections depend on the dielectric function, grain shape, and the
ratio $\aeff/\lambda$.

Polarization arises from asymmetry in the grain shape.  We can
characterize the asymmetry of a given shape by the ratio of a long
dimension to a short dimension.  We can also characterize the grain
asymmetry using the moment of inertia tensor, assuming the solid
material in the grain to have a single density $\rho$.  Define
\beq \label{eq:alpha_j}
\alpha_j\equiv \frac{I_j}{0.4 \rho V \aeff^2}
~~~,
\eeq
where $I_1\geq I_2\geq I_3$ are the eigenvalues of the moment of
inertia tensor.  A sphere has $\alpha_1=\alpha_2=\alpha_3=1$; all
other shapes have $\alpha_1>1$, and $\alpha_1+\alpha_2+\alpha_3>3$.

One measure of the asymmetry of the grain is provided by the ``asymmetry''
parameter
\beq \label{eq:R1}
\Rone \equiv \frac{\alpha_1}{\left(\alpha_2 \alpha_3\right)^{1/2}}\geq 1
~~~.
\eeq
A sphere or a cube has $\Rone=1$, but $\Rone$ becomes large for very
flattened or elongated shapes.  The twenty shapes in this study have
$1.19<\Rone<2.16$.

Flattening or elongation of the shape can be characterized by the 
``stretch'' parameter
\beq \label{eq:R2}
\Rtwo \equiv \frac{\alpha_2}{\left(\alpha_1 \alpha_3\right)^{1/2}}
~~~,
\eeq
with $\frac{1}{\sqrt{2}}<\Rtwo < 1$ for flattened shapes, and $\Rtwo >
1$ for elongated shapes.  Extreme flattening corresponds to
$\Rtwo\rightarrow\frac{1}{\sqrt{2}}$, and extreme elongation to
$\Rtwo\gg 1$.  The shapes in this study have $0.79<\Rtwo<2.16$.

Four shape classes are considered:

\begin{enumerate}

\item {\bf Spheroids}: Axial length $L$, diameter $D$,
  $\aeff=(D^2L/8)^{1/3}$.  We consider two flattened (oblate) spheroids
  ($D/L=1.4, 2.0$) and two elongated (prolate) spheroids ($L/D=1.5, 2.0$).

\item {\bf Cylinders}: Axial length $L$, diameter $D$,
  $\aeff=(3D^2L/16)^{1/3}$.  Elongated ($\Rtwo>1$) cylinders have
  $L/D>\sqrt{3/4}=0.866$; flattened ($\Rtwo<1$) cylinders have
  $L/D<0.866$.  We consider five flattened cylinders ($D/L=1.4, 1.6,
  1.67, 1.7, 2$) and three elongated cylinders ($L/D=1.5, 2.0, 2.5$).

\item {\bf Square Prisms}: Axial length $L$, width $W$,
  $\aeff=(3W^2L/4\pi)^{1/3}$, and fourfold rotational symmetry around
  the axis $\bahat$. We consider two flattened shapes ($W/L=1.5, 2.0$)
  and three elongated shapes ($L/W=1.5, 2.0, 2.5$).

\item {\bf Triaxial Ellipsoids}: Axial lengths $L_1<L_2<L_3$,
   $\aeff=(L_1 L_2 L_3/8)^{1/3}$.  We consider three examples:
  $L_1$:$L_2$:$L_3=$ 4:5:6, 10:13:15, and 3:4:5.  These all
  have $\Rtwo\approx 1$, and thus are neither very flattened nor
  very elongated.

\end{enumerate}
Twenty convex shapes, shown in Figure \ref{fig:shapes}, have been
studied.  Ten are ``flattened'' ($\Rtwo < 1$), and ten are
``elongated'' ($\Rtwo > 1$).  $\alpha_j$, $\Rone$, and $\Rtwo$ for the
shapes in this study can be found in Table \ref{tab:Phi table}.

\section{\label{sec:dielectric}
         Dielectric Function}

\begin{figure}
\begin{center}
\includegraphics[angle=0,width=8.0cm,
                 clip=true,trim=0.5cm 0.5cm 0.5cm 0.5cm]
{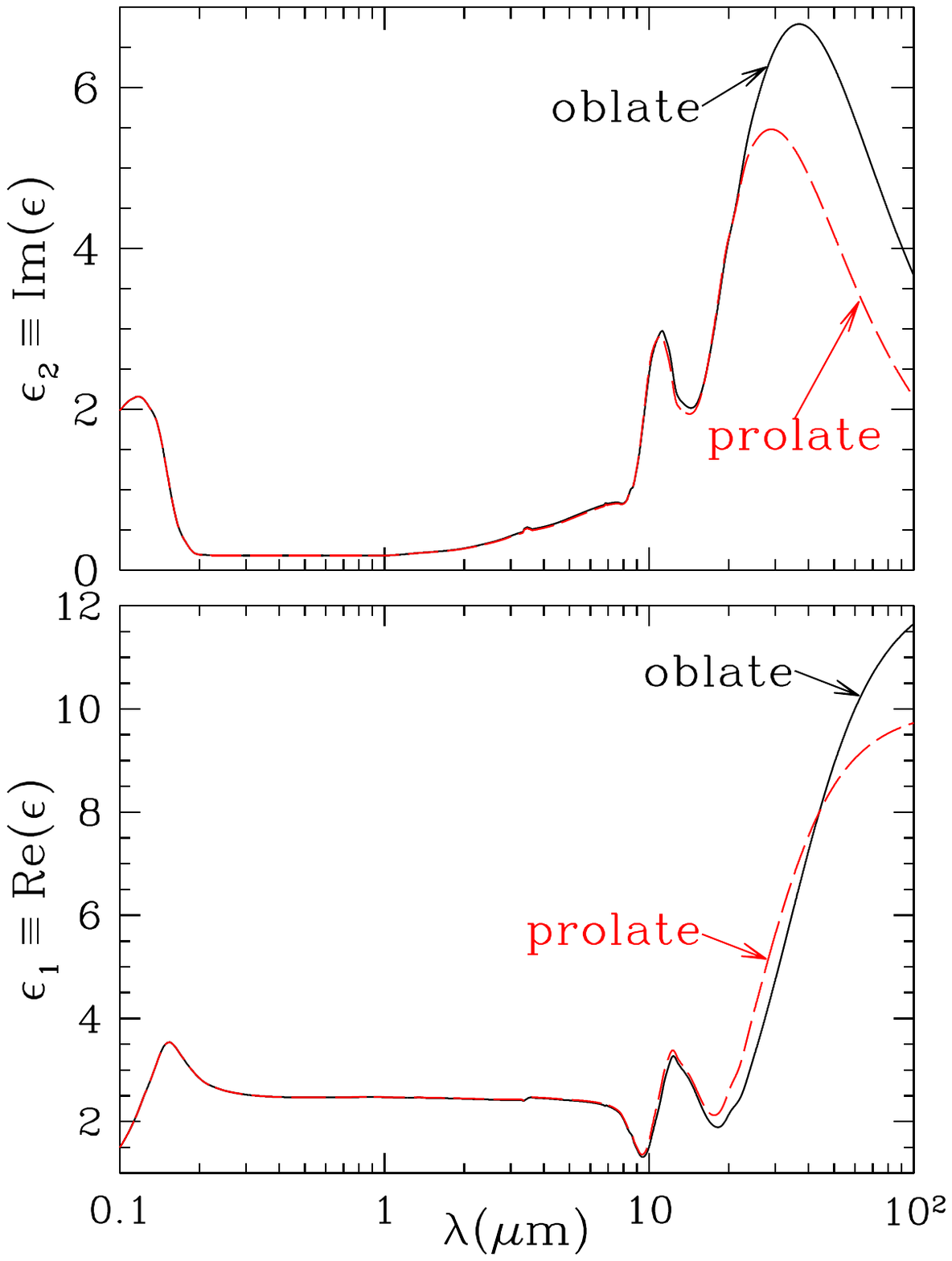}
\caption{\label{fig:diel}\footnotesize Black curves:
  $\epsilon(\lambda)$ used for flattened shapes.  Red curves:
  $\epsilon(\lambda)$ used for elongated shapes.  See text.
  }
\end{center}
\end{figure}

The complex dielectric function
$\epsilon(\lambda)=\epsilon_1+i\epsilon_2$ characterizes the response
of a substance to a local electric field oscillating at frequency
$c/\lambda$. The dielectric function of interstellar grain material
remains uncertain.  \citet{Draine+Hensley_2021a} (hereafter DH21a)
obtained dielectric functions for a hypothetical material
(``astrodust'') intended to represent the bulk of the interstellar
grain material.  The derived dielectric function depended on the
assumed grain shape and porosity; DH21a considered several different
spheroidal shapes and porosities.  The mid- and far-infrared
dielectric function was ``derived'' by requiring that
$\epsilon(\lambda)$ obey the Kramers-Kronig relations and that the
dust model reproduce the observed infrared and submm opacity of the
diffuse ISM.

Figure \ref{fig:diel} shows the effective dielectric function
$\epsilon(\lambda)$ derived by DH21a if astrodust is assumed to have
microporosity $\poromicro=0.2$, and the grains are taken to be either 5:7:7
oblate spheroids or 2:1:1 prolate spheroids.  The two dielectric
functions are very similar for $\lambda\ltsim 15\micron$ (see Figure
\ref{fig:diel}) but differ at longer wavelengths.  The material is
strongly absorptive ($\epsilon_2\gtsim 0.5$) in the vacuum UV and also
in the mid-IR and FIR.  The real part $\epsilon_1$ is large in the FIR

We use the ``oblate'' dielectric function in Figure \ref{fig:diel} for
the ``flattened'' shapes, and the ``prolate'' dielectric function from
Figure \ref{fig:diel} for the ``elongated'' shapes.

\section{\label{sec:axisymm}
         Scattering and Absorption}

The dimensionless extinction efficiency factor for randomly-oriented
particles is
\beq
\Qextran(\lambda) \equiv \frac{C_{\rm ext,ran}(\lambda)}{\pi\aeff^2}
~~~,
\eeq
where $C_{\rm ext,ran}(\lambda)$ is the extinction cross section
averaged over random orientations.
\begin{figure}
\begin{center}
\includegraphics[angle=0,width=10.0cm,
                 clip=true,trim=0.5cm 10.0cm 0.5cm 4.0cm]
{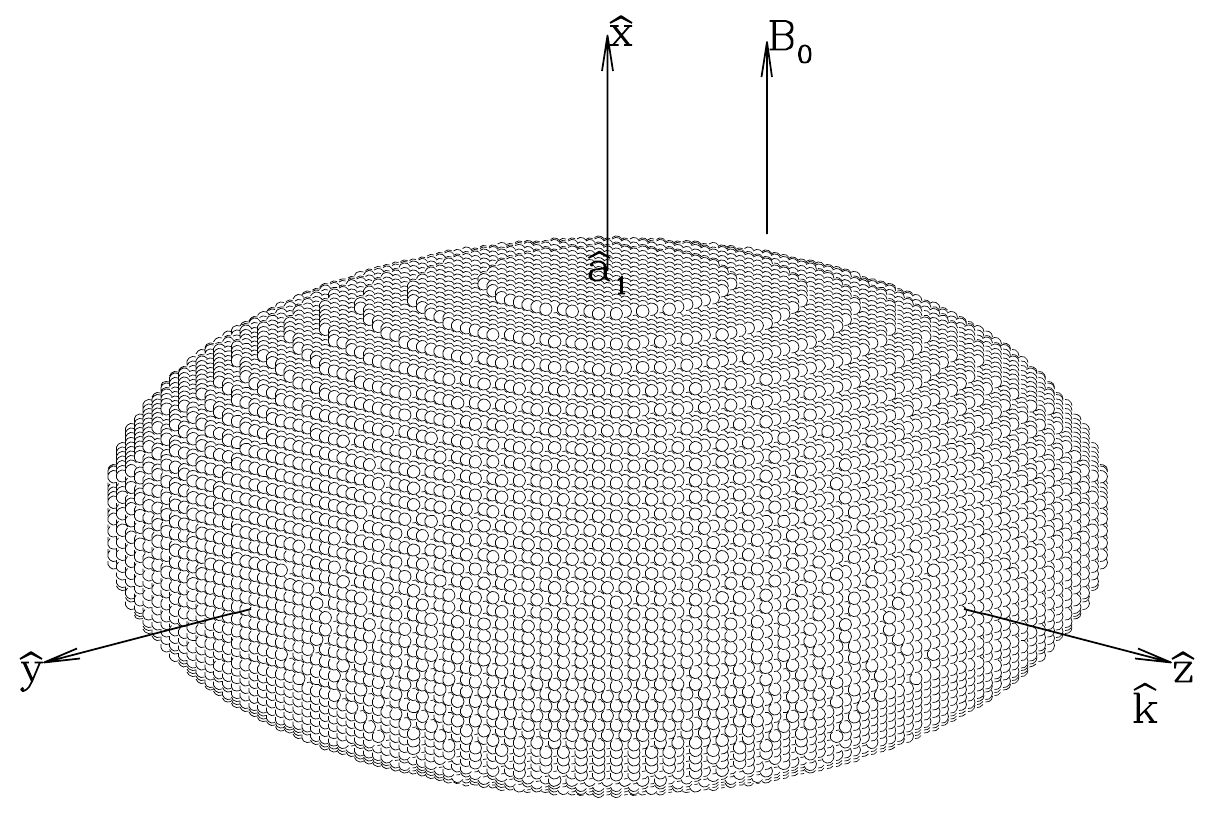}
\caption{\label{fig:PSA}\footnotesize Perfect spinning alignment (PSA)
  geometry.  The grain is spinning around principal axis $\bahat_1$,
  with $\bahat_1 \parallel \bB_0$, the local magnetic field.
  Polarization is maximum for radiation propagating with
  $\bkhat\perp\bB_0$.
  }
\end{center}
\end{figure}

Let $\bahat_1$ be the principal axis of largest moment of inertia, and
let $\bJ$ be the grain's angular momentum.  The idealized case of
``perfect spinning alignment'' (PSA) has the grains spinning with
$\bahat_1\parallel\bJ$, and $\bJ\parallel\bB_0$, where $\bB_0$ is the
local magnetic field (see Figure \ref{fig:PSA}).  The extinction cross
section will depend on the orientation of the rotation axis $\bahat_1$
relative to the line of sight.  Let $\bxhat$ and $\byhat$ be unit
vectors perpendicular to the line of sight.  We consider the limiting
case where $\bB_0$ is ``in the plane of the sky'', with
$\bB_0\parallel\bxhat$, thus $\bahat_1\parallel\bxhat$ (see Figure
\ref{fig:PSA}).  The dimensionless efficiency factors for polarization
and extinction for grains in PSA are defined to be
\beqa \label{eq:QpolPSA}
\QpolPSA(\lambda)
&\equiv&\frac{\langle C_{{\rm ext},y} (\lambda)- C_{{\rm
      ext},x}(\lambda)\rangle_{\rm PSA}} {2\pi\aeff^2}
\\ \label{eq:QextPSA}
\QextPSA(\lambda)
&\equiv&\frac{\langle C_{{\rm ext},y} (\lambda)+ C_{{\rm
      ext},x}(\lambda)\rangle_{\rm PSA}} {2\pi\aeff^2}
~~~,
\eeqa
where $C_{{\rm ext},x}$ and $C_{{\rm ext},y}$ are extinction cross sections
for radiation with $\bEinc\parallel\bxhat$ and
$\bEinc\parallel\byhat$, respectively, and $\langle...\rangle_{\rm PSA}$
denotes
averaging over rotations of the grain around $\bahat_1\parallel\bxhat$.

\section{\label{sec:DDA}
         Discrete Dipole Approximation}

The discrete dipole approximation (DDA) is a finite element
approximation for solving Maxwell's equations for a chosen target
geometry
\citep{Purcell+Pennypacker_1973,Draine_1988,Draine+Flatau_1994}.  The
public domain DDA program {\tt DDSCAT}\footnote{%
{\tt DDSCAT} version 7.3.3, available at \url{www.ddscat.org}.} 
is used to calculate scattering and absorption.  
The geometries considered here are among
the shape options available within {\tt DDSCAT}.

{\tt DDSCAT} approximates a target by an array of $\Ndip$ polarizable
points (referred to as ``dipoles'') located on a cubic lattice with
lattice spacing $d$.  Lattice points are included in the target array
if the point falls within the volume $V$ defined by the ideal target
shape; the choice of lattice spacing $d$ determines the number of
dipoles $N\approx V/d^3$.  For symmetric targets, {\tt DDSCAT} sets
the target symmetry axis to be parallel to one of the lattice axes,
with the target centroid offset by $(\delta_x,\delta_y,\delta_z)d$
from the nearest lattice point.  For all shapes considered here, {\tt
  DDSCAT} uses offsets $(\delta_x,\delta_y,\delta_z)=(0.5,0.5,0.5)$.
Each polarizable point has a prescribed complex polarizability
$\tilde{\alpha}_j(\lambda)$.  The ``lattice dispersion relation''
prescription \citep{Draine+Goodman_1993,Gutkowicz-Krusin+Draine_2004}
is used to determine the $\tilde{\alpha}_j(\lambda)$ appropriate for
modeling material with specified complex dielectric function
$\epsilon(\lambda)$.

For finite $\Ndip$, the ideal target geometry is imperfectly
reproduced by the dipole array, but the method converges to the ideal
geometry in the limit $\Ndip\rightarrow\infty$.  Figure
\ref{fig:shapes} shows DDA realizations using $\Ndip\approx 10^5$ for
the 20 convex targets.  Light scattering calculations are carried out
using even larger numbers of dipoles (see Table \ref{tab:N values}) in
order to more closely approximate the ideal target shapes, and to
adequately resolve the electromagnetic field even for vacuum
wavelengths as short as $0.1\micron$.

For an incoming polarized monochromatic plane wave, {\tt DDSCAT}
iteratively converges on the self-consistent solution for the $N$
oscillating dipole polarizations $\bP_j$.  From the solution $\bP_j$,
scattering and absorption cross sections are calculated.  

Let $\bkhat$ be the direction of propagation of the incident plane
wave, and let $\bahat$ be an axis fixed in the grain (if the grain has
rotational symmetry, it is convenient to choose $\bahat$ to be the
symmetry axis).  The electromagnetic scattering problem depends on the
angle $\Theta$ between $\bahat$ and $\bkhat$, and on the linear
polarization $\bEinc$ of the incident wave.  For non-axisymmetric
targets, the problem also depends on an angle $\beta$ specifying
rotation of the target around the target axis $\bahat$.  Let $C_{\rm
  E}(\Theta,\beta,\lambda)$ be the cross sections for $\bE_{\rm inc}$
in the $\bkhat-\bahat$ plane, and $C_{\rm H}(\Theta,\beta,\lambda)$
the cross section for $\bE_{\rm inc}$ perpendicular to the
$\bkhat-\bahat$ plane.  Cross sections are given in terms of
dimensionless efficiency factors $Q\equiv C/\pi\aeff^2$.

For each of the target shapes and sizes, the scattering problem is
solved for 151 values of $\lambda$, uniformly spaced in
$\log(\lambda)$ from $\lambda=0.1\micron$ to $100\micron$.  The 20
shapes considered here all have reflection symmetry through a plane
perpendicular to $\bahat$; thus we need to consider only
$\Theta\in[0,90^\circ]$.  We use 11 values of $\Theta$ (uniformly
spaced in $\cos\Theta$ from 0 to 1).  For the axisymmetric shapes,
$\beta$ is irrelevant.  For the rectangular prisms, we use 3 values of
$\beta\in(0,45^\circ)$: $7.5^\circ,22.5^\circ,37.5^\circ$.  For the
triaxial ellipsoids, we choose $\bahat=\bahat_1$, and use 3 values of
$\beta\in(0,90^\circ)$: $15^\circ, 45^\circ, 75^\circ$.

\subsection{\label{subsec:extrap}
            $\Ndip\rightarrow\infty$ Extrapolation and Uncertainties
            $|\Delta Q|$}

The errors in the DDA are primarily associated with the polarizations
of the dipoles near the target surface.  Let $D$ be some
characteristic dimension of the target.  Because the fraction of the
dipole sites that are within a distance $d$ of the surface scales as
$d/D$, the fractional error depends on $d/D$.  As discussed by
\citet{Collinge+Draine_2004}, because $\Ndip\propto(D/d)^3$, the
fractional error is expected to scale as $\Ndip^{-1/3}$.  Thus if
$Q_\infty$ is the exact result,
\beq \label{eq:Q(N)}
Q_N \approx Q_\infty + A \Ndip^{-1/3}
~~~,
\eeq
where $A$ is some constant.  If we calculate $Q_N$ for 
$\Ndip=N_1$ and $N_2$, we can extrapolate to estimate
$Q_\infty$ \citep{Shen+Draine+Johnson_2008}:
\beq \label{eq:extrap} 
  Q_\infty(N_1,N_2) \approx Q_{N_1} -
  \frac{Q_{N_1}-Q_{N_2}} {1-(N_1/N_2)^{1/3}}
  ~~~.  
\eeq 
In Appendix \ref{app:DDA accuracy} we test Equation (\ref{eq:extrap})
using a test case -- spheres -- where DDA calculations can be compared
to exact results obtained from Mie theory \citep[see,
  e.g.,][]{Bohren+Huffman_1983}.  Equation (\ref{eq:extrap}) is found
to give very accurate results.

Because Equation (\ref{eq:Q(N)}) is only an approximation to the actual
variation of $Q$ with $\Ndip$, the estimate (\ref{eq:extrap}) for
$Q_\infty$ will not be exact.  The uncertainty in this estimate can
itself be estimated by calculating $Q_N$ for three values
$N_1>N_2>N_3$, and comparing the extrapolation (\ref{eq:extrap}) using
$N_1$ and $N_2$ with the result using $N_2$ and $N_3$: the difference
is
\beq \label{eq:DeltaQ} 
\Delta Q \equiv Q_\infty(N_1,N_2)-Q_\infty(N_2,N_3)
=
\frac{(Q_{N_1}-Q_{N_2})}{1-(N_2/N_1)^{1/3}}
+
\frac{(Q_{N_2}-Q_{N_3})}{1-(N_2/N_3)^{1/3}} 
~~~.  
\eeq 
$|\Delta Q|$ is a reasonable estimate for the magnitude in the
uncertainty in the estimate of $Q_\infty$.

\begin{figure}
\begin{center}
\includegraphics[angle=0,width=4.4cm,
                 clip=true,trim=0.5cm 5.0cm 0.5cm 2.5cm]
{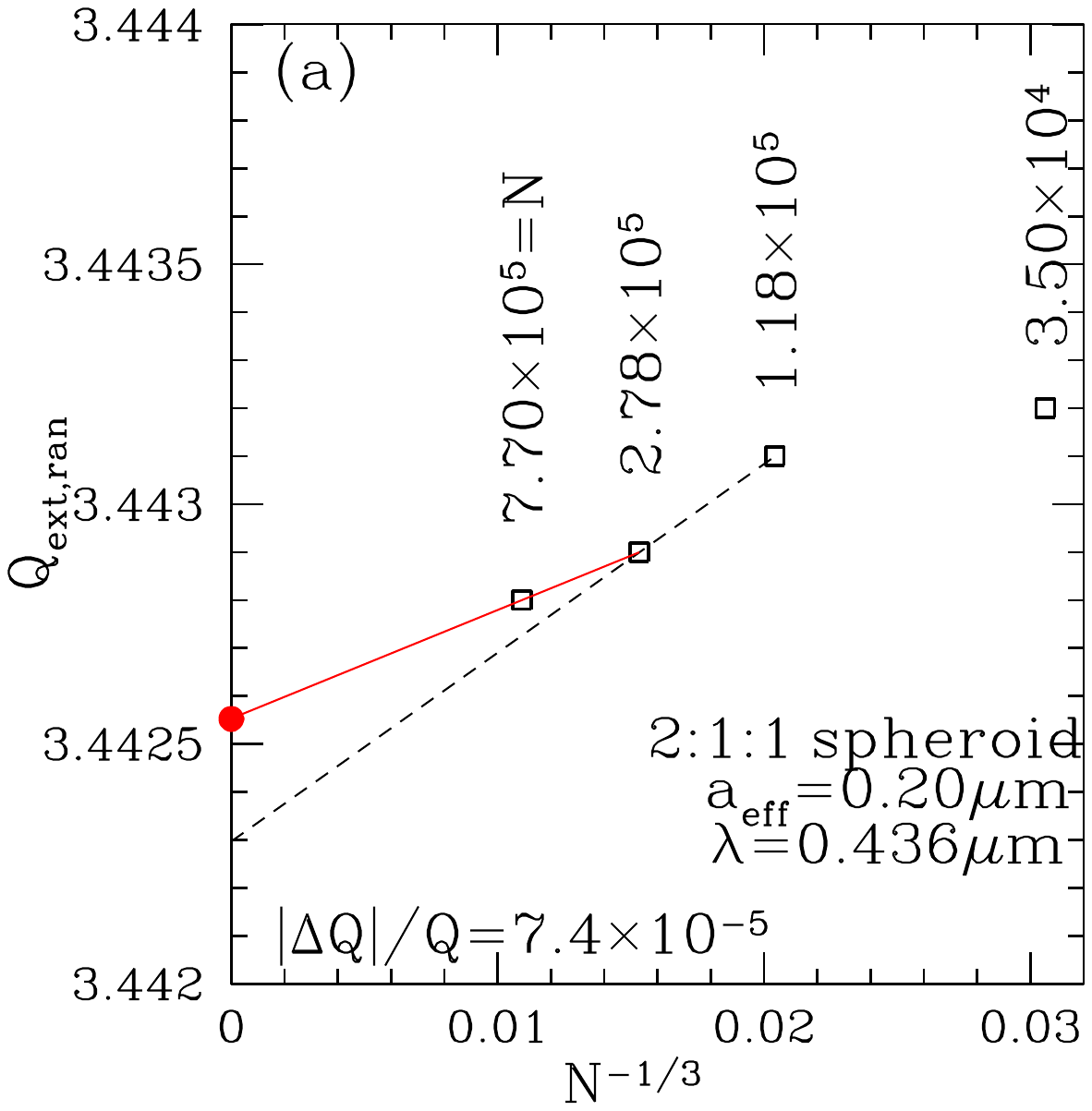}
\includegraphics[angle=0,width=4.4cm,
                 clip=true,trim=0.5cm 5.0cm 0.5cm 2.5cm]
{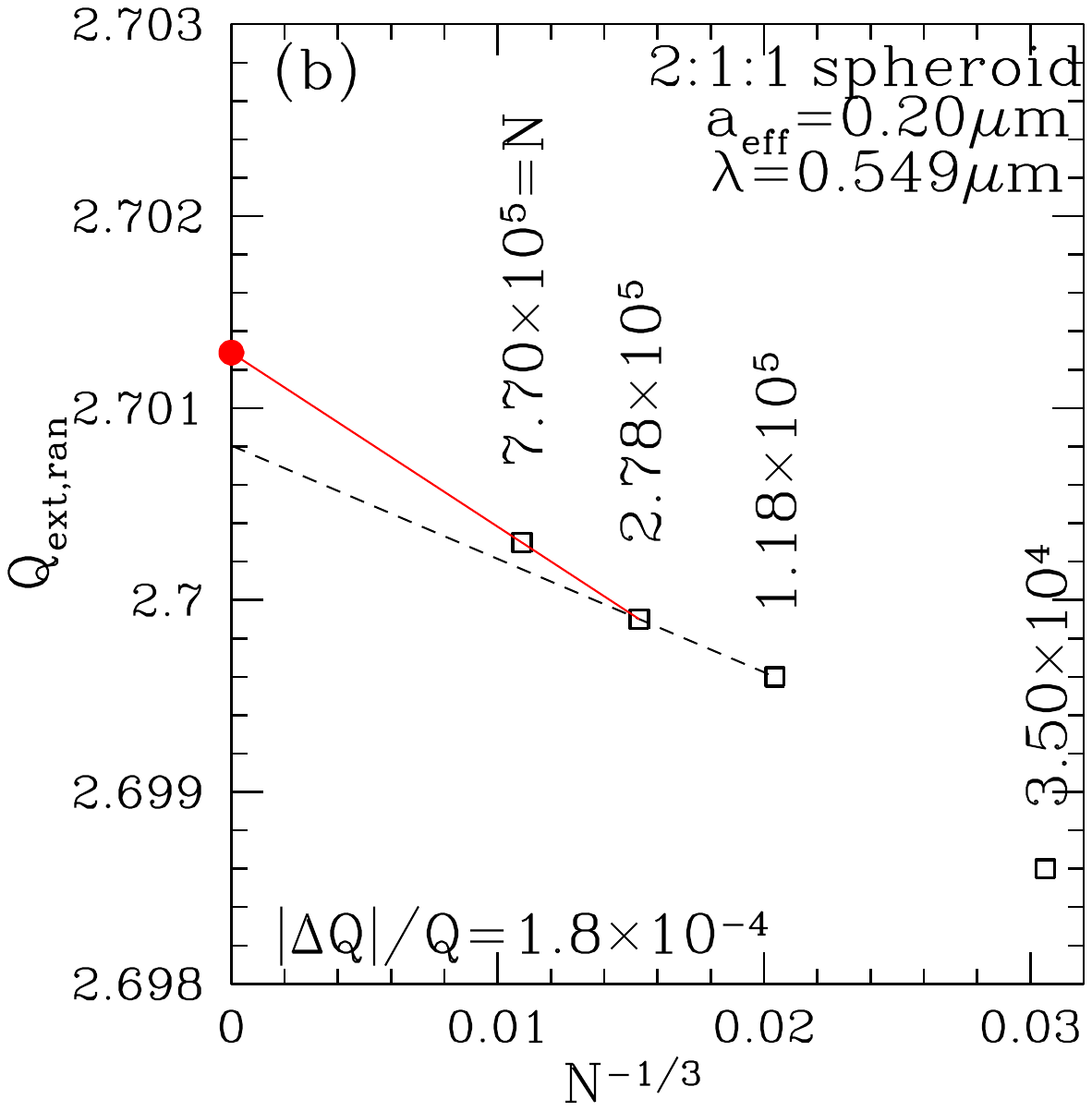}
\includegraphics[angle=0,width=4.4cm,
                 clip=true,trim=0.5cm 5.0cm 0.5cm 2.5cm]
{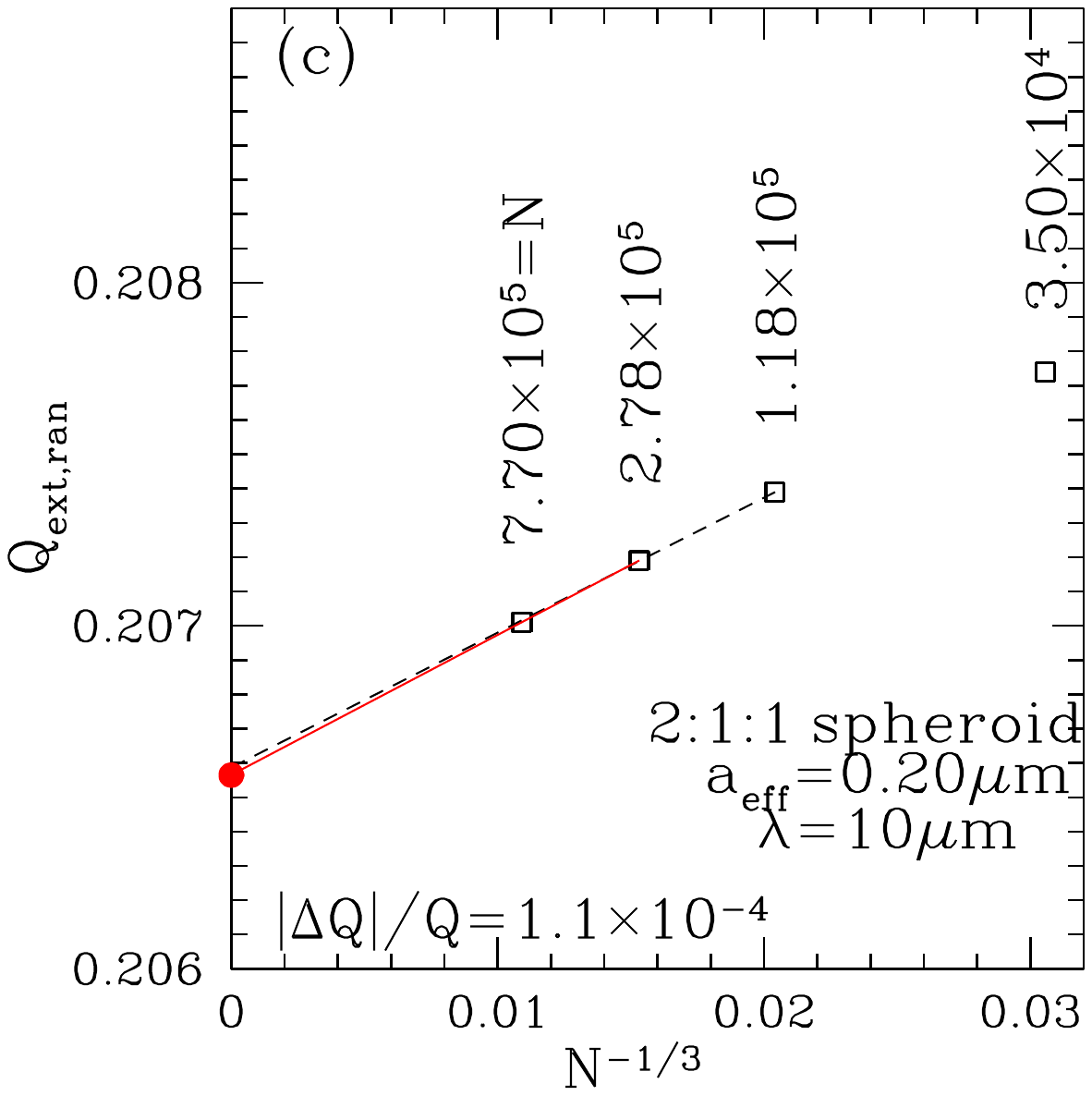}
\includegraphics[angle=0,width=4.4cm,
                 clip=true,trim=0.5cm 5.0cm 0.5cm 2.5cm]
{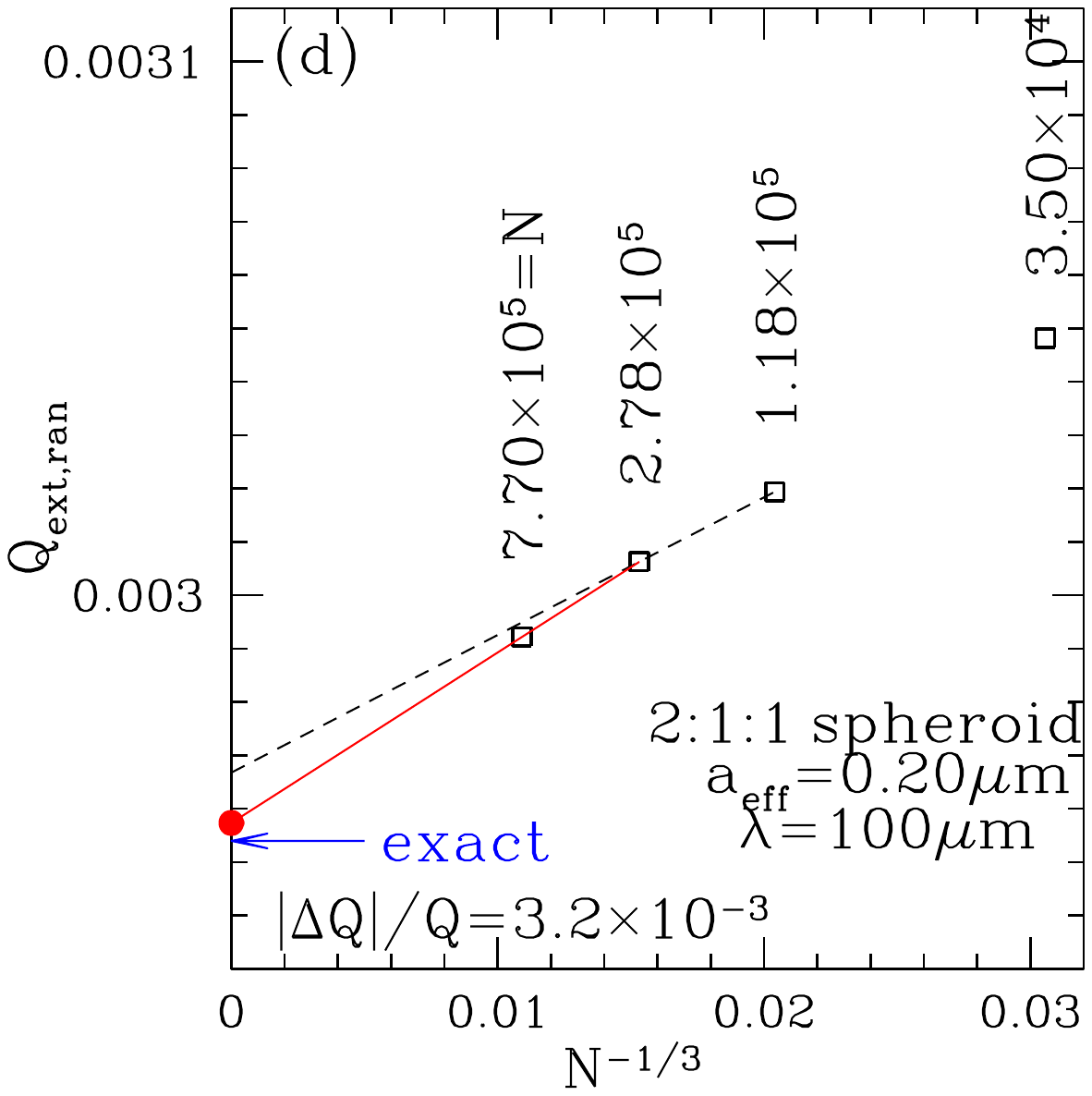}\\
\caption{\label{fig:qext_vs_N_sph}\footnotesize DDA calculations of
  $\Qextran$ for randomly-oriented 2:1:1 spheroids with volume-equivalent
  radii $\aeff=0.20\micron$ as a function of $\Ndip^{-1/3}$, where
  $\Ndip$ is the number of dipoles.  The ``prolate'' dielectric function from
  Figure \ref{fig:diel} is assumed.  Results are shown for four
  wavelengths $\lambda$.  For each $\lambda$, extrapolations to
  $Q_\infty$ are made using the results for the two largest values of
  $\Ndip$ (solid line) and using the second and third largest values
  of $\Ndip$ (dashed line).  The magnitude $|\Delta Q|$ of the
  difference between these two extrapolations is a reasonable estimate
  for the uncertainty in the estimate for $Q_\infty$.  The fractional
  uncertainty ranges from $\sim$$7\xtimes10^{-5}$ at
  $\lambda=0.44\micron$ to $\sim$$3\xtimes10^{-3}$ at
  $\lambda=100\micron$.  The exact result in the Rayleigh limit is
  shown for $\lambda=100\micron$ in (d); agreement is excellent.
  }
\end{center}
\end{figure}

Figure \ref{fig:qext_vs_N_sph} shows efficiency factors
$\Qextran(\lambda)$ for randomly-oriented 2:1:1 prolate spheroids as a
function of $\Ndip^{-1/3}$ for $\aeff=0.20\micron$ and 4 wavelengths
$\lambda$: $0.437\micron$, $0.550\micron$, $10\micron$, and
$100\micron$.  Results are shown for 4 values of $\Ndip$, ranging from
$\sim$$3.5\xtimes10^4$ to $\sim$$7.7\xtimes10^5$.  Extrapolations
using the largest two values of $\Ndip$ (solid line) and using the
second and third largest values of $\Ndip$ (dashed line) are shown.
The fractional uncertainty estimate $|\Delta Q|/Q$ in extrapolation of
$\Qextran$ to $\Ndip\rightarrow\infty$ (see Equation \ref{eq:DeltaQ})
is given for each example.  

At optical wavelengths the DDA results are very accurate, with
fractional uncertainties $|\Delta Q|/Q \sim 7\xtimes10^{-5}$ and
$2\xtimes10^{-4}$ at $\lambda=0.44\micron$ and $0.55\micron$.  At
$10\micron$ the \added{fractional} uncertainty is also small, $\sim$$1\xtimes10^{-4}$.
However, as the dielectric function $\epsilon$ becomes large in the
FIR (see Figure \ref{fig:diel}), the DDA becomes less accurate; for
the spheroidal shape, the estimated fractional uncertainty in
$\Qextran$ increases to $\sim3\xtimes10^{-3}$ at $\lambda=100\micron$.
However, at $\lambda=100\micron$ the grains are in the Rayleigh limit,
and we can compare with the analytic result for spheroids \citep[see,
  e.g.][]{Bohren+Huffman_1983}. Figure \ref{fig:qext_vs_N_sph}d shows
that the agreement is excellent; Equation (\ref{eq:DeltaQ}) evidently
overestimates the uncertainty.  To verify that this behavior is
general, similar results for 2:1:1 cylinders can be found in
Appendix \ref{app:DDA accuracy}.

Figure \ref{fig:qpol_vs_N_sph} shows $\QpolPSA$ as a function of
$\Ndip^{-1/3}$, calculated for $\aeff=0.2\micron$ 2:1:1 spheroids, for
four selected wavelengths.  In each case we extrapolate to
$\Ndip\rightarrow\infty$.  The fractional uncertainties $\Delta
\QpolPSA/\QpolPSA$ (given in each panel of Figure
\ref{fig:qpol_vs_N_sph}) are larger than the fractional uncertainties
in $\Qextran$ (given in each panel of Figure \ref{fig:qext_vs_N_sph}),
but are still relatively small at the optical wavelengths that are
important for starlight polarization: less than $1\%$ at
$\lambda=0.55\micron$ for most cases.  Similar plots of $\QpolPSA$
vs.\ $N^{-1/3}$ for 2:1:1 cylinders can be found in Appendix
\ref{app:DDA accuracy}.

\section{\label{sec:Q}
         Extinction and Polarization by Nonspherical Grains}

\subsection{Extinction Cross Sections}

\begin{figure}
\begin{center}
\renewcommand{\fwidth}{4.4cm}
\includegraphics[angle=0,width=\fwidth,
                 clip=true,trim=0.5cm 5.0cm 0.5cm 2.5cm]
{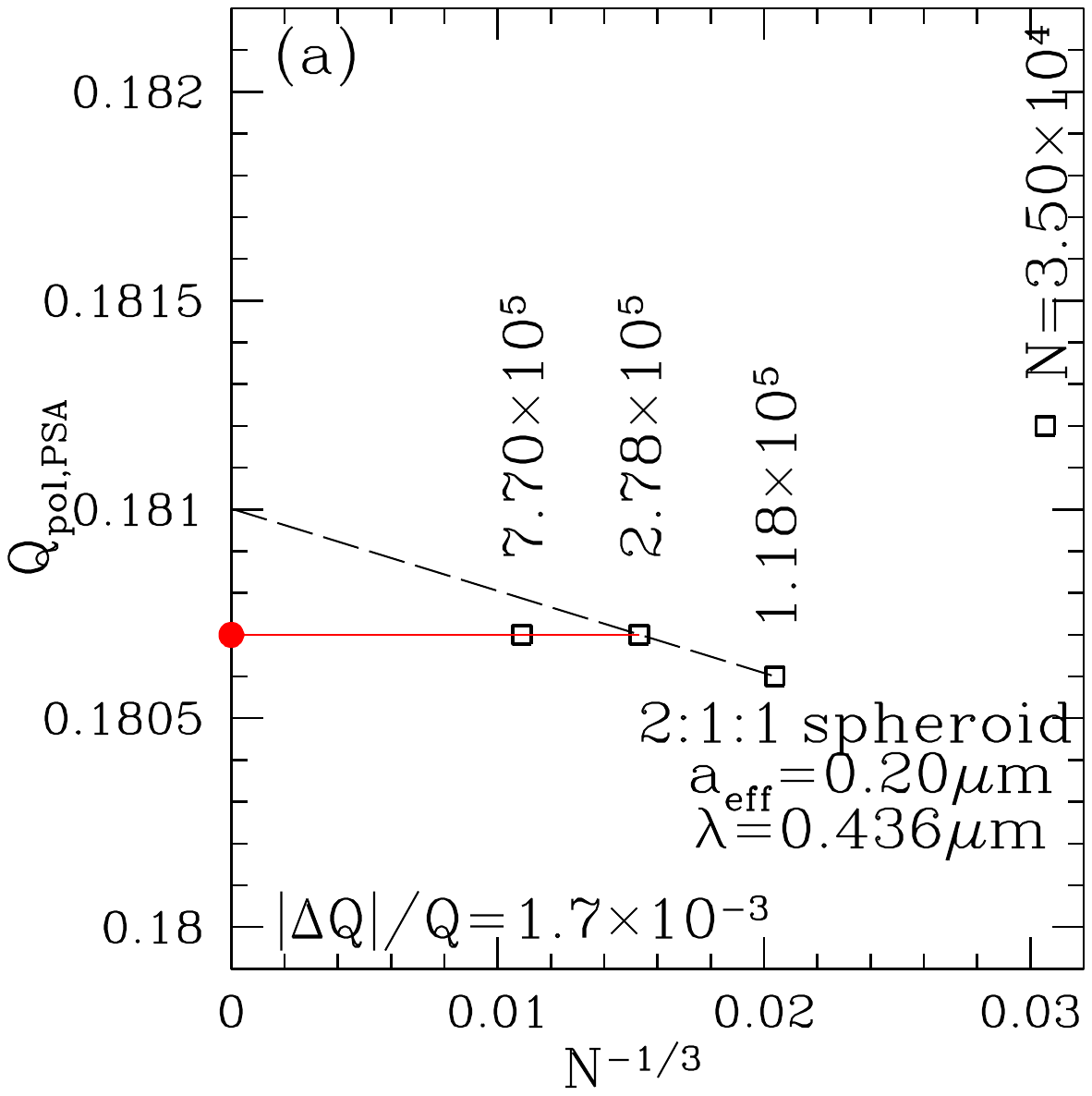}
\includegraphics[angle=0,width=\fwidth,
                 clip=true,trim=0.5cm 5.0cm 0.5cm 2.5cm]
{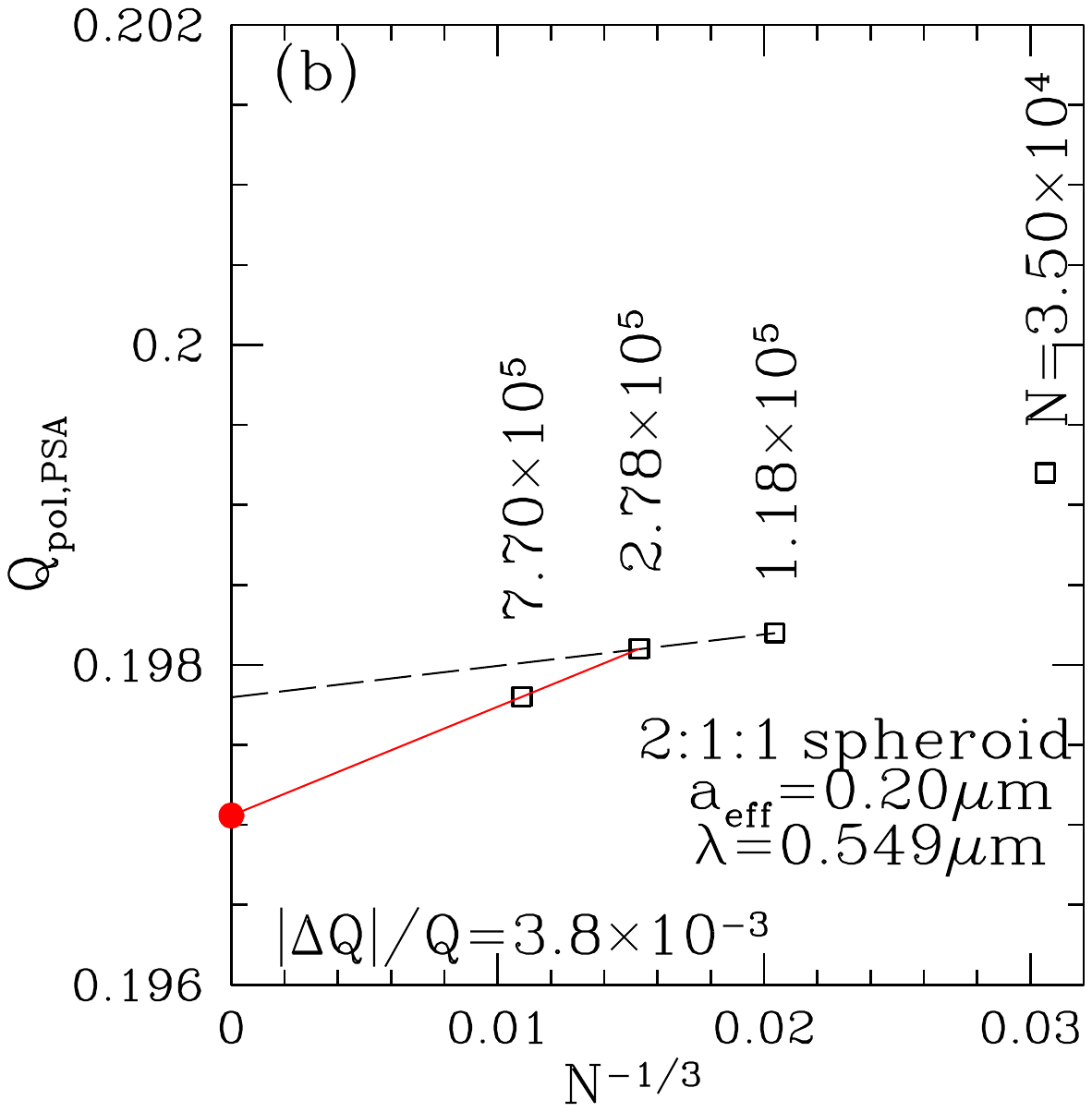}
\includegraphics[angle=0,width=\fwidth,
                 clip=true,trim=0.5cm 5.0cm 0.5cm 2.5cm]
{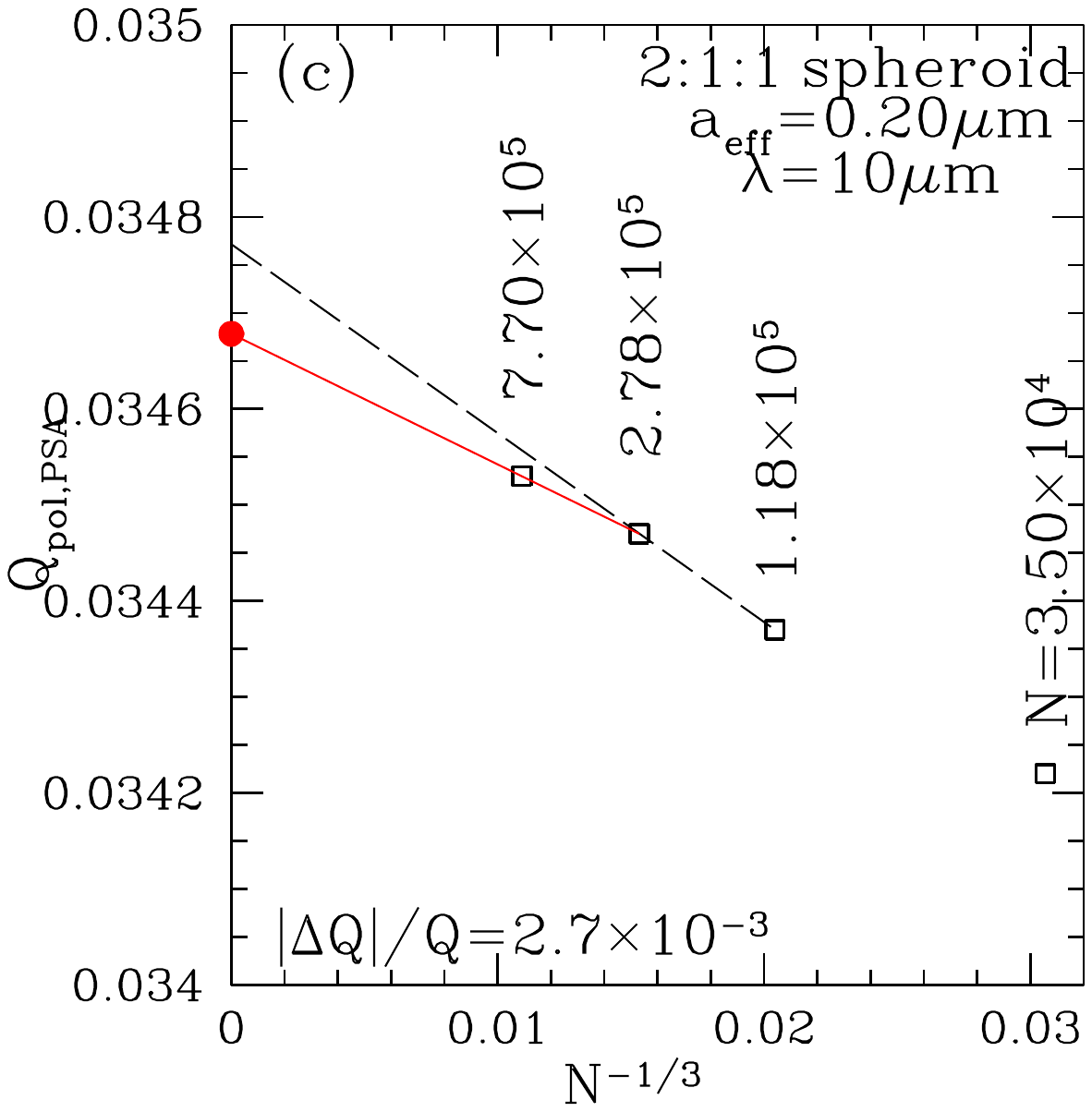}
\includegraphics[angle=0,width=\fwidth,
                 clip=true,trim=0.5cm 5.0cm 0.5cm 2.5cm]
{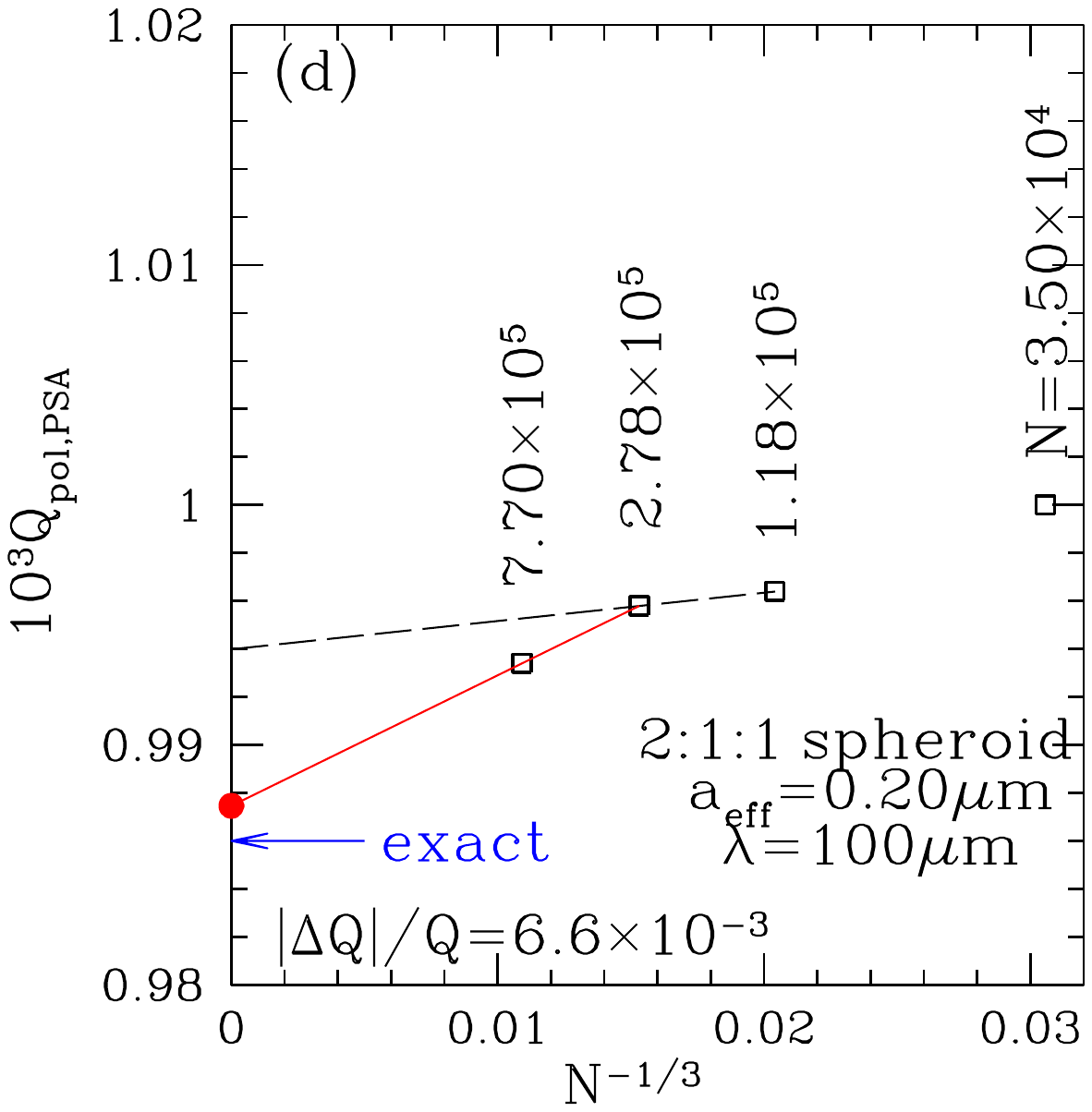}\\
\caption{\label{fig:qpol_vs_N_sph}\footnotesize DDA calculations of
  $\QpolPSA$ for 2:1:1 prolate spheroids with volume-equivalent radii
  $\aeff=0.20\micron$ as a function of $\Ndip^{-1/3}$, where $\Ndip$
  is the number of dipoles used to represent the target.  The
  ``prolate'' dielectric function from Figure \ref{fig:diel} is used.
  Results are shown for four wavelengths.  The exact result in the
  Rayleigh limit is shown for $\lambda=100\micron$ in (d); agreement
  is excellent. 
  }
\end{center}
\end{figure}
\begin{figure}
\begin{center}
\includegraphics[angle=0,width=8.5cm,
                 clip=true,trim=0.5cm 0.5cm 0.5cm 0.5cm]
{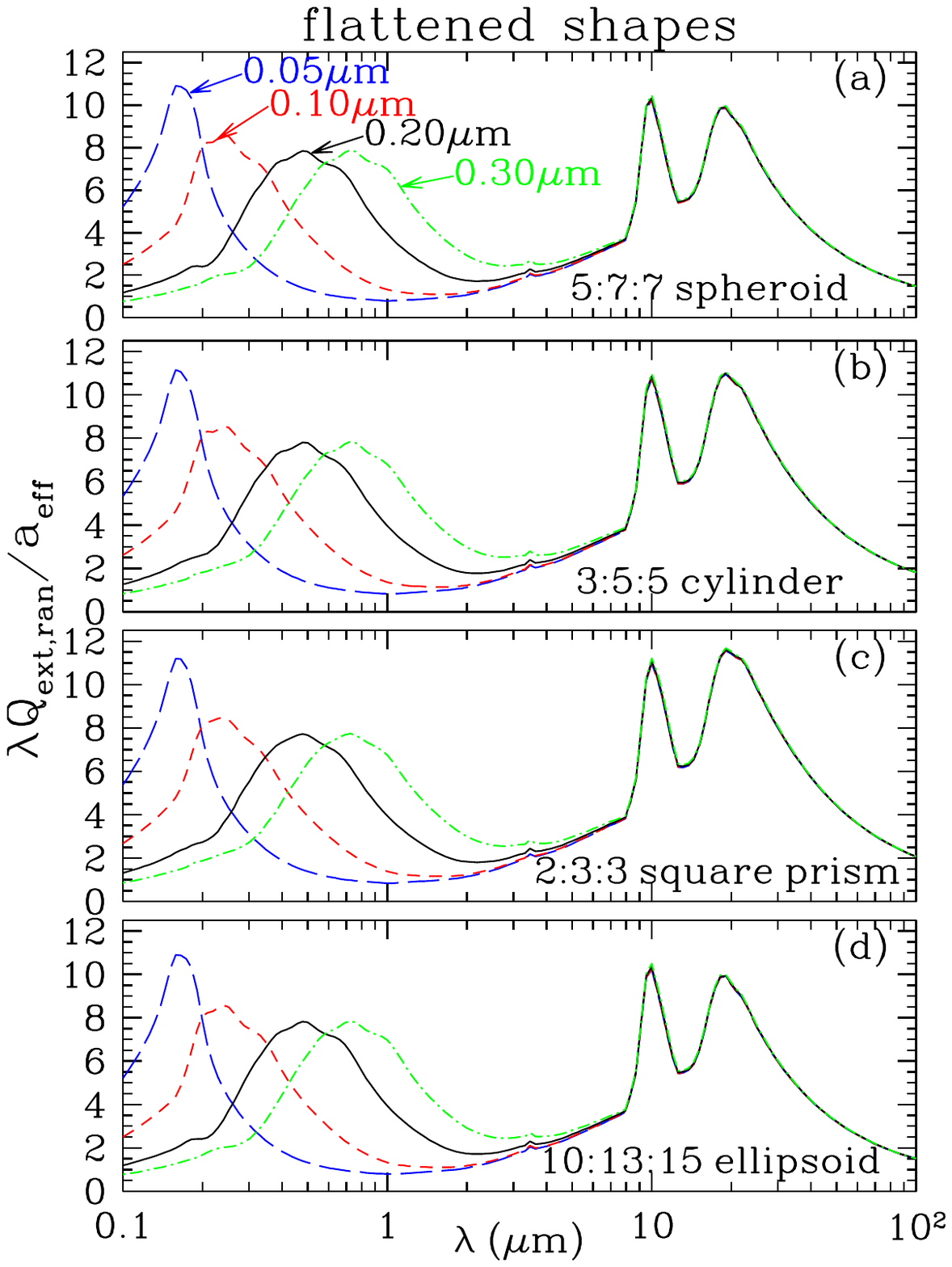}
\includegraphics[angle=0,width=8.5cm,
                 clip=true,trim=0.5cm 0.5cm 0.5cm 0.5cm]
{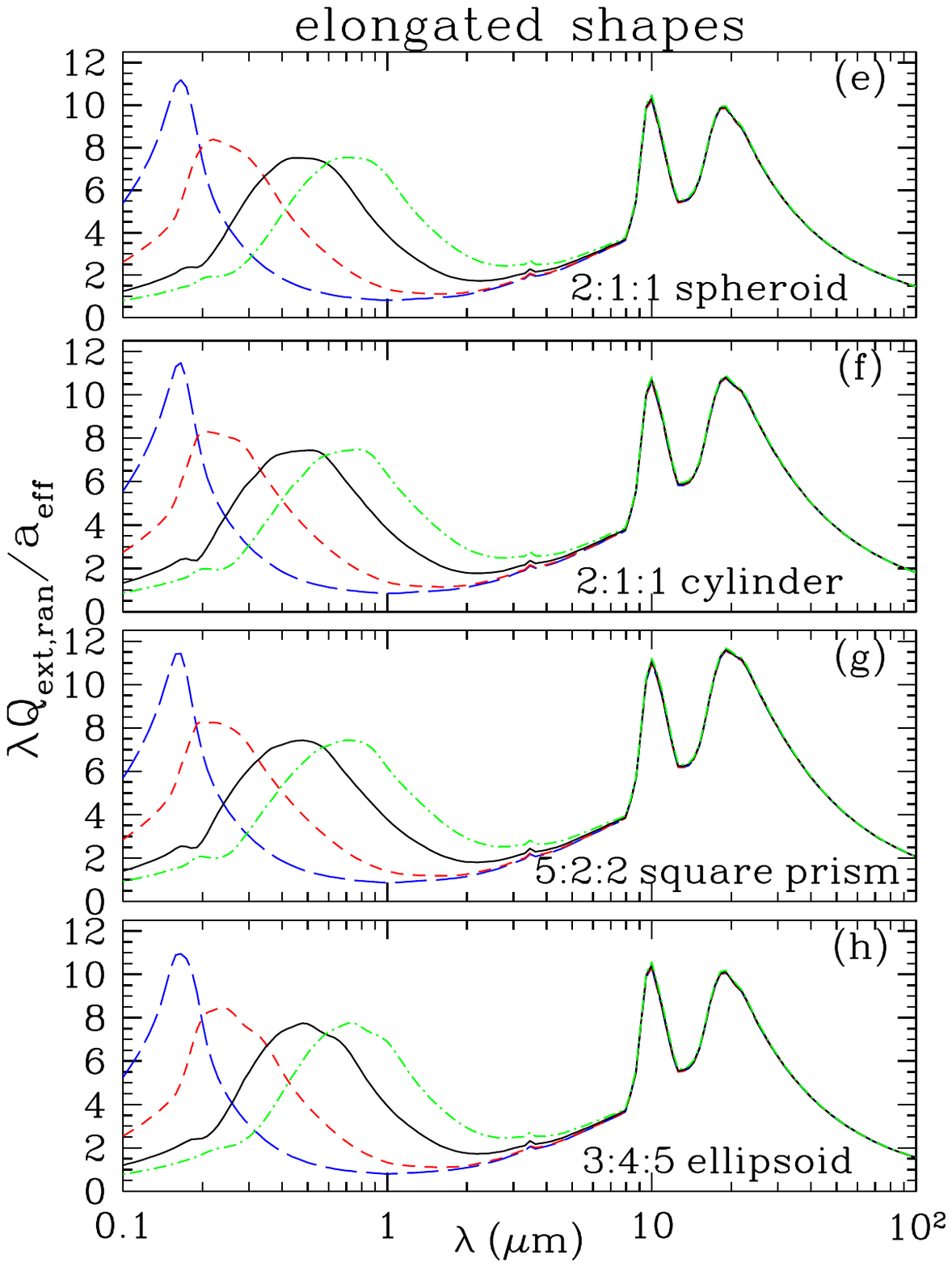}
\caption{\label{fig:wave*Qext/a_vs_lambda}\footnotesize 
  $\lambda\Qextran/\aeff$ as a function of wavelength $\lambda$ for 
  (a) 5:7:7 oblate spheroids, 
  (b) 3:5:5 cylinders, 
  (c) 2:3:3 square prisms, 
  (d) 10:13:15 ellipsoids,
  (e) 2:1:1 prolate spheroids, 
  (f) 2:1:1 cylinders, 
  (g) 5:2:2 square prisms, and 
  (h) 3:4:5 ellipsoids, 
  for $\aeff=0.05, 0.10, 0.20, 0.30\micron$.
  Flattened shapes use the ``oblate'' dielectric function shown in
  Figure \ref{fig:diel}; elongated shapes use the ``prolate''
  dielectric function from Figure \ref{fig:diel} (see text).  Cross
  sections are extrapolated using Equation (\ref{eq:extrap}) for
  $N_1,N_2$ in Table \ref{tab:N values}.
  }
\end{center}
\end{figure}

Figure \ref{fig:wave*Qext/a_vs_lambda} shows the full
$\lambda$-dependence of the dimensionless quantity
$\lambda\Qextran(\lambda)/\aeff$ for eight of the shapes from Figure
\ref{fig:shapes}, for four different grain sizes
($\aeff=0.05,0.1,0.2,0.3\micron$).  The eight selected shapes consist
of four flattened and four elongated geometries from Figure
\ref{fig:shapes}.  The chosen examples \added{all} have flattenings or
elongations sufficient to reproduce the observed polarization of
starlight.

For all cases, the quantity $\lambda \Qextran/\aeff$ peaks near
$\lambda \approx 2.5\aeff$, or $2\pi\aeff/\lambda \approx 2.5$.  For
$\lambda \gtsim 5\micron$, the grain sizes $a\leq 0.3\micron$
considered here are in the Rayleigh limit $\aeff\ll\lambda$,
absorption cross section per unit volume become independent of
particle size, and $\lambda\Qextran/\aeff$ becomes independent of
$\aeff$, as seen in Figure \ref{fig:wave*Qext/a_vs_lambda}.

\subsection{Polarization Cross Sections}

\begin{figure}
\begin{center}
\includegraphics[angle=0,width=8.5cm,
                 clip=true,trim=0.5cm 0.5cm 0.5cm 0.5cm]
{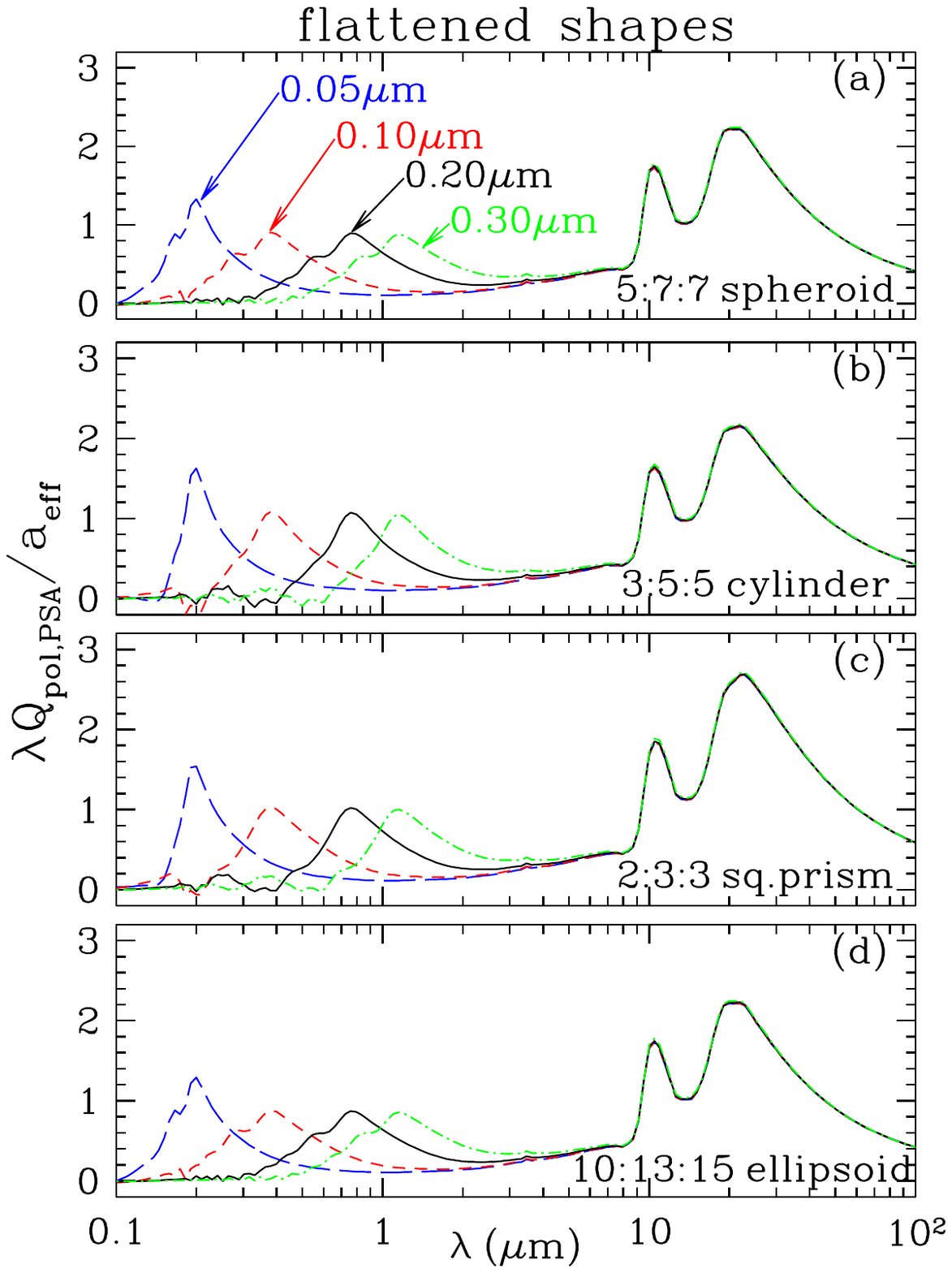}
\includegraphics[angle=0,width=8.5cm,
                 clip=true,trim=0.5cm 0.5cm 0.5cm 0.5cm]
{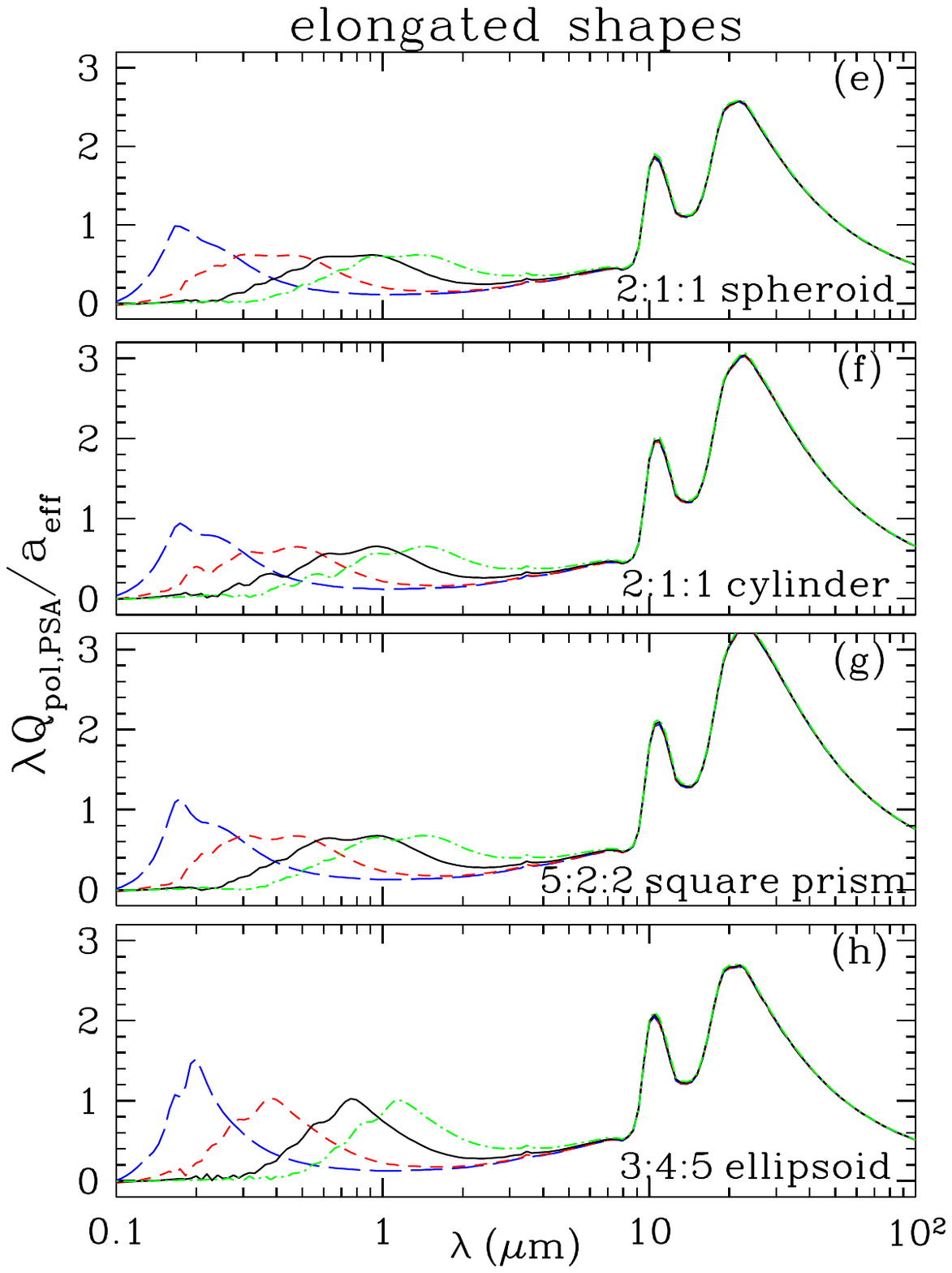}
\caption{\label{fig:wave*Qpol/a_vs_lambda}\footnotesize
    As in Figure \ref{fig:wave*Qext/a_vs_lambda}, but showing
    $\lambda \QpolPSA/\aeff$ as a function of wavelength
    $\lambda$ for perfect spinning alignment.
    }
\end{center}
\end{figure}

Figure \ref{fig:wave*Qpol/a_vs_lambda} shows the dimensionless
quantity $\lambda\QpolPSA/\aeff$ for the eight shapes in Figure
\ref{fig:wave*Qext/a_vs_lambda}, for four grain sizes.
$\lambda\QpolPSA/\aeff$ peaks at $\lambda\approx 3\aeff$, with
$\QpolPSA$ itself peaking near $\lambda \approx 2.5\aeff$ -- thus the
observed peak in starlight polarization near $\lambda\approx
0.55\micron$ must be due to grains with $\aeff\approx0.2\micron$.

The eight shapes considered here have similar extinction curves, as
seen in Figure \ref{fig:wave*Qext/a_vs_lambda}.  The polarization
profiles in Figure \ref{fig:wave*Qpol/a_vs_lambda} are broadly similar
to one another, but do differ from shape to shape.  For
example, it is clear from Figure \ref{fig:wave*Qpol/a_vs_lambda} that
the 2:1:1 cylinder (Figure \ref{fig:wave*Qpol/a_vs_lambda}f) has
enhanced polarization at FIR wavelengths relative to the
5:7:7 spheroid (Figure \ref{fig:wave*Qpol/a_vs_lambda}a).  We will
examine this further below.

\section{\label{sec:starlight_pol}
         Polarization of Starlight: Observational Constraints}

The observed wavelength dependence of starlight polarization is
approximated by the fitting function found by \citet{Serkowski_1973}:
\beq \label{eq:Serkowski}
p_\obs(\lambda) \approx \pmax
\exp\left[-K\left(\ln(\lambda/\lambdamax)\right)^2\right]
~~~,
\eeq
with $\pmax$ the polarization at the wavelength $\lambdamax$ where the
polarization peaks.  Typical sightlines have $\lambdamax\approx
0.55\micron$ and $K\approx 0.87$ \citep{Serkowski+Mathewson+Ford_1975,
  Whittet+Martin+Hough+etal_1992, Martin+Clayton+Wolff_1999,
  Whittet_2022}.  Starlight polarization is strongest and best
observed over the wavelength range $0.15\micron\ltsim \lambda \ltsim
2.5\micron$, and the ``Serkowski law'' (\ref{eq:Serkowski}) provides a
good approximation over this interval.  We concentrate attention on
the wavelength range $[\lambda_1,\lambda_2]$, with
$\lambda_1=0.15\micron$ and $\lambda_2=2.5\micron$.

Define an effective wavelength for the observed starlight polarization
over $[\lambda_1,\lambda_2]$:
\beq \label{eq:lambda_pobs}
\lambdapobs \equiv
\exp\left\{
\frac{\int_{\lambda_1}^{\lambda_2}\ln\lambda \,\, p_\obs(\lambda)\,d\ln\lambda}
{\int_{\lambda_1}^{\lambda_2} p_\obs(\lambda)\,d\ln\lambda}
\right\}
\approx 0.567\micron
\eeq
for $p_\obs(\lambda)$ given by Equation (\ref{eq:Serkowski}),
$\lambdamax=0.55\micron$, and $K=0.87$;
$\lambdapobs$ is close to (but not identical to) the wavelength
$\lambdamax$ of peak polarization.  The width of the starlight
polarization profile is characterized by
\beq \label{eq:sigmaobs}
\sigmaobs^2 \equiv
\frac{\int_{\lambda_1}^{\lambda_2} 
\left[\ln(\lambda/\lambdapobs)\right]^2 p_\obs(\lambda)\,d\ln\lambda}
{\int_{\lambda_1}^{\lambda_2} p_\obs(\lambda)\,d\ln\lambda}
\approx (0.64)^2
~~~.
\eeq
The strength of the starlight polarization reaches values as high
as\footnote{%
  For $[\pmax/E(B-V)]_{\rm max}=0.13\,{\rm mag}^{-1}$
  \citep{Panopoulou+Hensley+Skalidis+etal_2019,Planck_2018_XII} and
  $\NH/E(B-V)=8.8\xtimes10^{21}\cm^2\,{\rm mag}^{-1}$
  \citep{Lenz+Hensley+Dore_2017}.%
}
\beq \label{eq:obs_strength}
\left(\frac{\pmax}{\NH}\right)_{\rm max} = 1.48\times10^{-23}\cm^2\Ha^{-1}
~~~.
\eeq
From Equations (\ref{eq:Serkowski}) and (\ref{eq:obs_strength}),
the observed starlight polarization integral per H is
\beq \label{eq:OSPI}
\left(\frac{\Pi_{\rm obs}}{\NH}\right)_{\!\rm max}
\equiv \int_{\lambda_1}^{\lambda_2} 
\left(\frac{p_\obs(\lambda)}{\NH}\right)_{\!\rm max} d\lambda
\approx 1.23\micron ~ \left(\frac{\pmax}{\NH}\right)_{\!\rm max} \approx 1.82\xtimes10^{-27}\cm^3\Ha^{-1}
\eeq
\citep{Draine+Hensley_2021c}.  

At intermediate or high galactic latitudes, the volume of grain
material per H nucleon is estimated to be \citep{Draine+Hensley_2021a}
\beq
\Vgr\approx \frac{3.0\xtimes10^{-27}\cm^3\Ha^{-1}}{1-\poromicro}
~~~,
\eeq
where the ``microporosity'' $\poromicro$ is the vacuum fraction on
very small scales.  Observations of starlight polarization and
polarized submm emission both suggest that the porosity is not high
\citep[see][]{Draine+Hensley_2021c}.  Here we set $\poromicro=0.2$,
the value favored by \citet{Hensley+Draine_2023}.  Thus the
interstellar dust population must be consistent with
\beq \label{eq:Piobs/Vgr}
\left(\frac{\Pi_{\rm obs}}{\NH}\right)_{\rm \!\! max} \frac{1}{\Vgr} 
\approx 0.49 
~~~.  
\eeq
A viable model for interstellar dust should reproduce the full
observed wavelength dependence of extinction and polarization, both of
which are sensitive to the grain size distribution $dn_d/d\aeff$, with
polarization also sensitive to the fractional alignment $f_{\rm
  align}(\aeff)\leq 1$ of the particles.  Models of interstellar
polarization must reproduce the effective wavelength $\lambdapobs$
(Eq.\ \ref{eq:lambda_pobs}), the profile width $\sigma_\obs$
(Eq.\ \ref{eq:sigmaobs}), and the observed strength per grain volume
(Eq.\ \ref{eq:Piobs/Vgr}).  However, the starlight polarization has
some integrated properties that are relatively insensitive to grain
size -- the starlight polarization integral per grain volume $\Pi_{\rm
  obs}/\NH\Vgr$ (see Eq.\ \ref{eq:Piobs/Vgr}) and the fractional width
$\sigmaobs$ of the starlight polarization profile (see
Eq.\ \ref{eq:sigmaobs}).

\section{\label{sec:single-size}
         Effective Wavelength and Width of Polarization by Grains}

\begin{figure}
\begin{center}
\includegraphics[angle=0,width=8.0cm,
                 clip=true,trim=0.5cm 0.5cm 0.5cm 0.5cm]
{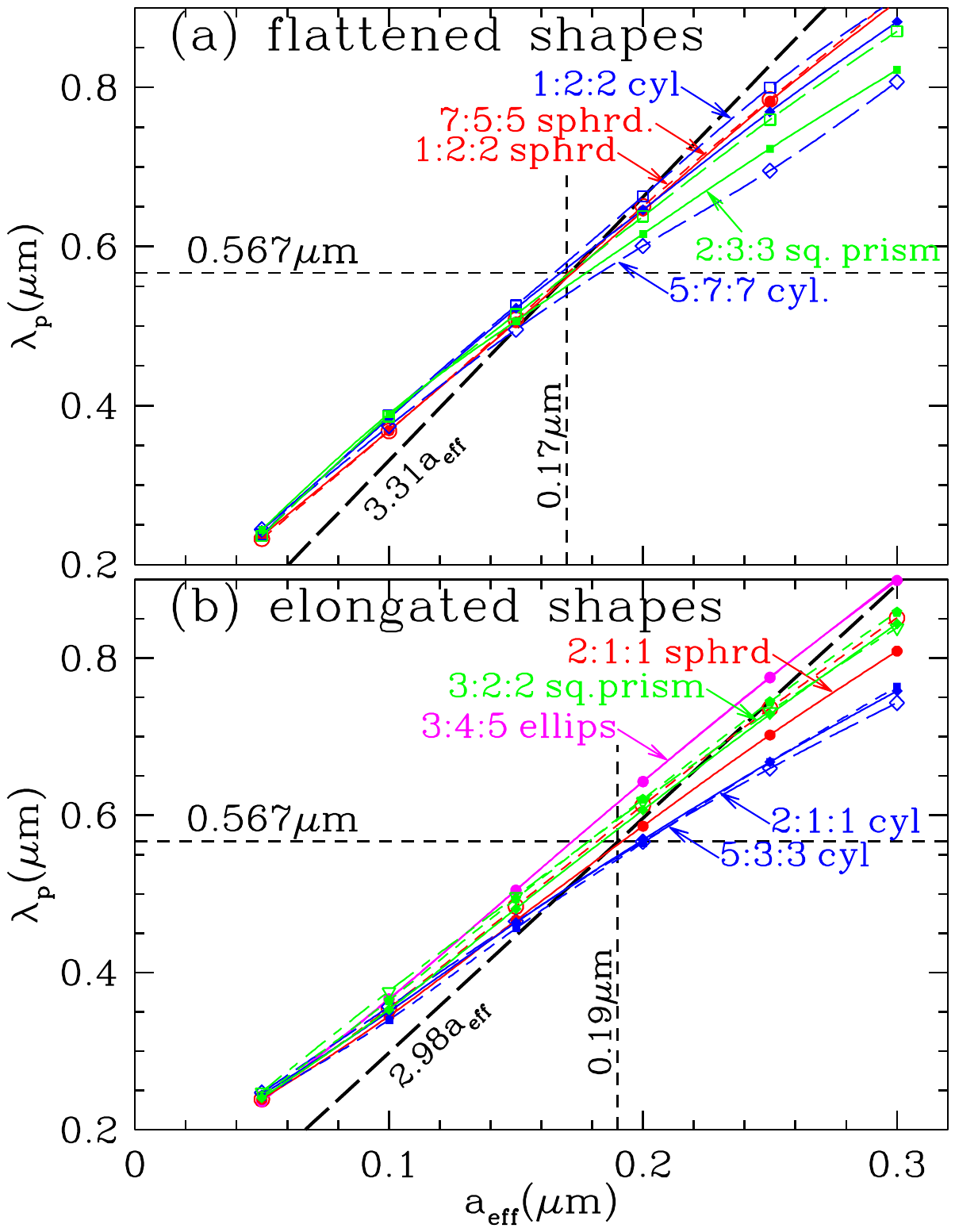}
\caption{\footnotesize \label{fig:lambdap} Effective wavelength for
  polarization $\lambdap(\aeff)$ for (a) 10 flattened and (b) 10
  elongated grain shapes studied here.  Some of the outliers are
  labelled.  The grains dominating starlight polarization should have
  $\lambdap\approx 0.567\micron$, corresponding to $\aeff$
  $\sim$$0.17\micron$ for the flattened shapes, or $\sim$$0.19\micron$
  for the elongated shapes.
  \btdnote{8.pdf}
  }
\end{center}
\end{figure}

The effective wavelength $\lambdap$ for the
polarization contribution of grains of a given shape and size $\aeff$
is \citep{Draine+Hensley_2021c}:
\beq \label{eq:lambdapeff}
\lambdap(\shape,\aeff)
\equiv 
\exp\left\{\frac{\int_{\lambda_1}^{\lambda_2} (\ln\lambda)\,
      \QpolPSA(\shape,\aeff,\lambda)\,d\ln\lambda} 
     {\int_{\lambda_1}^{\lambda_2} \QpolPSA(\shape,\aeff,\lambda) 
      \, d\ln\lambda}
\right\}
~~~.  
\eeq
Figure \ref{fig:lambdap} shows $\lambda_p$ as a function of $\aeff$
for the 20 shapes studied here.  The flattened ($\Rtwo<1$) shapes in
Figure \ref{fig:spei}a have
\beq
\lambdap
\approx 3.31\aeff ~~~,
\eeq
while the elongated ($\Rtwo>1$) shapes in Figure \ref{fig:spei}b have
\beq
\lambdap \approx 2.98\aeff ~~~.
\eeq
As discussed above, the grains responsible for starlight polarization
have $\lambdapobs\approx 0.567\micron$. From Figure \ref{fig:spei}a,b
we see that $\lambda\approx 0.567\micron$ corresponds to
$0.15\ltsim\aeff\ltsim 0.20\micron$ for the shapes considered here.
For a single shape and size, the width of the starlight polarization
profile can be characterized by
\beq \label{eq:sigma}
\sigmap^2(\shape,\aeff) \equiv
\frac{\int_{\lambda_1}^{\lambda_2} d\ln\lambda 
\left[\ln(\lambda/\lambdap)\right]^2 |\QpolPSA(\shape,\aeff,\lambda)|}
{\int_{\lambda_1}^{\lambda_2} d\ln\lambda \, |\QpolPSA(\shape,\aeff,\lambda)|}
~~~.
\eeq
Because $\QpolPSA$ can become negative on the short-wavelength wing of
the polarization profile, we introduce the absolute value operator in
Eq.\ (\ref{eq:sigma}) to avoid negative contributions to the
integrals.

Table \ref{tab:Phi table} gives $\sigmap$ for the shapes studied here.
For each case, results are interpolated to the value of $\aeff$
corresponding to $\lambdap=\lambdapobs=0.567\micron$.  The
uncertainties listed for $\sigmap$ are based on Eq.\ (\ref{eq:DeltaQ}),
using sizes $N_1,N_2,N_3$ in Table \ref{tab:N values}.

Individual grains do not reproduce the Serkowski law
(\ref{eq:Serkowski}) -- the Serkowski law arises after summing over a
range of grain sizes.  Because $\lambdap\propto\aeff$, a model with a
size distribution for the aligned grains will have
$\sigmaobs>\sigmap(\shape,\aeffp)$, where $\aeffp$ is the size with
$\lambdap=\lambdapobs$.  With $\sigmaobs\approx 0.64$ for the
Serkowski law (\ref{eq:Serkowski}) with the typical parameters
($K=0.87$, $\lambda_p=0.55\micron$), a conservative upper limit
$\sigmap< 0.60$ is chosen for compatibility with the observed
starlight polarization.  \citet{Hensley+Draine_2023} showed that a
model using 5:7:7 oblate spheroids ($\sigmap=0.516$: see Table
\ref{tab:Phi table}) with a suitable size distribution and alignment
function can reproduce both the average extinction law and the
observed starlight polarization.

After summing over a size distribution, shapes with $\sigmap > 0.60$
will result in a model polarization profile broader than observed.
The condition $\sigmap<0.60$ appears to rule out the 5:7:7 cylinder,
the 2:3:3 square prism, the 5:3:3 cylinder, and the 2:1:1 cylinder
(see Figure \ref{fig:spei}).

\newcommand \fna a
\newcommand \fnb b
\newcommand \fnc c
\newcommand \fnd d
\begin{table}
\begin{center}
{\footnotesize
\caption{\label{tab:Phi table}
         Starlight Polarization Properties of Different Shapes}
\begin{tabular}{c c c c c c c}
\hline
\multicolumn{7}{c}{Flattened Shapes ($\Rtwo<1$)}\\
shape & $\Rone\,^\fna$ & $\Rtwo\,^\fnb$ & $\aeff(\micron)$ & $\sigma_p~^\fnc$ & $\PhiPSA~^\fnd$ & comment \\
\hline
5:7:7 cylinder     & 1.1903 & 0.9166 & 0.184 & $0.610\pm0.002$ & $0.365\pm0.000$ & $\PhiPSA$ too small, $\sigma_p$ too large\\
5:8:8 cylinder     & 1.3151 & 0.8720 & 0.168 & $0.549\pm0.008$ & $0.634\pm0.020$ & $\PhiPSA$ too small\\
5:7:7 spheroid     & 1.3243 & 0.8690 & 0.172 & $0.516\pm0.001$ & $0.726\pm0.004$ \\
10:13:15 ellipsoid & 1.3325 & 0.9983 & 0.172 & $0.516\pm0.002$ & $0.718\pm0.001$ \\
3:5:5 cylinder     & 1.3514 & 0.8602 & 0.165 & $0.538\pm0.001$ & $0.701\pm0.003$ \\
10:17:17 cylinder  & 1.3686 & 0.8548 & 0.163 & $0.524\pm0.003$ & $0.735\pm0.011$ \\
2:3:3 square prism & 1.3846 & 0.8498 & 0.178 & $0.603\pm0.000$ & $0.738\pm0.000$ & $\sigma_p$ too large\\
1:2:2 cylinder     & 1.5    & 0.8165 & 0.165 & $0.514\pm0.004$ & $1.048\pm0.019$ \\
1:2:2 spheroid     & 1.6    & 0.7906 & 0.171 & $0.498\pm0.001$ & $1.461\pm0.002$ \\
1:2:2 square prism & 1.6    & 0.7906 & 0.171 & $0.540\pm0.000$ & $1.255\pm0.001$ \\
\hline\\
\multicolumn{7}{c}{Elongated Shapes ($\Rtwo>1$)}\\
shape & $\Rone\,^\fna$ & $\Rtwo\,^\fnb$ & $\aeff(\micron)$ &$\sigma_p~^\fnc$ & $\PhiPSA~^\fnd$ & comment \\
\hline
3:2:2 square prism & 1.2748 & 1.2748 & 0.179 & $0.585\pm0.001$ & $0.364\pm0.002$ & $\PhiPSA$ too small\\
3:2:2 spheroid     & 1.2748 & 1.2748 & 0.182 & $0.544\pm0.002$ & $0.430\pm0.000$ & $\PhiPSA$ too small\\
4:5:6 ellipsoid    & 1.3211 & 1.0398 & 0.173 & $0.518\pm0.001$ & $0.674\pm0.003$ & $\PhiPSA$ too small\\
3:4:5 ellipsoid    & 1.4063 & 1.0620 & 0.173 & $0.514\pm0.001$ & $0.856\pm0.006$ \\
5:3:3 cylinder     & 1.5336 & 1.5336 & 0.200 & $0.630\pm0.008$ & $0.561\pm0.009$ & $\PhiPSA$ too small, $\sigma_p$ too large\\
2:1:1 square prism & 1.5811 & 1.5811 & 0.179 & $0.552\pm0.000$ & $0.605\pm0.001$ & $\PhiPSA$ too small\\
2:1:1 spheroid     & 1.5811 & 1.5811 & 0.192 & $0.560\pm0.000$ & $0.709\pm0.002$ \\
2:1:1 cylinder     & 1.7795 & 1.7795 & 0.199 & $0.609\pm0.007$ & $0.712\pm0.011$ & $\sigma_p$ too large\\
5:2:2 square prism & 1.9039 & 1.9039 & 0.184 & $0.546\pm0.000$ & $0.773\pm0.001$ \\
5:2:2 cylinder     & 2.1602 & 2.1602 & 0.201 & $0.586\pm0.025$ & $0.880\pm0.014$ \\
\hline
\multicolumn{6}{l}{$\fna$ Asymmetry parameter (Eq.\ \ref{eq:R1}).}\\
\multicolumn{7}{l}{$\fnb$ Stretch parameter (Eq.\ \ref{eq:R2}).}\\
\multicolumn{6}{l}{$\fnc$ Polarization broadening parameter (Eq.\ \ref{eq:sigma}).}\\
\multicolumn{6}{l}{$\fnd$ Polarization efficiency integral (Eq.\ \ref{eq:Phi_PSA}).}\\
\end{tabular}
}
\end{center}
\end{table}

\section{\label{sec:SPEI}
         The Starlight Polarization Efficiency Integral}

\begin{figure}
\begin{center}
\includegraphics[angle=0,width=8.5cm,
                 clip=true,trim=0.5cm 0.5cm 0.5cm 0.5cm]
{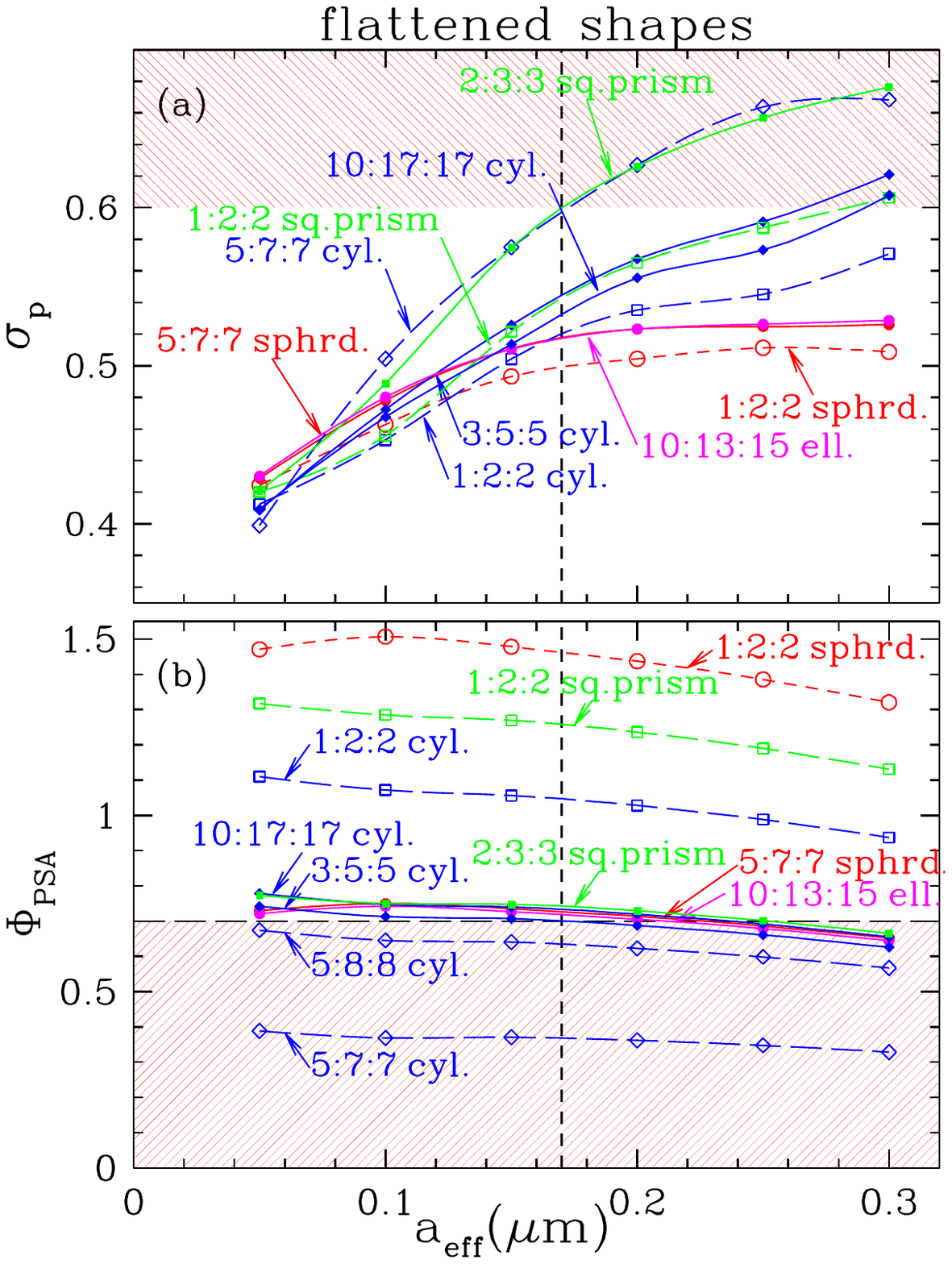}
\includegraphics[angle=0,width=8.5cm,
                 clip=true,trim=0.5cm 0.5cm 0.5cm 0.5cm]
{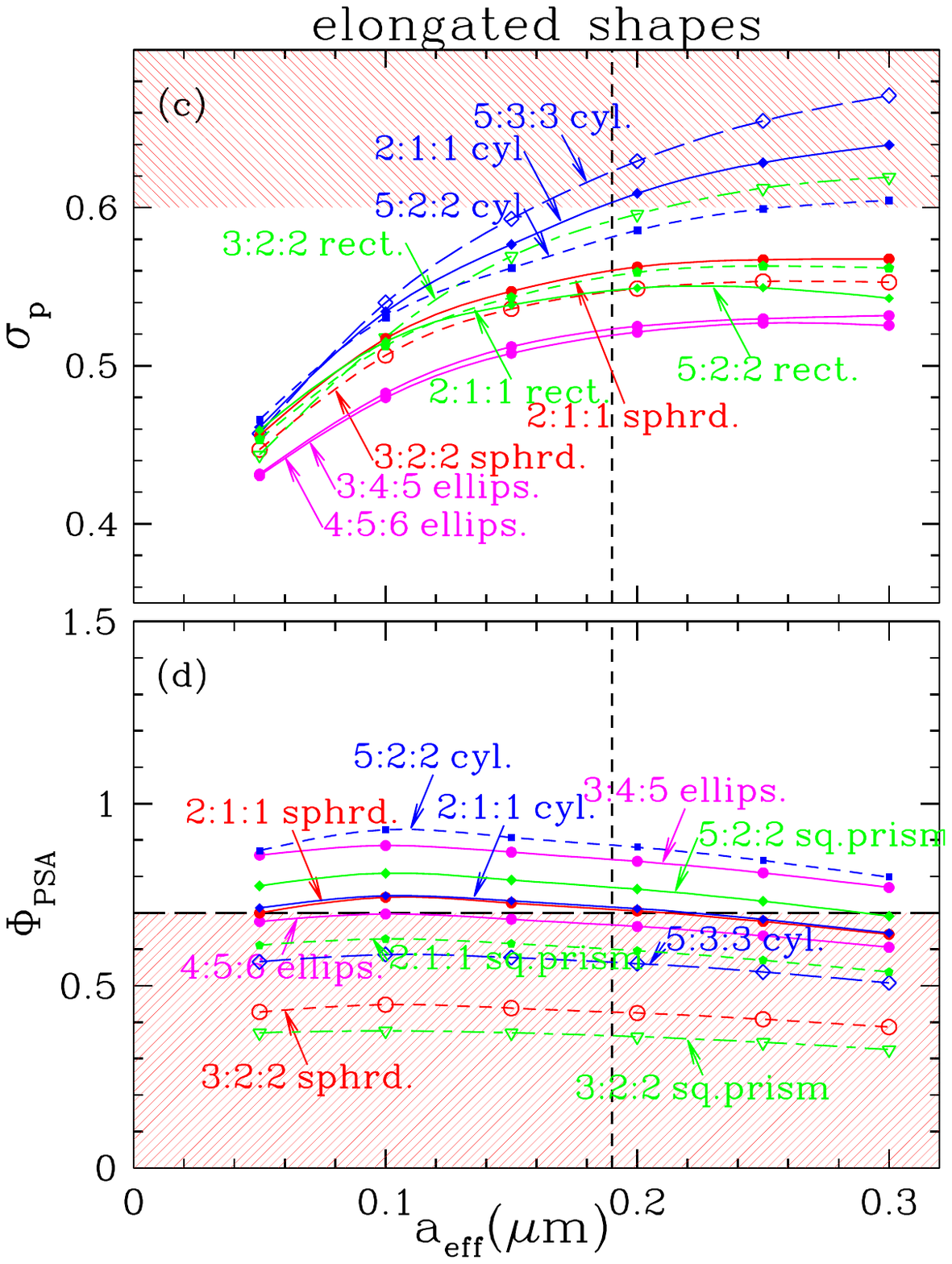}
\caption{\label{fig:spei}\footnotesize (a,c) Characteristic width
  $\sigmap$ of the polarization profile (see Equation \ref{eq:sigma})
  for different grain shapes, as a function of $\aeff$.  The shaded
  region is forbidden by the observed width of the interstellar
  polarization (see text).  (b,d) Polarization efficiency integral
  $\PhiPSA$ as a function of grain size $\aeff$.  The shaded region is
  forbidden by Equation
  (\ref{eq:minPhiPSA}).
  }
\end{center}
\end{figure}
\begin{figure}
\begin{center}
\includegraphics[angle=0,width=7.0cm,
                 clip=true,trim=0.5cm 5.0cm 0.5cm 2.5cm]
{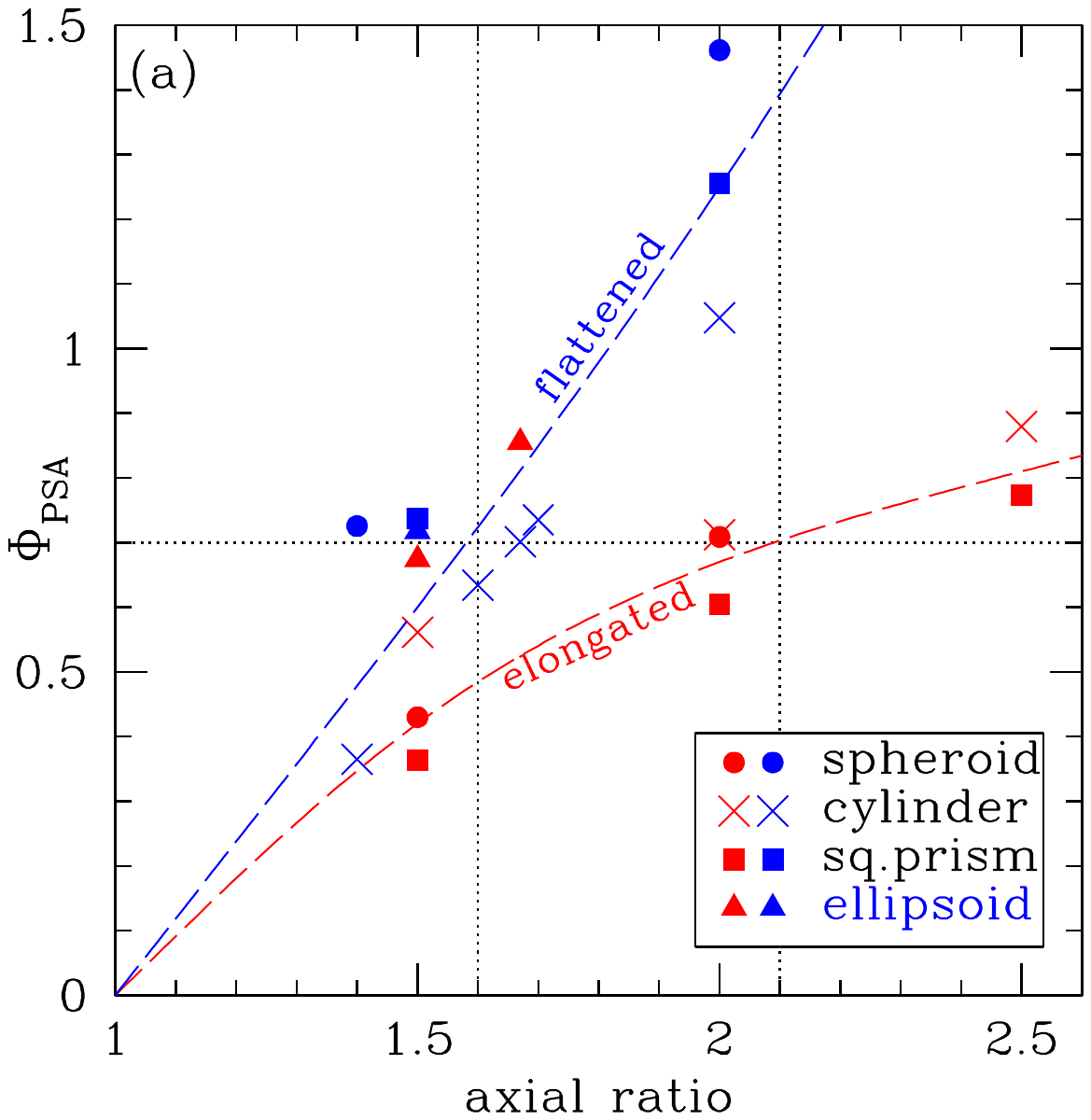}
\includegraphics[angle=0,width=7.0cm,
                 clip=true,trim=0.5cm 5.0cm 0.5cm 2.5cm]
{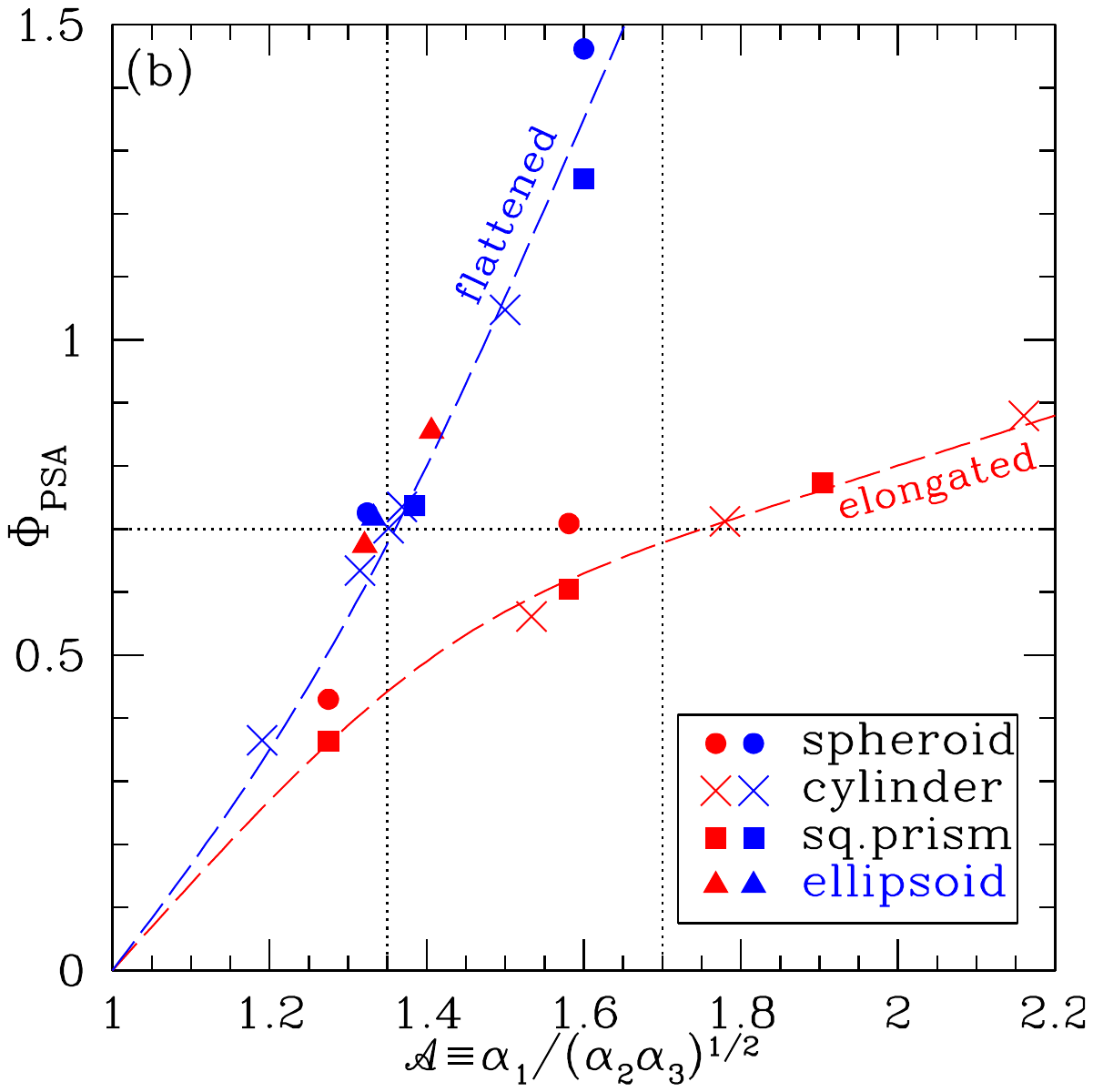}
\caption{\label{fig:phi}\footnotesize Starlight polarization
  efficiency $\PhiPSA$ vs.\ (a) axial ratio, and (b) $\Rone\equiv
  \alpha_1/(\alpha_2 \alpha_3)^{1/2}$ (Eq.\ \ref{eq:R1}).  Values
  shown are for $\aeff=0.2\micron$ grains.  Dashed curves show trends
  for flattened and elongated spheroids.
  }
\end{center}
\end{figure}
 
Consider now the strength of the observed polarization.
\citet{Draine+Hensley_2021c} defined the starlight
polarization efficiency integral:
\beq \label{eq:Phi}
\Phi(\shape,\aeff,\gamma) \equiv
\frac{1}{f_{\rm align}(\aeff)\sin^2\gamma}
\int_{\lambda_1}^{\lambda_2}
\frac{\langle C_{{\rm ext},y}(\lambda)\rangle-
      \langle C_{{\rm ext},x}(\lambda)\rangle}{2V} ~d\lambda
~~~,
\eeq
for radiation propagating in the $\bzhat$ direction and $\bB_0$ in the
$\bxhat-\bzhat$ plane; $\gamma$ is the angle between the magnetic
field and the $\bzhat$ axis, $\falign(\aeff)$ is the fractional
alignment,\footnote{%
   $f_{\rm align}(\aeff)\equiv
   \langle (\bahat_1\cdot\bbhat)^2\rangle$,
   where $\bahat_1$ is the principal axis of largest moment of inertia,
   and $\bbhat\equiv\bB_0/|\bB_0|$.}
and $\langle...\rangle$ denotes orientational average.  

The starlight polarization efficiency integral $\Phi$ is a
dimensionless measure of the ability of a dust grain to contribute to
starlight polarization. In principle, $\Phi$ depends on both the
magnetic field orientation $\gamma$ and the distribution of grain
alignments, but \citet{Draine+Hensley_2021c} showed that $\Phi$ is
nearly independent of both $\gamma$ and the degree of grain alignment,
both factors being effectively compensated for by the denominator
$\falign\sin^2\gamma$ in Equation (\ref{eq:Phi}).  Here we compute
$\PhiPSA$ for perfect spinning alignment ($\falign=1$) and $\bB_0$ in
the plane of the sky ($\sin^2\gamma=1$):
\beq \label{eq:Phi_PSA}
\PhiPSA(\shape,\aeff) \equiv 
\int_{\lambda_1}^{\lambda_2}
\frac{3\QpolPSA(\shape,\aeff,\lambda)}{4\aeff}\,d\lambda
~~~,
\eeq
with $\QpolPSA$ defined by Equation (\ref{eq:QpolPSA}).  Figure
\ref{fig:spei} shows $\PhiPSA$ as a function of $\aeff$ for the 20
shapes considered here, for the dielectric functions of Figure
\ref{fig:diel}.  As previously noted by \citet{Draine+Hensley_2021c}
for spheroids, $\PhiPSA$ is sensitive to the grain shape, but is
nearly independent of size $\aeff$ for $0.1\micron <\aeff <
0.3\micron$ -- the range of grain sizes primarily responsible for the
observed polarization of starlight.

Can one estimate $\PhiPSA$ from a simple measure of the geometric
shape, such as the axial ratio, or the asymmetry parameter $\Rone$
defined by Eq.\ (\ref{eq:R1})?  Figure \ref{fig:phi}a shows how
$\PhiPSA$ depends on axial ratio, while Figure \ref{fig:phi}b shows
$\PhiPSA$ versus $\Rone$.  $\PhiPSA$ tends to increase with increasing
axial ratio, or increasing asymmetry parameter $\Rone$.  If we
separate the shapes into flattened ($\Rtwo<1$) and elongated
($\Rtwo>1$), we see that the axial ratio is a fair predictor of
$\PhiPSA$, but $\PhiPSA$ seems to show an even tighter correlation
with $\Rone$.\footnote{%
   The two ``outliers'' in Figure \ref{fig:phi}b
   are triaxial ellipsoids with $\Rtwo=1.04$ and $1.06$ that
   seem to fall on the``flattened'' track.}

$\PhiPSA$ is calculated for grains spinning around the principal axis
of largest moment of inertia (i.e., ``shortest'' axis), as expected for
suprathermal rotation.  For a given axial ratio, flattened shapes
(with one short axis and two long axes) tend to have larger $\PhiPSA$
than elongated shapes (with two short axes and one long axis).

\section{\label{sec:discuss}
         Discussion}

\subsection{Constraints on Grain Shape from Starlight Polarization}

\begin{table}[b]
{\footnotesize
\begin{center}
\caption{\label{tab:allowed}Infrared Polarization Properties for Shapes with $\PhiPSA>0.7$, $\sigmap<0.6$}
\begin{tabular}{c c c c c c c}
\hline

\vspace*{-0.7em}\\

shape  & $\PhiPSA$$^a$ & $\sigmap$\,$^a$ & $[p_{\rm em}({\rm FIR})]_{\rm max}\,^b$ 
& $\left[\frac{\overline{\CpolPSA}(10\!-\!11\mu{\rm m})}{V\PhiPSA}\right]^c$
& $\left[\frac{\Delta\CpolPSA({\rm sil.})}{V\PhiPSA}\right]^d$ & comment \\
       &               &                &  & $(\cm^{-1})$ & $(\cm^{-1})$ &       \\            
\hline
3:5:5 cylinder     
       & 0.701 & 0.538 & $0.1746\pm0.0000$  & $1632\pm5$ & $1025\pm2$ &
         $\left[p_{\rm em}({\rm FIR})\right]_{\rm max}$ too small\\
{\bf 2:1:1 spheroid}
       & 0.709 & 0.560 & $0.2500\pm0.0001$  & $1830\pm2$ & $1225\pm2$ & OK \\
{\bf 10:13:15 ellipsoid} 
       & 0.718 & 0.516 & $0.2050\pm0.0001$  & $1688\pm5$ & $1082\pm5$ & OK \\
10:17:17 cylinder
       & 0.735 & 0.524 & $0.1744\pm0.0002$  & $1627\pm8$ & $1023\pm8$ &
         $\left[p_{\rm em}({\rm FIR})\right]_{\rm max}$ too small\\
{\bf 5:7:7 spheroid}
       & 0.736 & 0.516 & $0.2015\pm0.0003$  & $1682\pm1$ & $1075\pm3$ & OK \\
{\bf 5:2:2 square prism}
       & 0.773 & 0.546 & $0.2495\pm0.0003$  & $1835\pm1$ & $1246\pm1$ & OK \\
{\bf 3:4:5 ellipsoid}
       & 0.856 & 0.514 & $0.2014\pm0.0006$ & $1677\pm18$ & $1060\pm37$ & OK\\
5:2:2 cylinder
       & 0.880 & 0.586 & $0.2652\pm0.0008$  & $1897\pm9$ & $1305\pm3$ &
         $\left[p_{\rm em}({\rm FIR})\right]_{\rm max}$ too large\\
1:2:2 cylinder
       & 1.048 & 0.514 & $0.1710\pm0.0001$ & $1641\pm1$  & $1037\pm1$  &
        $\left[p_{\rm em}({\rm FIR})\right]_{\rm max}$ too small\\
1:2:2 square prism
       & 1.255 & 0.540 & $0.1773\pm0.0000$ & $1695\pm1$ & $1096\pm1$ &
        $\left[p_{\rm em}({\rm FIR})\right]_{\rm max}$ too large\\
1:2:2 spheroid
       & 1.461 & 0.498 & $0.1691\pm0.0001$ & $1641\pm2$ & $1034\pm10$ &
        $\left[p_{\rm em}({\rm FIR})\right]_{\rm max}$ too small\\
\hline
\multicolumn{7}{l}{$a$ $\PhiPSA$ and $\sigma_p$ from 
                   Table \ref{tab:Phi table}.}\\
\multicolumn{7}{l}{$b$ Maximum FIR polarization fraction
                   (Eq.\ \ref{eq:pfir_max}).}\\
\multicolumn{7}{l}{$c$ $(\CpolPSA/V)/\PhiPSA$ averaged over 10--11$\mu$m
                   (see Eq.\ \ref{eq:Cpol/VPhi}).}\\
\multicolumn{7}{l}{$d$ Silicate feature (see Eq.\ \ref{eq:DeltaCpol/Vphi}).}\\
\end{tabular}
\end{center}
}
\end{table}
The observed polarization integral $\Pi_{\rm obs}$ is proportional to
the mass-weighted fractional alignment $\langle\falign\rangle$, the
dust volume per H $\Vgr$, and the polarization efficiency integral
$\PhiPSA$:
\beq \label{eq:<falign>} 
\Pi_{\rm obs} = 
\langle \falign \rangle \Vgr \PhiPSA
~~~.
\eeq
Combining Equations (\ref{eq:Piobs/Vgr}) and (\ref{eq:<falign>}), we
obtain
\beq \label{eq:falign*Phi}
\langle\falign\rangle \PhiPSA \approx 0.49
~~~.
\eeq
The observed extinction curve rises steeply into the FUV; models
seeking to reproduce this rise
\citep[e.g.,][]{Mathis+Rumpl+Nordsieck_1977} require $\sim$30\% of the
dust mass in grains with $\aeff<0.1\micron$.  However, starlight
polarization is weak in the FUV; the Serkowski law
(\ref{eq:Serkowski}) can only be reproduced if grains with
$\aeff\ltsim 0.1\micron$ contribute minimally to the polarization.  If
all grains have the same shape, then the small ($\aeff\ltsim
0.1\micron$) grains cannot be appreciably aligned; this implies that
\beq
\langle\falign\rangle \ltsim 0.7
~~~.
\eeq
Thus, the grains responsible for
starlight polarization must have
\beq \label{eq:minPhiPSA}
\PhiPSA = \frac{0.49}{\langle\falign\rangle} \gtsim 0.70
~~~.
\eeq
From Figures \ref{fig:spei}b and \ref{fig:spei}d (or Table
\ref{tab:Phi table}), we see that 13 of the shapes considered here --
8 of the flattened shapes, and 5 of the elongated shapes -- have
$\PhiPSA>0.7$, and are therefore allowed by the \emph{strength} of
starlight polarization.  Figures \ref{fig:phi}a,b show how $\PhiPSA$
depends on axial ratio and asymmetry parameter $\Rone$.  Asymmetry
parameter $\Rone$ appears to be the better predictor of $\PhiPSA$.  In
order to produce sufficient starlight polarization, flattened shapes
($\Rtwo < 1$) should have $\Rone\gtsim 1.35$, and elongated shapes
($\Rtwo > 1$) should have $\Rone\gtsim 1.7$.

Seven of the 20 considered shapes have $\PhiPSA$ that is too small, and
therefore are ruled out by Equation (\ref{eq:minPhiPSA}).
The low values of $\PhiPSA$ imply that such shapes cannot contribute
substantially to the interstellar extinction.

In addition to satisfying (\ref{eq:minPhiPSA}), acceptable shapes must
have $\sigmap<0.60$ in order to reproduce the average Serkowski law
(\ref{eq:Serkowski}) with $K\approx0.87$.
Of the 20 shapes studied, only
11 (see Table \ref{tab:allowed}) are compatible with the Serkowski law
(\ref{eq:Serkowski}) with the strength described by
(\ref{eq:Piobs/Vgr}).

Recent work \citep{Angarita+Versteeg+Haverkorn+etal_2023} finds
$[\pmax/E(B-V)]_{\rm max}=0.158_{-0.009}^{+0.013}\,{\rm mag}^{-1}$,
i.e., 22\% more polarization per unit reddening that the value
$0.13\,{\rm mag}^{-1}$
\citep{Panopoulou+Hensley+Skalidis+etal_2019,Planck_2018_XII} adopted
in the present study.  If the larger value were to be adopted, then the
minimum $\PhiPSA$ would increase from 0.70 (Eq.\ \ref{eq:minPhiPSA}) to
0.85, ruling out some of the grain shapes (e.g. 2:1:1 spheroid, or
5:7:7 spheroid) which are allowed by $\PhiPSA>0.7$.  Further
observational work to better establish $[\pmax/E(B-V)]_{\rm max}$ is
needed.

\subsection{Constraints from FIR and Submm Polarization}

The thermal emission from aligned grains is polarized.  The grains are
in the Rayleigh limit $a\ll\lambda$, with $\Qext\approx\Qabs$.  The
fractional polarization of the thermal emission is maximized if
$\bB_0$ is in the plane of the sky, with
\beq
\left[p_{\rm em}(\lambda)\right]_{\rm max} = 
\frac{\langle\falign\rangle \QpolPSA(\lambda)}
     {\langle\falign\rangle \QextPSA(\lambda) 
     + (1-\langle\falign\rangle)\Qran(\lambda)}
~~~.
\eeq
Combining this with (\ref{eq:falign*Phi}) and the condition
$\langle\falign\rangle<0.7$ we obtain the predicted
maximum polarization of thermal emission:
\begin{figure}
\begin{center}
\includegraphics[angle=0,width=8.0cm,
                 clip=true,trim=0.5cm 0.5cm 0.5cm 0.5cm]
{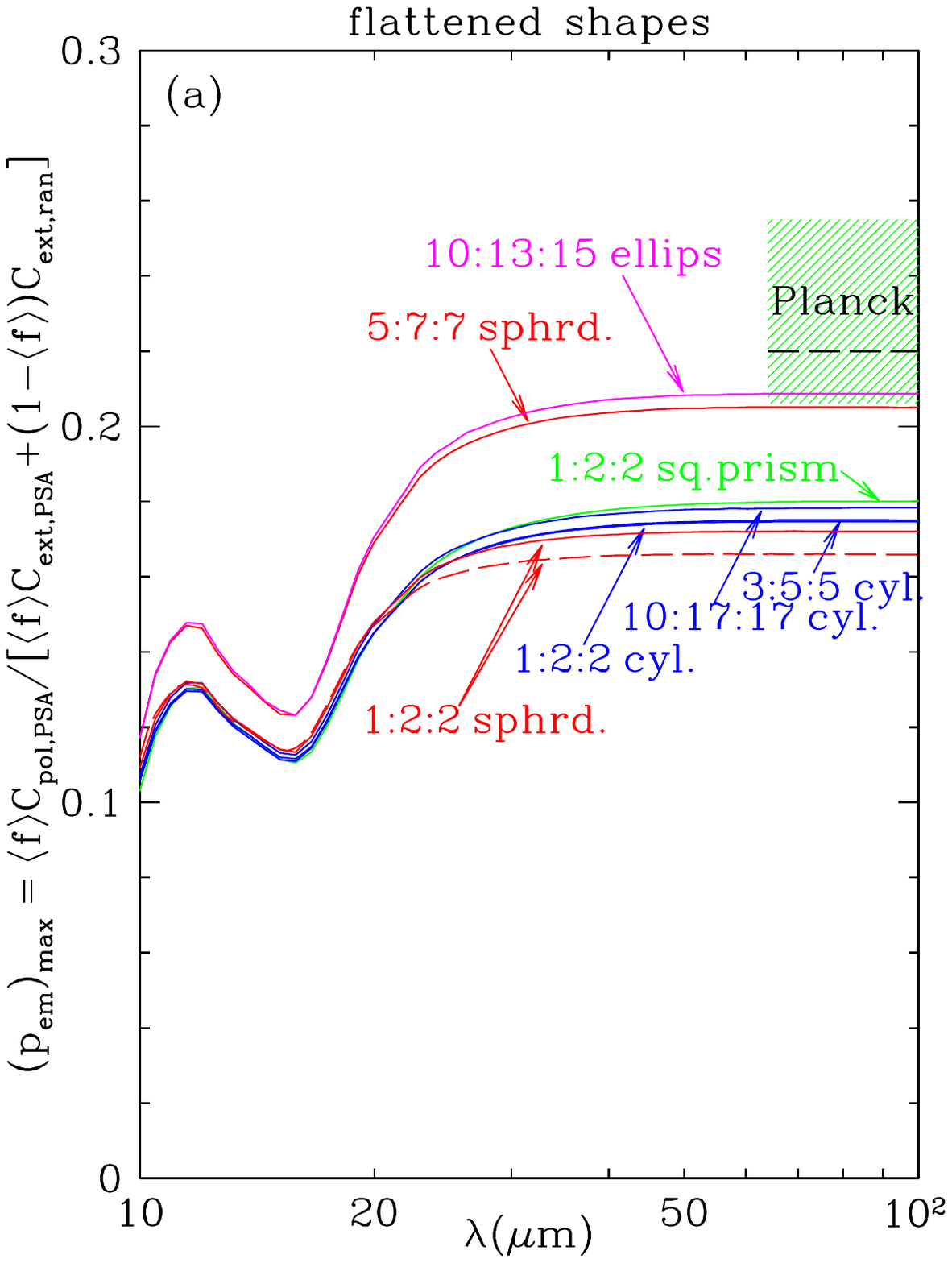}
\includegraphics[angle=0,width=8.0cm,
                 clip=true,trim=0.5cm 0.5cm 0.5cm 0.5cm]
{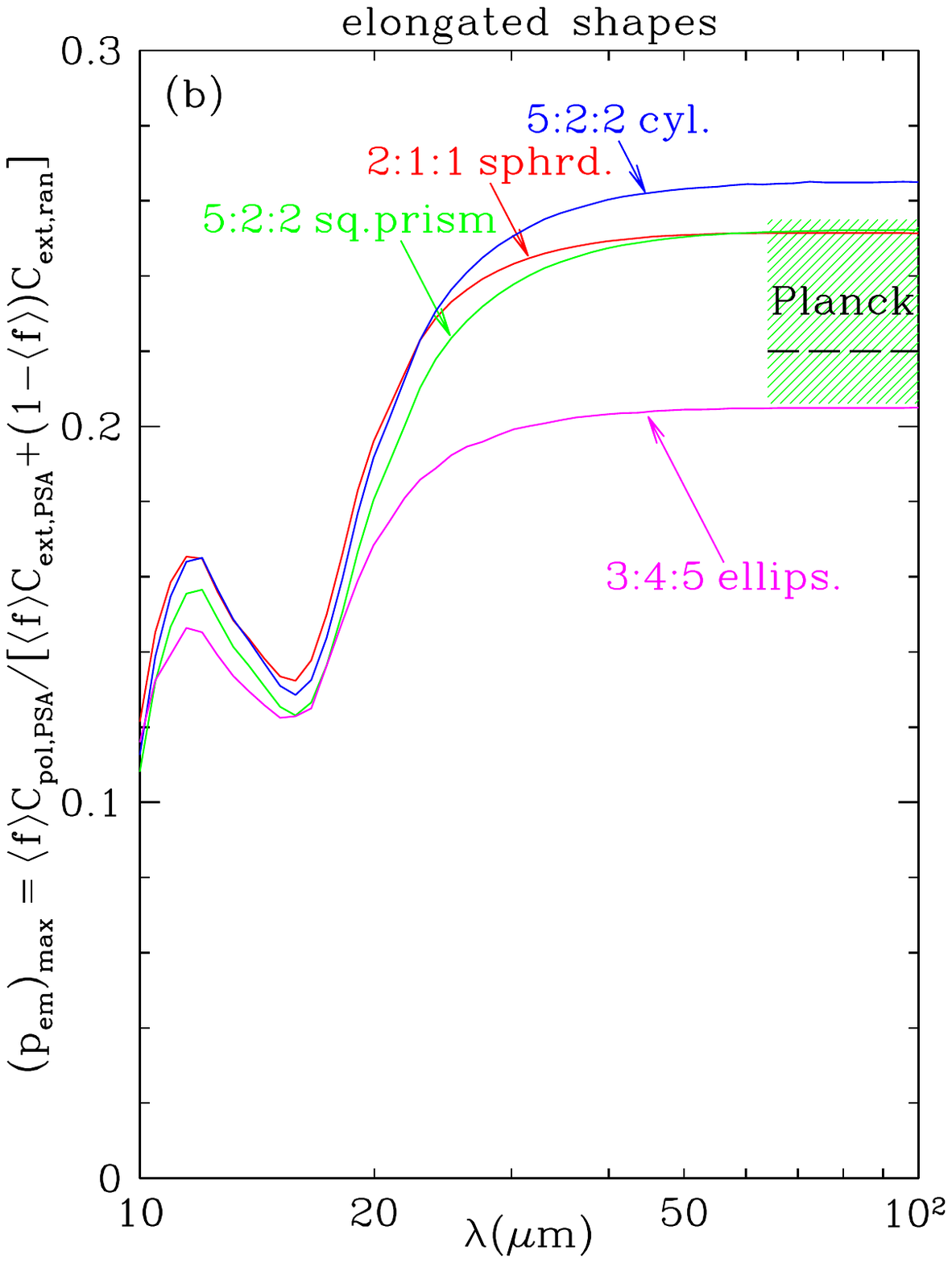}
\caption{\label{fig:pFIR/Phi}\footnotesize The maximum fractional
  polarization $(p_{\rm em})_{\rm max}$ for thermal emission as a
  function of $\lambda$ for (a) flattened shapes and (b) elongated
  shapes.  Results are shown for $\aeff=0.2\micron$, but $\CpolPSA/V$,
  $\CextPSA/V$, and $\PhiPSA$ are essentially independent of $\aeff$
  for the grain sizes of interest (see Figs.\
  \ref{fig:wave*Qext/a_vs_lambda}, \ref{fig:wave*Qpol/a_vs_lambda} and
  \ref{fig:spei}), and therefore so is $(\CpolPSA/\CextPSA)/\PhiPSA$.
  The green shaded zone shows the range $(p_{\rm em})_{\rm
    max}=0.220_{-0.014}^{+0.035}$ allowed by \citet{Planck_2018_XII}.
  (a) Results for the ``oblate'' dielectric function in Fig.\
  \ref{fig:diel}, except for the broken curve, calculated for the
  1:2:2 spheroid using a self-consistent dielectric function (see
  text) -- the reduction in FIR polarization is small.
  (b) Results for the ``prolate'' dielectric function in Fig.\ \ref{fig:diel}.
  }
\end{center}
\end{figure}
%
\beq \label{eq:pfir_max}
\left[p_{\rm em}(\lambda)\right]_{\rm max} =
\frac{\QpolPSA(\lambda)}
     {\QextPSA(\lambda) + 
      \left[{\rm max}\left(\frac{\PhiPSA}{0.49},\frac{1}{0.7}\right) - 1\right]
     \Qran(\lambda)}
~~~,
\eeq
plotted in Figure \ref{fig:pFIR/Phi} for the 11 shapes with
$\PhiPSA>0.7$ and $\sigmap<0.6$.  For all shapes, it is evident from
Figure \ref{fig:pFIR/Phi} that $\left[p_{\rm em}(\lambda)\right]
\rightarrow \left[p_{\rm em}({\rm FIR})\right]_{\rm max}= const$ for
$\lambda\gtsim50\micron$.  Thus we can compare the predicted
$\left[p_{\rm em}(100\micron)\right]_{\rm max}$ to the maximum
polarization fraction at $850\micron$ determined by
\citet{Planck_2018_XII}: $\left[p_{\rm em}(850\micron)\right]_{\rm
  max}\approx 0.220_{-0.014}^{+0.035}$, shown as the green shaded zone
in Figure \ref{fig:pFIR/Phi}.

Five shapes appear to be consistent with both starlight polarization
\emph{and} the Planck polarization constraint: 5:7:7 (oblate)
and 2:1:1 (prolate) spheroids\footnote{%
  \citet{Draine+Hensley_2021c} (DH21b) also found that 5:7:7 spheroids
  with porosity $\poromicro=0.2$ were viable, but concluded that 2:1:1
  spheroids produced too much submm polarized emission per unit
  optical polarization.  The slightly different conclusion of the
  present study is attributable to different choices of observational
  constraints.  DH21b used the Planck determination of $850\micron$
  polarized intensity per unit visual polarization in diffuse clouds
  \citep{Planck_2018_XII}, which requires an assumed grain
  temperature.  Here we instead use the \citet{Planck_2018_XII}
  estimates of the peak fractional polarization at $850\micron$, which
  does not require an assumed temperature, but does require accurate
  removal of unpolarized extragalactic backgrounds by the {\it Planck}
  team.};
10:13:15 and 3:4:5 triaxial ellipsoids; and 5:2:2 square prisms.

Some of the other shapes fall below or above the Planck constraints:

{\bf\emph{1:2:2 Oblate Spheroid:}} Using the ``oblate'' dielectric
function from Figure \ref{fig:diel}, this shape \emph{underpredicts}
the Planck fractional polarization, as previously found by
\citet{Draine+Hensley_2021c}.  Thus this shape is ruled out, despite
being compatible with the starlight polarization (see Table
\ref{tab:allowed}).

For the ``oblate'' dielectric function used here, the FIR-submm
opacity calculated for the 1:2:2 spheroid is too large by $\sim$20\%
(see Figure \ref{fig:figfir}a), which raises concern about the
calculated fractional polarization.  A dielectric function consistent
with the observed FIR-submm opacity reduces $\epsilon_2$ at long
wavelengths (to lower the opacity), which reduces $\epsilon_1$ at long
wavelengths, resulting in a \emph{reduction} in the fractional
polarization of the submm emission relative to what was calculated
using the ``oblate'' dielectric function from Figure \ref{fig:diel}.
For the 1:2:2 spheroid, a self-consistent $\epsilon(\lambda)$ was
obtained by \citet{Draine+Hensley_2021c}.  The polarization calculated
using this $\epsilon(\lambda)$ is shown by the dashed curve in Figure
\ref{fig:pFIR/Phi}a: $[p_{\rm em}(100\micron)]_{\rm max}$ drops, but
only slightly, from 17.1\% to 16.6\%.

{\bf\emph{Flattened Cylinders:}} The 10:17:17, 3:5:5, and 1:2:2
flattened cylinders all \emph{underpredict} the Planck fractional
polarization, and therefore are ruled out.  We note that with the
adopted dielectric function, these shapes have a FIR-submm opacity
that is too large by 22--30\% (see Figure \ref{fig:figfir}).  As
discussed above, a self-consistent calculation for these shapes would
\emph{further} reduce the submm fractional polarization.  Thus these
shapes are ruled out, despite being compatible with the starlight
polarization ($\PhiPSA>0.7$ and $\sigma_p<0.6$: see Table
\ref{tab:allowed}).

{\bf\emph{5:2:2 Elongated Cylinder:}} This shape \emph{overpredicts}
the submm fractional polarization.  With the adopted dielectric
functions, this shape also overpredicts the FIR opacity (see Figure
\ref{fig:figfir}).  As discussed above, a self-consistent
$\epsilon(\lambda)$ would lead to a small \emph{reduction} in the the
polarization fraction at long wavelengths, but we expect the change to
be small, as found for the 1:2:2 oblate spheroid.

\added{For simplicity, the present discussion has presumed that all
  grains have the same shape.  However, a mixture of shapes including
  some with $[p_{\rm em}]_{\rm max}<0.206$ and some with $[p_{\rm
      em}]_{\rm max}>0.255$ could collectively satisfy the
  \emph{Planck} constraint $[p_{\rm em}]_{\rm
    max}=0.220_{-0.014}^{+0.035}$\,.}

\subsection{On the Utility of Spheroids for Modeling Polarization}

\begin{figure}
\begin{center}
\includegraphics[angle=0,width=8.0cm,
                 clip=true,trim=0.5cm 5.0cm 0.5cm 2.5cm]
{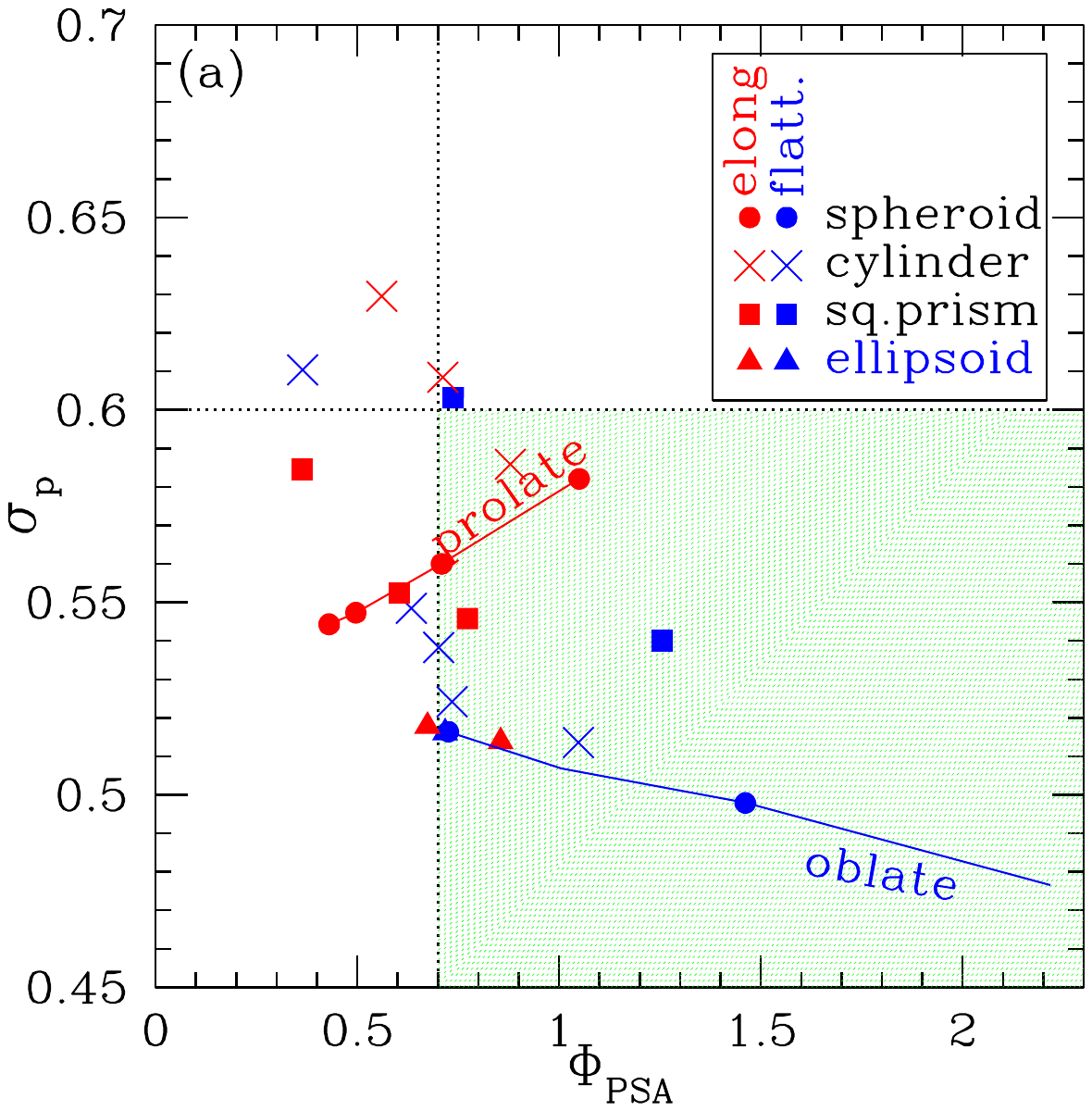}
\includegraphics[angle=0,width=8.0cm,
                 clip=true,trim=0.5cm 5.0cm 0.5cm 2.5cm]
{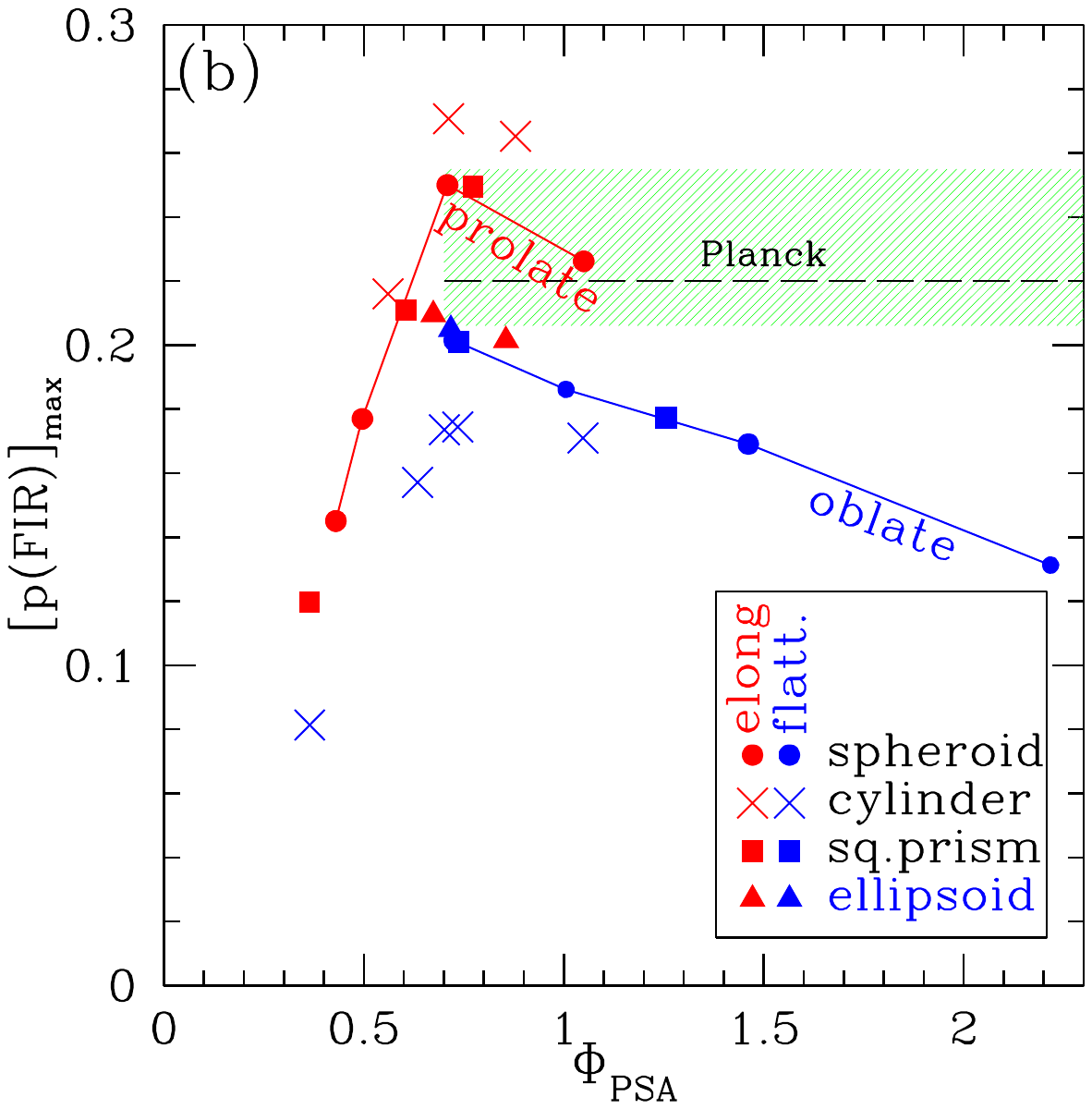}
\caption{\label{fig:sigma vs phi}\footnotesize (a) Polarization
  profile width parameter $\sigmap$ vs.\ starlight polarization
  efficiency integral $\PhiPSA$ for different shapes, for $\aeff$
  values such that $\lambdap=0.567\micron$.  (b) Maximum FIR-submm
  polarization (Eq.\ \ref{eq:pfir_max}) vs.\ $\PhiPSA$.  Symbols show
  the 20 shapes studied here using the DDA.  Solid curves show results
  for prolate spheroids with axial ratios 1.5 -- 3, and oblate
  spheroids with axial ratios 1.4 -- 3.
  }
\end{center}
\end{figure}
For the first time, we have studied the polarization properties of
\added{finite} nonspheroidal grain shapes \added{over a wide range of
wavelengths.\footnote{\citet{Hong+Greenberg_1980} and others
used results for infinite cylinders to estimate cross sections for finite
cylinders}}  Cross sections were calculated for a
variety of convex shapes, both flattened and elongated, from the FUV
($\lambda=0.1\micron$) to the FIR ($\lambda=100\micron$).

Some grain geometries can be ruled out, because they are unable
to generate sufficient starlight polarization (e.g., 3:2:2 spheroid),
because they produce a polarization profile that is broader than
observed (e.g., 2:3:3 square prism), or because they overpredict
(e.g., 5:2:2 cylinder) or underpredict (e.g., 3:5:5 cylinder) the
submm polarization fraction.  However, some of the considered
shapes remain viable.

For a given dielectric function $\epsilon(\lambda)$, we see from
Figure \ref{fig:wave*Qext/a_vs_lambda} that different shapes with the
same $\aeff$ have similar (although not identical) extinction cross
sections $\Cran(\lambda)$.  The polarization cross sections
$\CpolPSA(\lambda)$ exhibit more shape-to-shape variation (see Figure
\ref{fig:wave*Qpol/a_vs_lambda}).

Figure \ref{fig:sigma vs phi}a shows $\PhiPSA$ and $\sigmap$ for the
twenty shapes studied; the allowed region ($\PhiPSA>0.7$,
$\sigmap<0.6$) is shaded green.  For each shape (see Table
\ref{tab:Phi table}) $\aeff$ is taken to be such that
$\lambdap=0.567\micron$ as in the observed starlight polarization.
Figure \ref{fig:sigma vs phi}b shows the maximum FIR-submm
polarization vs.\ $\PhiPSA$ for the twenty convex shapes in this
study.  The allowed region $\PhiPSA>0.7$ and $[p({\rm FIR})]_{\rm
  max}=0.220_{-0.014}^{+0.035}$ \citep{Planck_2018_XII} is shaded
green.

It is much easier to calculate absorption and scattering cross
sections for spheroids than for any other finite nonspherical shape.
In the Rayleigh limit $\lambda\gg \aeff$, exact results for spheroids
are easily evaluated.  For larger $\aeff/\lambda$, methods based on
spheroidal wave functions \citep{Voshchinnikov+Farafonov_1993} are
much faster than the DDA for comparable levels of accuracy.

Prolate and oblate spheroids fall on the two tracks shown in Figure
\ref{fig:sigma vs phi}a.  For axial ratios $\leq3$, the spheroids have
$\sigmap<0.6$, consistent with the observed polarization; prolate and
oblate spheroids have $\PhiPSA > 0.7$ for axial ratio $>2$, and
$>1.4$, respectively.  Figure \ref{fig:sigma vs phi}a shows that
spheroids with suitable aspect ratios have $(\PhiPSA,\sigmap)$ values
passing through the allowed domain $\PhiPSA>0.7$, $\sigmap<0.6$.
Figure \ref{fig:sigma vs phi}b shows the corresponding tracks in the
$\PhiPSA-[p(\rm FIR)]_{\rm max}$ plane, where we see that the only
oblate shape that is allowed is $b/a=1.4$ (less flattened spheroids do
not provide enough starlight polarization; more flattened spheroids do
not have a large enough fractional polarization in the FIR).  It
appears that prolate shapes with axial ratios $a/b\geq 2$ are allowed.
 
Spheroids cannot exactly reproduce the light scattering properties of
specific nonspheroidal shapes, but can be used to approximate the
scattering properties of submicron convex shapes that have $\PhiPSA$,
$\sigmap$, and $[p({\rm FIR})]_{\rm max}$ in the allowed range.
While the detailed shape matters,
spheroids continue to be useful to explore the effects of grain
shape on polarized extinction and emission by interstellar grains.

\subsection{Which Axial Ratios are Favored?}

Models seeking to reproduce the wavelength dependences of both
extinction and polarization require that small
($\aeff\ltsim0.05\micron$) grains are essentially nonpolarizing, while
grains with $\aeff\gtsim 0.15\micron$ are substantially polarizing
(i.e., both significantly nonspherical \emph{and} appreciably
aligned), with $\falign(\aeff)\PhiPSA\approx 0.7$ for $\aeff\gtsim
0.15\micron$ (see, e.g., Figure 6 of \citet{Draine+Fraisse_2009} and
Figure 1 of \citet{Hensley+Draine_2023}).  As discussed above, in
order to reproduce the highest values of
$\pmax/E(B-V)\approx0.13{\,\rm mag}^{-1}$ observed for starlight
polarization, the grains must have starlight polarization efficiency
integral $\PhiPSA> 0.70$.  If a mixture of shapes is present, as
seems likely, the $\aeff \gtsim 0.1\micron$ grains must have
mass-weighted $\langle\PhiPSA\rangle > 0.7$.

The mechanisms responsible for alignment of interstellar grains remain
incompletely understood, but it is plausible that above some critical
size, starlight torques
\citep{Draine+Weingartner_1996,Draine+Weingartner_1997} maintain a
large fraction of the grains in suprathermal rotation, with high
degrees of alignment resulting from a combination of starlight torques
and paramagnetic or super-paramagnetic dissipation.  If
$\falign(0.2\micron)\approx 1$, then $\langle\falign\rangle \approx
0.7$, and the observed starlight polarization can be accounted for by
grains with $\PhiPSA\approx 0.49/\langle\falign\rangle \approx 0.7$.

For this reason, it seems likely that the ``typical'' grains with
$\aeff\gtsim 0.1\micron$ will have shapes corresponding to
$\PhiPSA\approx 0.7$.  From Figure \ref{fig:spei} and Table
\ref{tab:allowed} we see that this corresponds to an axial ratio
$\sim$$1.4$ for oblate spheroids, or $\sim$$2$ for prolate spheroids.
Other viable shapes (e.g., cylinders, square prisms, or trisphere)
will have axial ratios $\sim$1.5 if flattened, or $\sim$$2$ if
elongated.  However, the more extreme flattened shapes (e.g., 1:2:2
spheroid) are unable to account for the fractional polarization
observed by Planck at submm wavelengths (see Figure
\ref{fig:pFIR/Phi}), and, therefore, cannot be dominant
constituents.

\subsection{Polarization of the 10$\mu${\rm m} Silicate Feature}

\begin{figure}
\begin{center}
\includegraphics[angle=0,width=8.0cm,
                 clip=true,trim=0.5cm 5.0cm 0.5cm 2.5cm]
{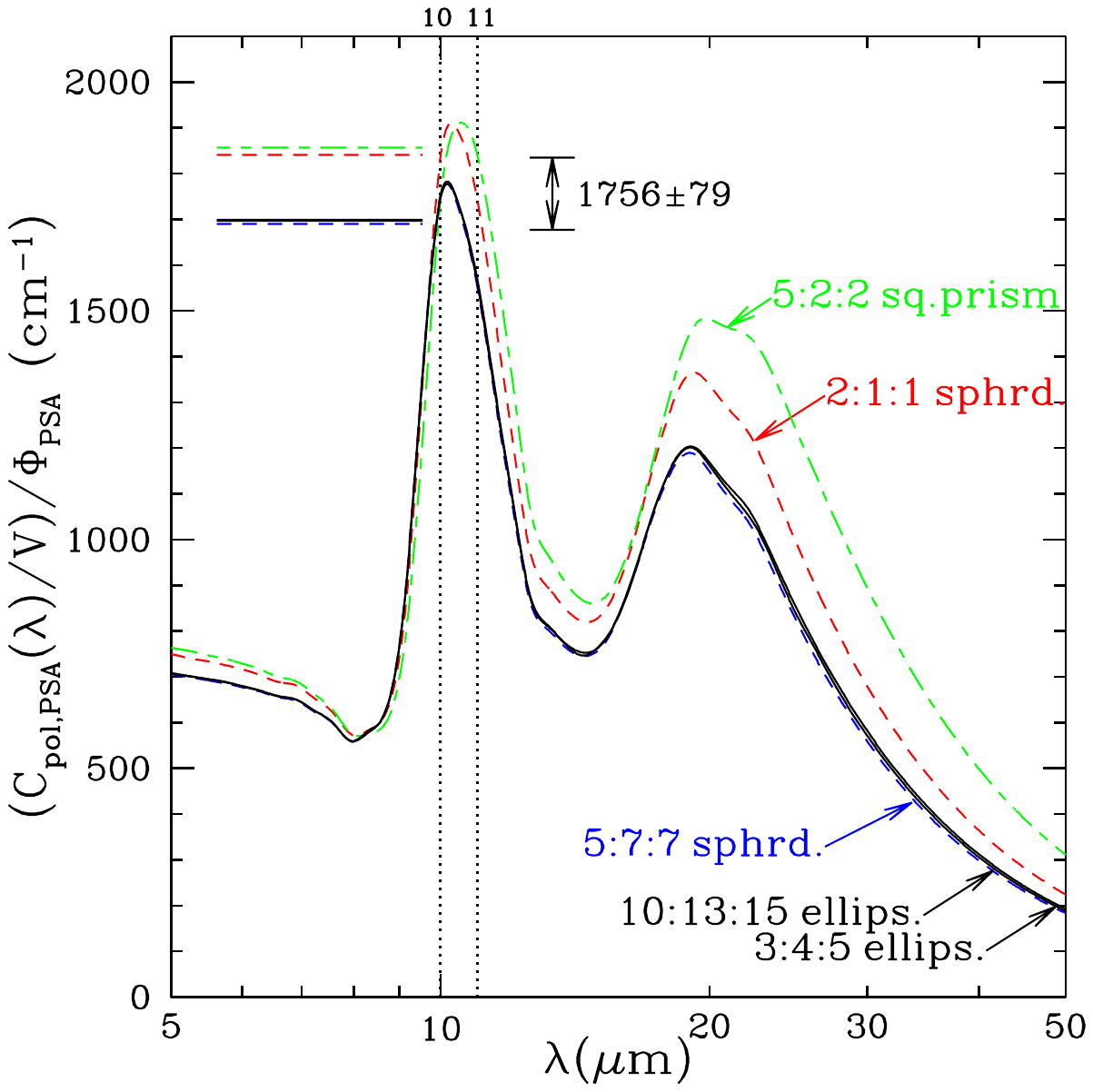}
\caption{\label{fig:Qpol/SPEI}\footnotesize $(\CpolPSA/V)/\PhiPSA$ as
  a function of $\lambda$ for 11 shapes allowed by starlight
  polarization (see text).  Results are shown for $\aeff$ values from
  Table \ref{tab:Phi table}, but $(\CpolPSA/V)/\PhiPSA$ is essentially
  independent of $\aeff$ for $\lambda > 5\micron$ and the grain sizes
  of interest (see Figures \ref{fig:wave*Qext/a_vs_lambda} and
  \ref{fig:spei}). Flattened shapes (solid curves) generally have
  lower values of $[\CpolPSA(10-11\micron)/V]/\PhiPSA$ than elongated
  shapes (broken curves).
  }
\end{center}
\end{figure}

For $\lambda\gtsim10\micron$, submicron grains are in the Rayleigh
limit $\aeff\ll\lambda$, and $\Cpol/V$ is independent of
$\aeff$. Initially unpolarized starlight will acquire a polarization
\beq \label{eq:p_in_Rayleigh_limit}
p(\lambda) = \NH \Vgr \langle\falign\rangle \sin^2\gamma 
\left[\frac{\CpolPSA(\lambda)}{V}\right]
~~~,
\eeq
provided $p\ltsim 0.1$ so that the polarization is 
$\sim$linear in the optical depth.
The observed starlight polarization integral (\ref{eq:OSPI}) is
\beq
 \Pi_{\rm obs} \approx 1.23\xtimes10^{-4} \pmax \cm =
\NH\Vgr\langle\falign\rangle \sin^2\gamma \PhiPSA
~~~.
\eeq
Thus
\beq \label{eq:NH_Vgr_falign_sin2gamma}
\NH\Vgr\langle\falign\rangle\sin^2\gamma = 
\frac{1.23\xtimes10^{-4}\pmax\cm}{\PhiPSA}
~~~.
\eeq
Substituting (\ref{eq:NH_Vgr_falign_sin2gamma}) into
(\ref{eq:p_in_Rayleigh_limit}) one obtains
\beq \label{eq:pir/pmax}
\frac{p(\lambda)}{\pmax} \approx 1.23\xtimes10^{-4}
\left[\frac{\CpolPSA(\lambda)/V}{\PhiPSA}\right] \cm
~~~.
\eeq
Figure \ref{fig:Qpol/SPEI} shows $[\CpolPSA/V]/\PhiPSA$ for $\lambda
\gtsim 5\micron$.  The grains are in the Rayleigh limit, hence
$\CpolPSA/V$ is independent of $\aeff$ (see Figure
\ref{fig:wave*Qpol/a_vs_lambda}).  We have already seen (Figure
\ref{fig:spei}) that $\PhiPSA$ is nearly independent of $\aeff$ for
the grain sizes ($0.10-0.30\micron$) that dominate polarization.  Thus
$[\CpolPSA/V]/\PhiPSA$ in Figure \ref{fig:Qpol/SPEI} is nearly
independent of $\aeff$ for $0.10\ltsim\aeff\ltsim0.30\micron$.
Averaging over the $10\micron\!-\!11\micron$ interval, the 5 shapes
with $\PhiPSA>0.7$, $\sigmap<0.6$, and $[p_{\rm em}({\rm FIR})]_{\rm max}
\in [0.20,0.25]$ have
\beq \label{eq:Cpol/VPhi}
\frac{\overline{\CpolPSA}(10\!-\!11\micron)/V}{\PhiPSA} \approx 1756\pm79 \cm^{-1} ~~.
\eeq 
Figure \ref{fig:Qpol/SPEI} shows that the flattened shapes have
lower values of $[\CpolPSA(10\!-\!11\micron)/V]/\PhiPSA$
compared to the elongated shapes.
The polarization associated with the silicate feature itself can be
characterized by the difference in mean polarization in the 10-11$\micron$
and the 8-9$\micron$ intervals:
\beq \label{eq:DeltaCpol/VPhi}
\frac{\Delta\CpolPSA({\rm sil.})}{V \PhiPSA} \equiv 
\frac{\overline{\CpolPSA}(10\!-\!11\micron)-\overline{\CpolPSA}(8\!-\!9\micron)}{V\,\PhiPSA}
= 1153\pm93\cm^{-1}
~~~.
\eeq

The shapes in Figure \ref{fig:Qpol/SPEI} range from smooth spheroids
to square solids (with sharp edges), and from flattened to elongated.
The fact that \emph{all} of the 11 convex shapes in Figure
\ref{fig:Qpol/SPEI} make very similar predictions for
$p(10-11\micron)/\pmax$ suggests that other convex or nearly-convex
shapes are likely to also fall in the range given by
Eq.\ (\ref{eq:Cpol/VPhi}).  From (\ref{eq:pir/pmax}) and
(\ref{eq:Cpol/VPhi}) one obtains
\beq \label{eq:DeltaCpol/Vphi}
\left(\frac{\overline{p}(10\!-\!11\micron)}{\pmax}\right)_{\rm theory}
\approx 0.216\pm0.010
~~~,
\eeq
[in agreement with the value 0.22 found for spheroids by
\citet{Draine+Hensley_2021c}] and
\beq \label{eq:theoryDelta}
\left(\frac{\Delta p_{\rm sil.}}{\pmax}\right)_{\rm theory} 
\approx 0.142\pm0.011
~~~.
\eeq

The sightline to Cyg OB2-12 is the only sightline for which both optical
and 10$\micron$ polarization have been measured \citep{Telesco+Varosi+Wright+etal_2022},
with $p(10\micron)=(1.24\pm0.28)\%$, 
$\overline{p}(10\!-\!11\micron)=(1.14\pm0.26)\%$ and 
$\overline{p}(8\!-\!9\micron)\approx (0.44\pm0.24)\%$.
With $\pmax=0.0967\pm0.0010$
\citep{Whittet+Martin+Hough+etal_1992}, we have
\beq \label{eq:p10/pmax}
\left(\frac{\overline{p}(10\!-\!11\micron)}{\pmax}\right)_{\rm \!observed} = 
0.118\pm
0.029
~~~,
\eeq
a $\sim$$3.0\sigma$ discrepancy with the smallest value ($0.206$)
allowed by Equation (\ref{eq:DeltaCpol/Vphi}) for the 5 shapes
considered here.  Similarly,
\beq \label{eq:Deltap/pmax}
\left(\frac{\Delta p_{\rm sil}}{p_{\rm max}}\right)_{\rm \!observed} 
= 0.072\pm0.025
~~~,
\eeq
a $2.4\sigma$ discrepancy with the smallest value (0.131) allowed by
Eq.\ (\ref{eq:theoryDelta}).

\medskip

It is striking that \emph{none} of the grain shapes considered here
can account for the observed low values of $p(10\!-\!11\micron)/\pmax$
and $\Delta p_{\rm sil}/p_{\rm max}$ toward Cyg OB2-12.  What does
this tell us?

\medskip

If the ratio $\overline{p}(10-11\micron)/\pmax=0.118\pm0.029$ found by
\citet{Telesco+Varosi+Wright+etal_2022} toward Cyg OB2-12 is confirmed
to apply to interstellar dust, then the following possibilities arise:
\begin{enumerate}
\item The actual grain shape may differ from the convex shapes
  considered here.
\item The adopted dielectric function may be incorrect.
\item The fundamental assumption of the astrodust model -- that
  extinction and polarization at both optical and infrared wavelengths
  is dominated by a single type of grain -- may be incorrect.
\end{enumerate}
We consider these in turn:

\subsubsection{Grain Shape?}

None of the convex shapes studied in this paper can reproduce the low
value of $\Cpol(10\micron)/\pmax$ observed toward Cyg OB2-12, but one
can ask whether non-convex shapes might be able to.  Paper II examines
a number of very different, and very non-convex, shapes: coagulates.
None of the coagulates studied in Paper II polarize strongly enough to
account for the observed starlight polarization, and thus cannot
explain the $10\micron$ polarization.  Given the wide range of shapes
considered here and in Paper II, it seems unlikely that merely
changing the grain shape can resolve the discrepancy between predicted
and observed values of $\overline{p}(10-11\micron)/\pmax$.

\subsubsection{Dielectric Function?}

The predicted $10\micron$ polarization for ``astrodust'' was based on
the dielectric function derived by DH21a from
the infrared extinction law determined by \citet{Hensley+Draine_2020}
for the sightline to Cyg OB2-12.  The excess extinction
$\Delta\tau_{\rm sil}\equiv\tau(9.7\micron)-\tau(8\micron)$ in the
$10\micron$ silicate feature can be reliably determined from the
observed spectrum of Cyg OB2-12 \citep{Hensley+Draine_2020}.  However,
the underlying continuum extinction in the 5--8$\micron$ region is
controversial, requiring both accurate absolute photometry and an
accurate estimate for the unobscured stellar flux.  A 20\% increase in
the calibration of the Spitzer IR Spectrograph
\citep{Houck+Roellig+vanCleve+etal_2004} (or a 20\% decrease in the
model for the flux from the photosphere + stellar wind at 8$\micron$)
would result in a factor of $\sim$2 reduction in the estimate for
$\tau(8\micron)$ toward Cyg OB2-12. The dielectric function used here
predicts that $\sim$$30\%$ of the polarization at $10\micron$ arises
from the underlying ``continuum'' opacity (i.e.,
$\CpolPSA(8\micron)\approx 0.30\CpolPSA(10\micron)$ in Figure
\ref{fig:Qpol/SPEI}).

A recent study by \citet{Gordon+Misselt+Bouwman+etal_2021} estimated
the 5--8$\micron$ extinction to be significantly lower than found by
\citet{Hensley+Draine_2020}. If the dielectric function used by the
astrodust model were to be revised to be consistent with the
extinction found by \citet{Gordon+Misselt+Bouwman+etal_2021},
the predicted polarization at $10\micron$ would be
reduced.

Because the physics is intrinsically nonlinear,\footnote{
   $p(10\micron)$ depends nonlinearly on both the real and imaginary 
   parts of the dielectric function.
} recalculation of $p(10\micron)/\pmax$ requires a new self-consistent
dielectric function, which will be the subject of future work.
However, a rough estimate can be made by assuming the polarization to
simply scale with the extinction.  If the 8$\micron$ extinction
$\tau(8\micron)$ were to be reduced by 50\% (say), with the feature
strength $\Delta\tau_{\rm sil}\equiv\tau(9.7\micron)-\tau(8\micron)$
unchanged, the resulting $p(10\micron)/\pmax$ might be reduced by
$\sim$$0.30\times50\%\approx 15\%$, so that the predicted
$p(10\micron)\approx 0.85\times0.022 = 0.018$.  The discrepancy with
the observed value $0.0124\pm0.0028$, would become $\sim$$2\sigma$ --
still appreciable.  Future work will derive a new dielectric function
consistent with the \citet{Gordon+Misselt+Bouwman+etal_2021}
extinction curve, but, based on the present discussion, this does not
seem likely to bring the model into agreement with the
\citet{Telesco+Varosi+Wright+etal_2022} result for
$p(10\micron)/\pmax$.

\subsubsection{The Astrodust Model?}

The astrodust model \citep{Hensley+Draine_2023}
postulates that most of the interstellar grain mass is provided by
grains with a single composition, incorporating both silicate and
non-silicate material in the same particles.  As we have seen above,
the model prediction for $p(10\micron)/\pmax$ appears to be
inconsistent with the measurement by
\citet{Telesco+Varosi+Wright+etal_2022} of $p(10\micron)/\pmax$ toward
Cyg OB2-12.

It is not clear how the model could be modified to accomodate the
measured value of $p(10\micron)/\pmax$.  One might try to concentrate
the silicate material in grains that are less aspherical, or less well
aligned, but the high ratio $[\pmax/E(B-V)]_{\rm max}=0.13$ for
starlight polarization \citep{Panopoulou+Hensley+Skalidis+etal_2019}
as well as submm polarization fractions exceeding 20\%
\citep{Planck_2018_XII} require that a large fraction of the dust
grains be both substantially nonspherical and highly aligned.  Given
that the silicate material must account for a major fraction of the
dust mass in the diffuse ISM, it is hard to see how a substantial
fraction of the silicate mass could be sequestered in grains that are
relatively inefficient polarizers.  In addition, if the silicate and
non-silicate materials are in separate grain populations (with
different opacities, shapes, and degrees of alignment), the
polarization fraction at FIR and submm wavelengths will be
wavelength-dependent, as in the models of \citet{Draine+Fraisse_2009},
whereas the observed polarization fraction appears to be nearly
independent of wavelength \citep[see][]{Hensley+Draine_2021}.

The 10$\micron$ polarization is evidently a powerful test of
interstellar grain models.  Further measurements of 10$\micron$
polarization (with CanariCam or other instruments) toward Cyg OB2-12
and other sightlines, and additional studies of the 4--8$\micron$
extinction (with JWST) will be of great value.

\subsubsection{Is Cyg OB2-12 A Suitable Target?}

Studies of the polarizing
properties of the dust toward Cyg OB2-12 have assumed that the light
from Cyg OB2-12 is unpolarized, so that the observed polarization is
entirely the result of extinction by aligned dust grains along the
line-of-sight.  Is this correct?

Cyg OB2-12 is a very unusual star: a blue hypergiant, classified as
B5Ia$^+$ by \citet{Humphreys_1978}, and as B3Iae by
\citet{Kiminki+Kobulnicky+Kinemuchi+etal_2007}.  At an assumed
distance $D=1.75\kpc$, it is one of the most luminous stars in the
Galaxy, with luminosity $L=1.4\xtimes10^6\Lsol$
\citep{Clark+Najarro+Negueruela+etal_2012, Hensley+Draine_2020}. Cyg
OB2-12 varies in brightness \citep[$\Delta B\approx
  0.3$mag:][]{Gottlieb+Liller_1978, Laur+Tuvikene+Eenmae+etal_2012,
  Salas+Maiz-Apellaniz+Barba_2015}, and in spectrum, with spectral
class changing from B3 to B8 in 12 months
\citep{Kiminki+Kobulnicky+Kinemuchi+etal_2007}.  The stellar wind
produces absorption and emission features
\citep{Clark+Najarro+Negueruela+etal_2012, Hensley+Draine_2020}.  The
ionized wind is detected at cm wavelengths, showing unexpected
variability at 21cm \citep{Morford+Fenech+Prinja+etal_2016}.

Cyg OB2-12 appears to be a member of a triple system, with a companion
at projected separation $\sim$$100\AU$, and a second companion at
projected separation $\sim$$2000\AU$
\citep{Caballero-Nieves+Nelan+Gies+etal_2014,Maryeva+Chentsov+Goranskij+etal_2016}.
\citet{Klochkova+Islentieva+Panchuk_2022} report radial velocity
variations of amplitude $\sim$$8\kms$; if periodic, the period must be
$\ltsim2\yr$.  \citet{Oskinova+Huenemoerder+Hamann+etal_2017} argue
that Cyg OB2-12 is a colliding-wind binary.

The polarization position angle PA is wavelength-dependent.  Pre-2015
observations were consistent with
\beq \label{eq:PAtrend}
{\rm PA} \approx 115^\circ + 1.7^\circ (\micron/\lambda) ~~{\rm for}~~ 
0.44\micron < \lambda < 2.2\micron ~~, 
\eeq
\citep[see Figure 4 of][]{Whittet_2015} which can be explained by a
model with two separate dust layers along the line of sight, with
different alignment directions (PA=56$^\circ$ and PA=125$^\circ$), and
different wavelength dependence for the polarization
\citep{McMillan+Tapia_1977,Whittet_2015}. Such a two-layer model is
consistent with the observed circular polarization
\citep{Martin+Angel_1976}.  However,
\citet{Telesco+Varosi+Wright+etal_2022} observe ${\rm PA}= 126^\circ$
at $\lambda=10\micron$, inconsistent with (\ref{eq:PAtrend}), and
inconsistent with the two-layer model of \citet{Whittet_2015}, unless
the ``foreground'' dust layer (assumed to contribute $A_V=3.6$mag and
$p_V=0.0335$) for some reason contributes zero
polarization at $10\micron$.

Furthermore, there are indications that the linear polarization may be
time-variable.  In the J band, \citet{Dyck+Jones_1978} found $p({\rm
  J})=0.0331 \pm 0.0017$, \citet{Wilking+Lebofsky+Martin+etal_1980}
found $p({\rm J})=0.0410 \pm 0.0013$, and \citet{Bailey+Hough_1982}
found $p({\rm J})=0.0370 \pm 0.0008$.  In the K band,
\citet{Wilking+Lebofsky+Martin+etal_1980} found $p({\rm K})=0.0133 \pm
0.0005$, and \citet{Bailey+Hough_1982} found $p({\rm K})=0.0113 \pm
0.0003$.  \citet{Blinov+Maharana+Bouzelou+etal_2023} report
substantial variations of $U/I$ and $Q/I$ over 2400 days of
observations.

If the polarization of Cyg OB2-12 is actually variable on a
$\sim$$1\yr$ time scale, the variations must be due to changing
polarization of the star itself, in which case Cyg OB2-12 becomes much
less useful for studying polarization by interstellar dust, and
Eqs.\ (\ref{eq:p10/pmax}, \ref{eq:Deltap/pmax}) need not apply to
interstellar dust.  Intrinsic stellar polarization in Cyg OB2-12 
could arise from electron scattering in the ionized outflow if it 
departs from spherical symmetry.\footnote{%
   The Cyg OB2-12 wind model
   \citep{Clark+Najarro+Negueruela+etal_2012,Hensley+Draine_2020} has
   $\sigma_{\rm T} \int_{R_*}^\infty n_e(r)dr = 0.43$, where
   $\sigma_{\rm T}$ is the Thompson scattering cross section.  Thus a
   large fraction of the photons emitted by the photosphere will be
   scattered in the wind, leading to net polarization if the wind is
   asymmetric.}
It is important to remeasure both optical and mid-IR polarization
toward Cyg OB2-12, to test for time-dependence.  If Cyg OB2-12
has intrinsic polarization, there will be a strong need for
measurements of both optical and $10\micron$ polarization for other
sightlines.

\added{

\subsection{Polarization of the 3.4$\mu$m Feature}

The ``astrodust'' dielectric function used here includes an absorption
feature at $3.4\micron$ arising from the C-H stretching mode in
aliphatic hydrocarbons assumed to be present in ``astrodust''.  DH21a
noted that this model predicts a polarization feature at $3.4\micron$
that exceeds observational upper limits.  DH21a proposed that if the
$3.4\micron$ absorbers are present only near the grain surface, rather
than throughout the grain volume, the $3.4\micron$ polarization will
be weakened because a larger fraction of the $3.4\micron$ absorption
will then be contributed by minimally-aligned $a\ltsim 0.1\micron$
grains.  DH21a estimated that if the $3.4\micron$ absorbers are
located only in a surface layer of thickness $\sim$$0.010\micron$, the
predicted polarization in the $3.4\micron$ feature becomes comparable
to current upper limits.

If $3.4\micron$ C-H absorption is concentrated near the grain surface,
the astrodust model should be modified to allow for spatial variation
in the dielectric function within the grain, at least near
$3.4\micron$; elaboration along these lines will be considered in
future work.  Polarization in the $3.4\micron$ feature can be weakened
but not eliminated: further improvements in mid-IR spectropolarietry
should detect polarization of the $3.4\micron$ feature.

Just as for the silicate feature, polarization of the $3.4\micron$
feature relative to optical polarization (not shown here) is
relatively insensitive to shape for the compact convex shapes
considered here.

}

\section{\label{sec:summary}
         Summary}

The principal results are as follows:
\begin{enumerate}

\item The starlight polarization efficiency integral $\PhiPSA$ and the
  polarization width $\sigmap$ are integral properties of grains that
  can be used to test grain candidates.  Interstellar grains should
  have $\PhiPSA>0.7$ and $\sigmap<0.6$.

\item An additional test is provided by the submillimeter
  polarization: some shapes predict too much submm polarization
  (relative to optical polarization), while others predict too little.

\item Twenty convex shapes were studied using the DDA: spheroids,
  cylinders, square prisms, and triaxial ellipsoids.  Only 11 of these
  shapes have $\PhiPSA(\aeff)> 0.7$, and $\sigma_p<0.6$, as required
  to reproduce the polarization of starlight.  Only 5 of the 20 shapes
  studied are compatible with both starlight polarization \emph{and}
  submm polarization (see Table \ref{tab:Phi table}, Table
  \ref{tab:allowed}, and Figure \ref{fig:sigma vs phi}).

\item While the extinction and polarization cross sections per volume
  $\Cran(\lambda)/V$ and $\Cpol(\lambda)/V$ vary from shape to shape,
  shapes that are consistent with observational constraints can be
  well-approximated by spheroids with appropriate axis ratios.

\item All shapes have similar $[\CpolPSA(\lambda)/V]/\PhiPSA$ at
  mid-infrared wavelengths 8--15$\micron$.  The different shapes
  therefore predict similar ratios $p(10\micron)/\pmax$, where $\pmax$
  is the peak starlight polarization.

\item The discrepancy between the predicted
  \citep{Draine+Hensley_2021c} and observed
  \citep{Telesco+Varosi+Wright+etal_2022} 10$\micron$ polarization of
  Cyg OB2-12 does not seem likely to be resolved by varying the
  assumed grain shape.

\item Cyg OB2-12 may be intrinsically polarized, which would call into
  question its use to study interstellar polarization in the infrared.
  It is important to measure $p(10\micron)/\pmax$ on other sightlines.

\end{enumerate}
\begin{acknowledgements}

This work was supported in part by NSF grant AST-1908123.  I thank
C.\ Telesco and F.\ Varosi for kindly providing machine-readable data
for Cyg OB2-12, Robert Lupton for availability of the SM package,
\added{and the anonymous referee for helpful comments.}

\end{acknowledgements}

\bibliography{/u/draine/work/libe/btdrefs}

\appendix

\section{\label{app:DDA}
         Applying the Discrete Dipole Approximation}

We seek to calculate scattering and absorption of monochromatic
radiation by targets composed of material characterized by a complex
dielectric function $\epsilon(\lambda)$.  The discrete dipole
approximation (DDA) consists of replacing the continuum target by an
array of polarizable points
\citep{Purcell+Pennypacker_1973,Draine_1988,Draine+Flatau_1994}, and
solving the scattering problem for this array.  The calculations are
greatly accelerated by employing fast Fourier transforms
\citep{Goodman+Draine+Flatau_1991}.  


The dipoles are located on a cubic lattice with interdipole separation
$d$.  We employ the public domain code {\tt DDSCAT 7.3.3}$^1$.  The
{\tt DDSCAT} error tolerance parameter is set to {\tt
  TOL}=$1\xtimes10^{-5}$.

\begin{table}[b]
\begin{center}
{\footnotesize
\caption{\label{tab:N values}
         Numbers of dipoles used}
\begin{tabular}{l c c c}
\hline
shape              & $N_1$   & $N_2$   & $N_3$\\
\hline
5:7:7 cylinder     & 1538000 &  788480 & 404480 \\
5:8:8 cylinder     & 1031360 &  433920 & 129120 \\
5:7:7 spheroid     & 2199512 &  793312 & 336264 \\
10:13:15 ellipsoid & 1031224 &  306912 & 130162 \\
3:5:5 cylinder     &  813888 &  241152 & 101180 \\
10:17:17 cylinder  &  847584 &  491040 & 206685 \\
2:3:3 square prism & 1152000 &  589824 & 248832 \\
1:2:2 cylinder     &  825088 &  374136 & 103296 \\
1:2:2 spheroid     & 1082288 &  458136 & 136680 \\
1:2:2 square prism & 1048576 &  442368 & 131072 \\
\hline
3:2:2 square prism &  768000 &  324000 & 165888 \\
3:2:2 spheroid     &  793152 &  407144 & 172640 \\
4:5:6 ellipsoid    &  509216 &  295316 & 125206 \\
3:4:5 ellipsoid    &  856744 &  439792 & 108120 \\
5:3:3 cylinder     &  956400 &  488160 & 206280 \\
2:1:1 square prism &  524288 &  221184 &  65536 \\
2:1:1 spheroid     &  769792 &  277888 & 117640 \\
2:1:1 cylinder     & 1147680 &  413184 & 173769 \\
5:2:2 square prism &  933120 &  274680 &  81920 \\
5:2:2 cylinder     &  730800 &  424200 & 216480 \\
\hline
\end{tabular}}
\end{center}
\end{table}
Accuracy of the DDA requires that the dipole array provide a good
geometric approximation to the actual target, \emph{and} that the
dipole separation $d$ be small enough that the phase shift between
lattice sites be small.  \citet{Draine+Flatau_1994} recommended
\beq
|m|kd \ltsim 0.5
\eeq
which corresponds to a requirement on the number of dipoles:
\beq \label{eq:N required}
N \gtsim 8300 |m|^3 \left(\frac{\aeff}{\lambda}\right)^3
~~~.
\eeq
At $\lambda=0.1\micron$, the astrodust dielectric function
$m=1.41+0.70i$.  Eq.\ (\ref{eq:N required}) becomes
\beq
N(\lambda=0.1\micron) > 3.3\xtimes10^4 \left(\frac{\aeff}{0.1\micron}\right)^3
~~~.
\eeq
For $\aeff=0.2\micron$ this becomes $N > 2.6\xtimes10^5$.  For each of
the 20 shapes considered here we repeat the DDA calculations for three
values $N_1>N_2>N_3$ (see Table \ref{tab:N values}) with
$N_1>5\xtimes10^5$ for every shape to ensure accuracy.  For cases with
$N>5\xtimes10^5$, the {\tt DDSCAT} calculations are carried out using
64-bit arithmetic to avoid round-off error.  Parameter files ({\tt
  ddscat.par}) as well as files specifying the dielectric functions
used in the {\tt DDSCAT} calculations are available online at
\dataset[Replication Data for: Polarization and Grain Shape: Convex
  Shapes]{https://doi.org/10.7910/DVN/QALLG8}.

\section{Accuracy of the DDA \label{app:DDA accuracy}}
\subsection{Spheres}

\begin{figure}
\begin{center}
\includegraphics[angle=0,width=4.4cm,
                 clip=true,trim=0.5cm 5.0cm 0.5cm 2.5cm]
{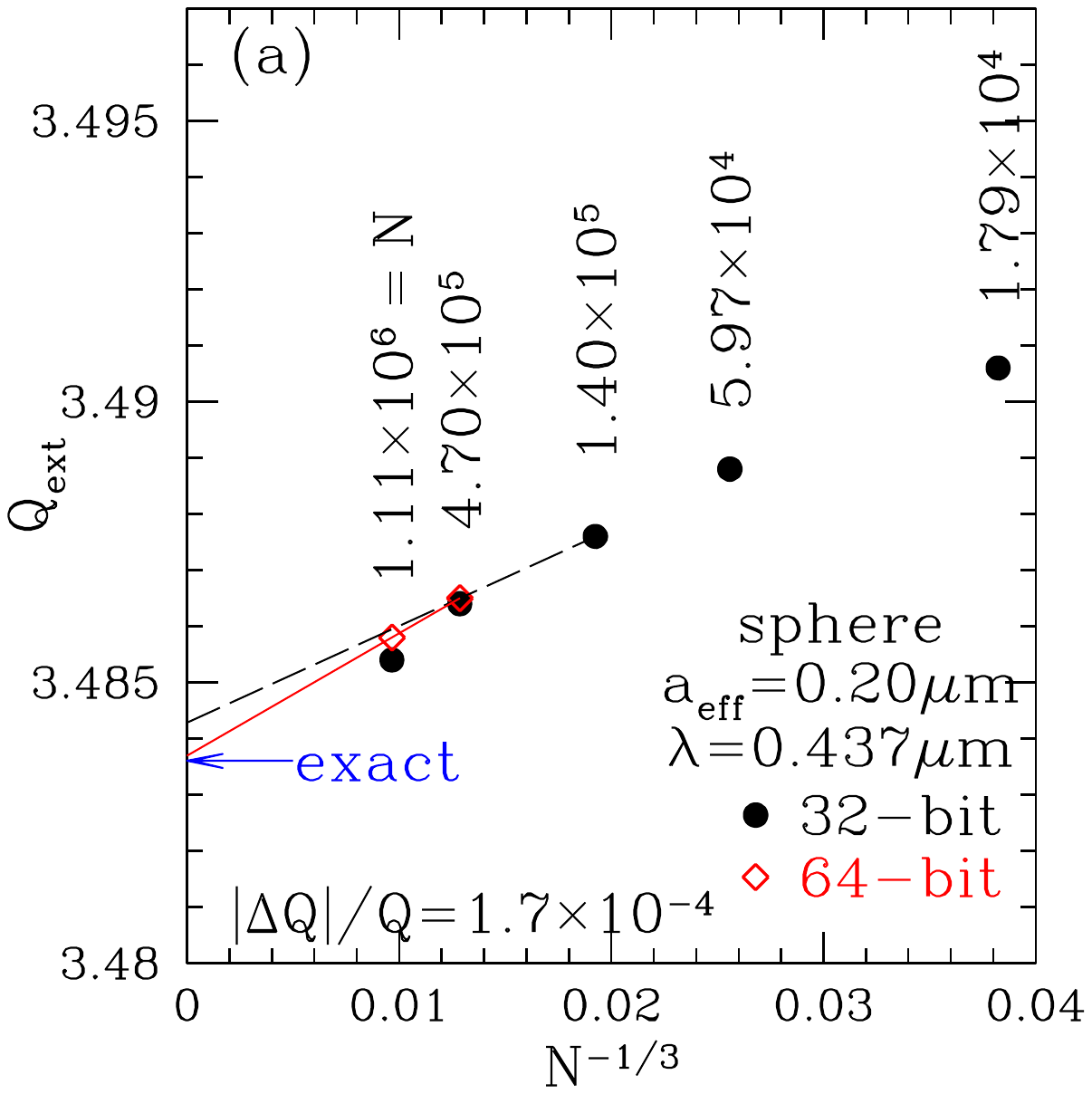}
\includegraphics[angle=0,width=4.4cm,
                 clip=true,trim=0.5cm 5.0cm 0.5cm 2.5cm]
{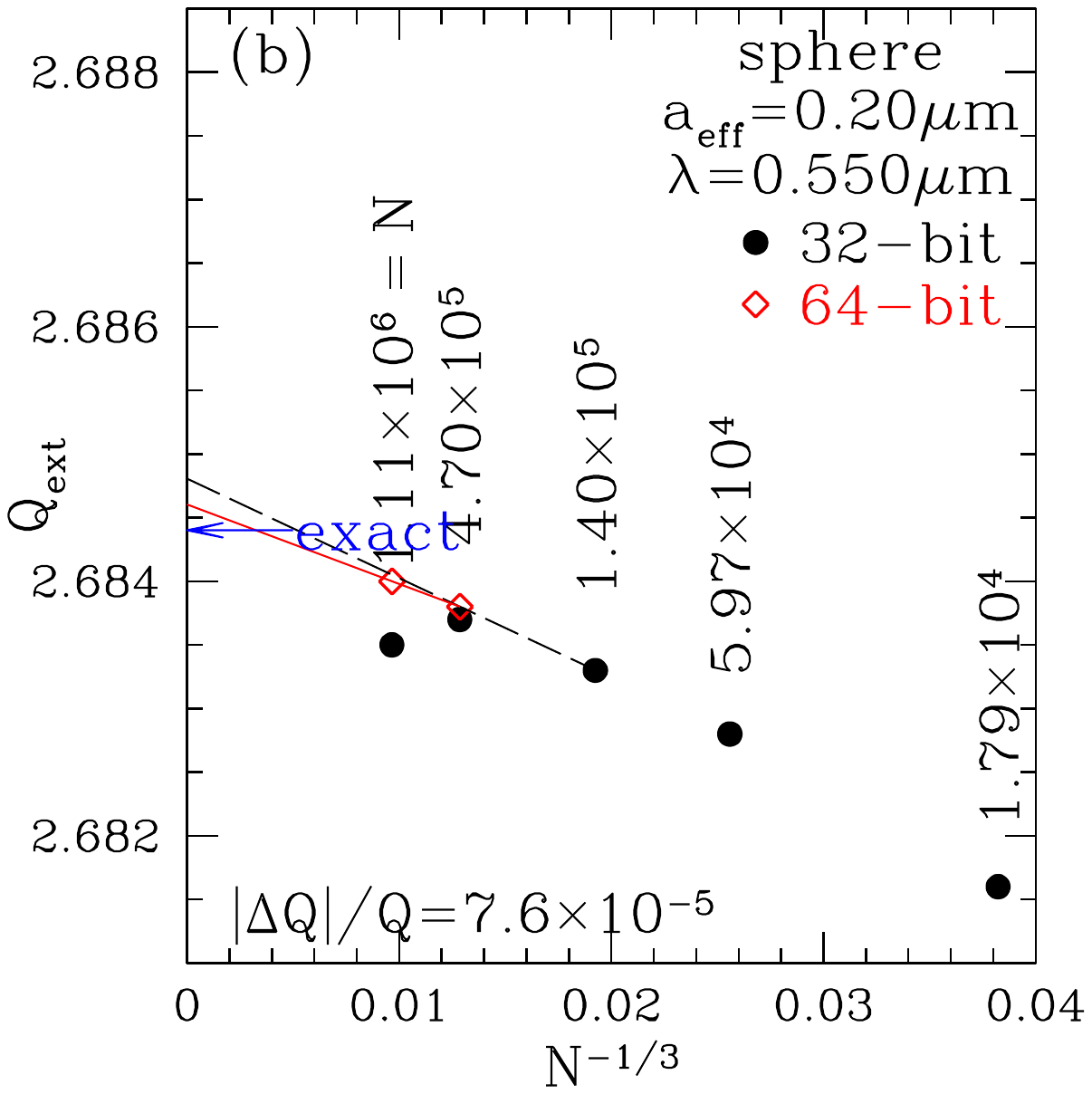}
\includegraphics[angle=0,width=4.4cm,
                 clip=true,trim=0.5cm 5.0cm 0.5cm 2.5cm]
{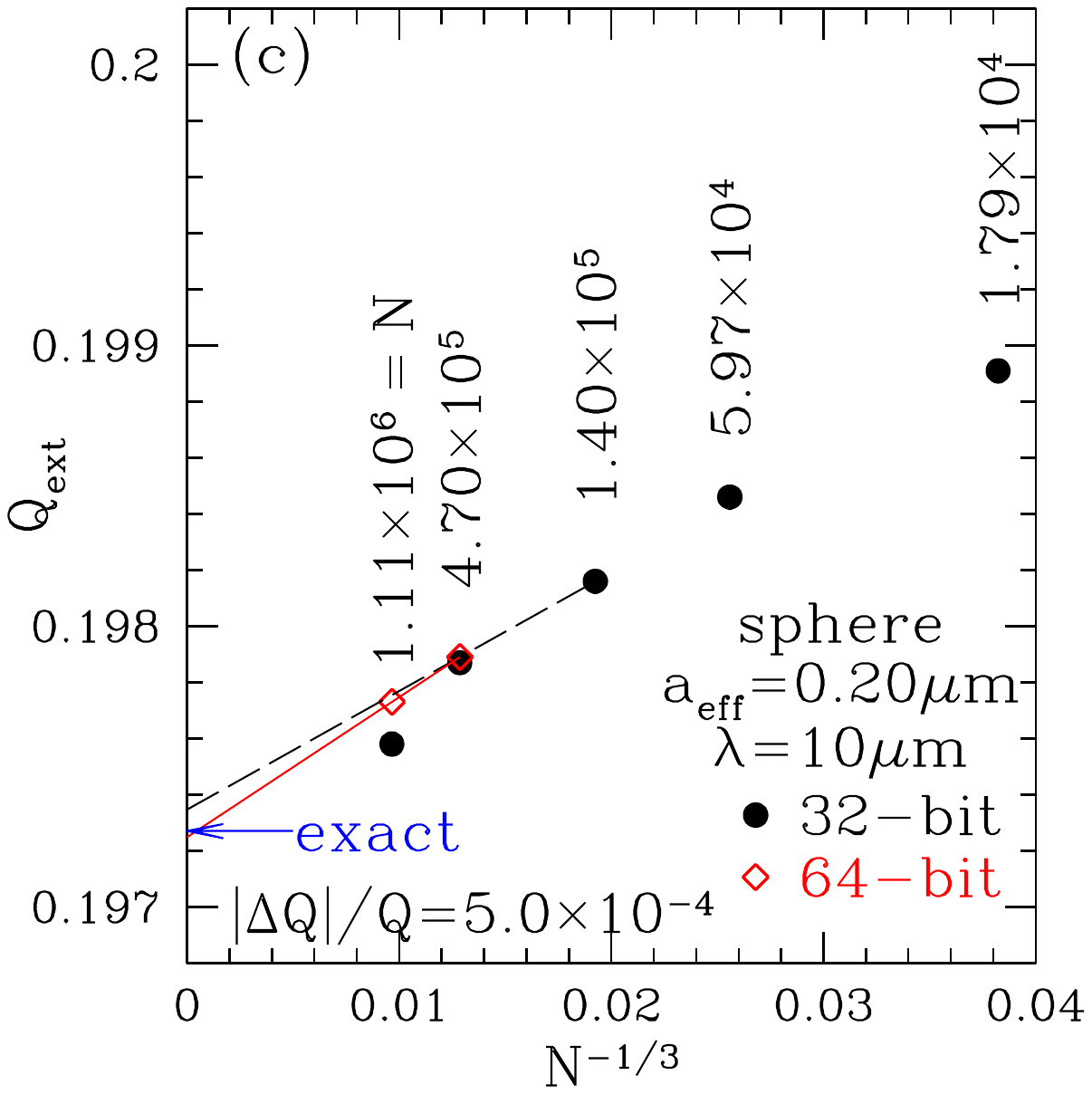}
\includegraphics[angle=0,width=4.4cm,
                 clip=true,trim=0.5cm 5.0cm 0.5cm 2.5cm]
{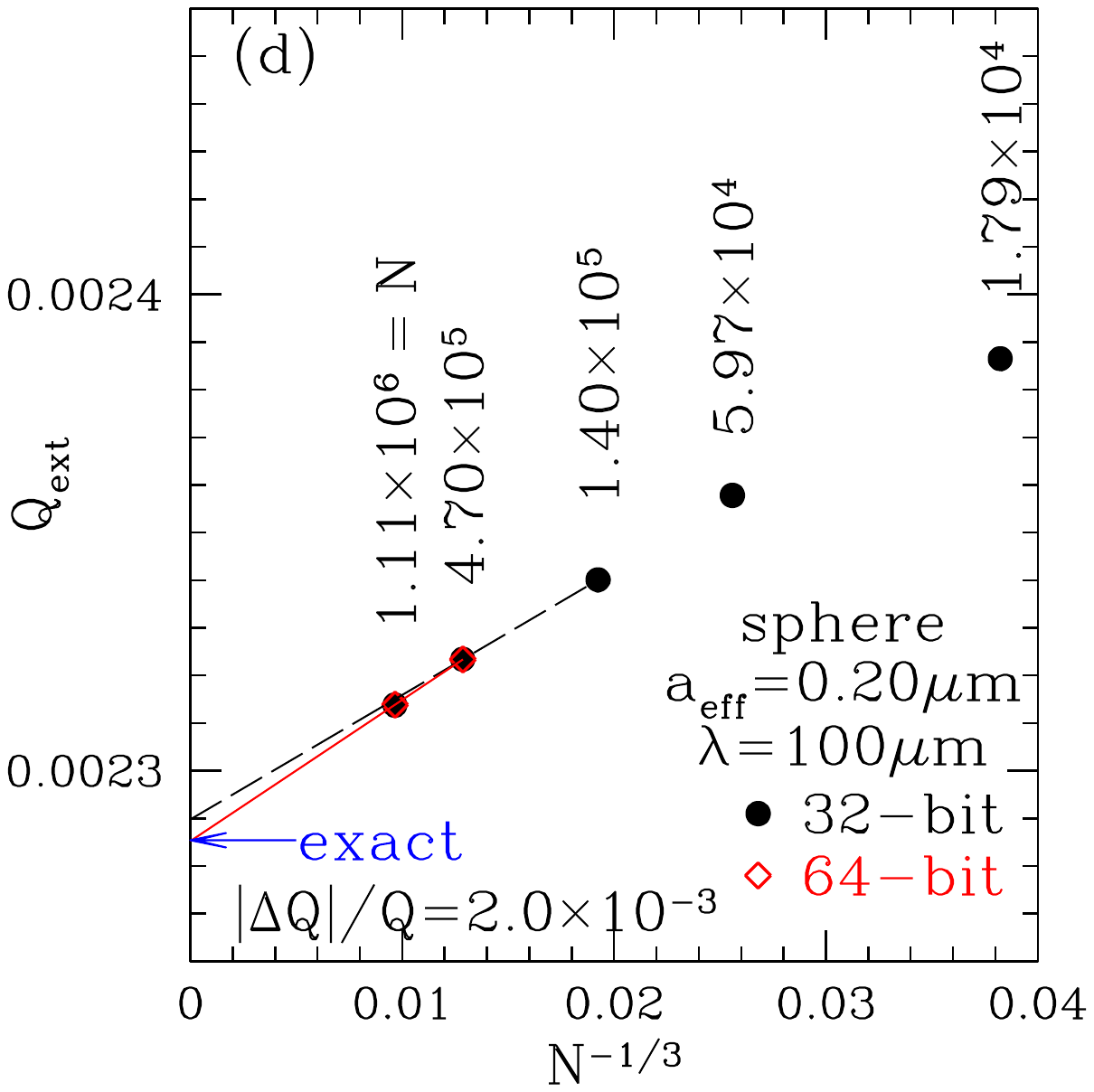}\\
\caption{\label{fig:qext_sphere}\footnotesize DDA calculations of
  $\Qextran$ for $a=0.10\micron$ spheres as a function of
  $\Ndip^{-1/3}$, where $\Ndip$ is the number of dipoles used to
  represent the sphere.  The dielectric function in Figure
  \ref{fig:diel} is employed.  Results are shown for four wavelengths.
  In each case the exact result calculated with Mie theory is shown.
  Extrapolation using Equation (\ref{eq:extrap}) (solid and dashed lines)
  is seen to yield very accurate results.
  }
\end{center}
\end{figure}

To verify the accuracy of the DDA, we use the DDA to calculate
scattering and absorption by spheres.  As discussed in section
\ref{subsec:extrap}, results calculated for two different numbers of
dipoles $\Ndip$ are used to extrapolate to $\Ndip\rightarrow\infty$,
and compared with exact cross sections calculated using Mie theory.
Figure \ref{fig:qext_sphere} shows results for spheres with radii
$a=0.2\micron$ and 4 selected wavelengths, from the vacuum ultraviolet
to the FIR.  Extrapolation to $\Ndip\rightarrow\infty$ gives results
that are in excellent agreement with the exact solutions.  In all 4
cases shown, the estimate $\Delta Q/Q$ for the fractional uncertainty
is seen to be very conservative: the actual errors in the extrapolated
$\Qext$ are in all cases significantly smaller than $|\Delta Q|$ from
Equation (\ref{eq:DeltaQ}).

\subsection{Cylinders and Square Prisms}

In this section similar studies are carried out for cylinders and
square prisms, to confirm that the behavior found for spheroids
(see Figures \ref{fig:qext_vs_N_sph} and \ref{fig:qpol_vs_N_sph}) is
general.

\begin{figure}[t]
\begin{center}
\includegraphics[angle=0,width=4.4cm,
                 clip=true,trim=0.5cm 5.0cm 0.5cm 2.5cm]
{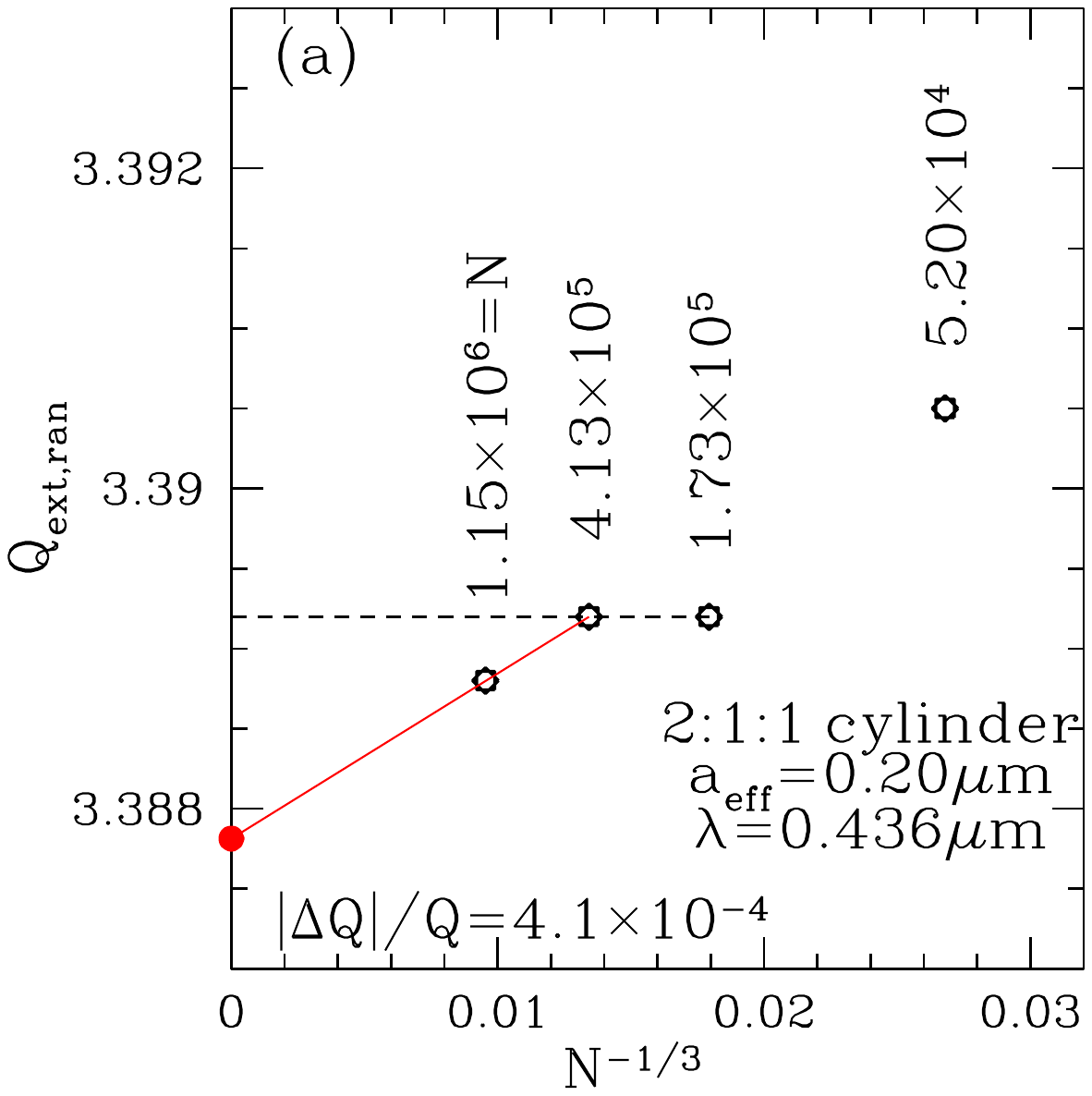}
\includegraphics[angle=0,width=4.4cm,
                 clip=true,trim=0.5cm 5.0cm 0.5cm 2.5cm]
{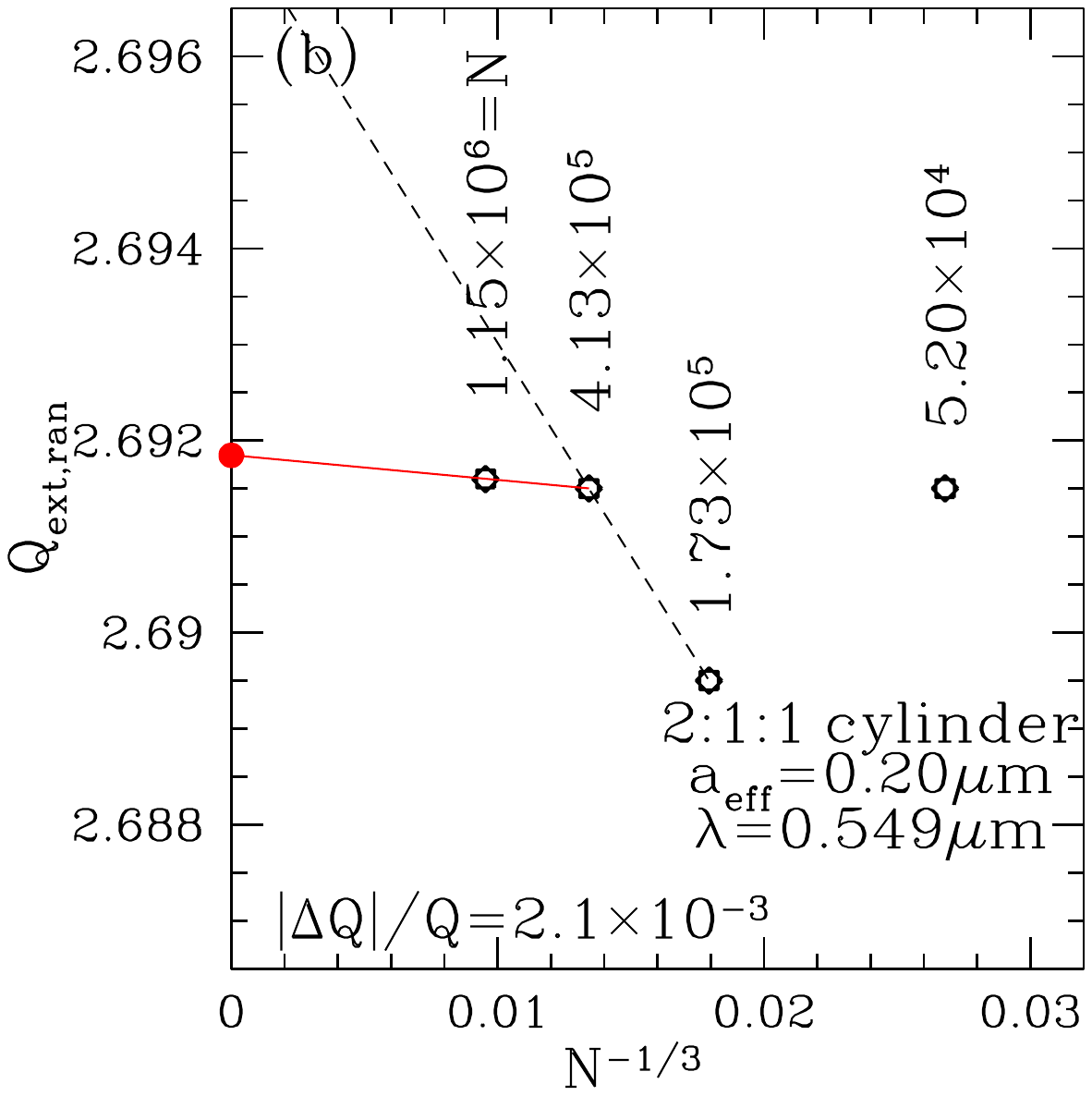}
\includegraphics[angle=0,width=4.4cm,
                 clip=true,trim=0.5cm 5.0cm 0.5cm 2.5cm]
{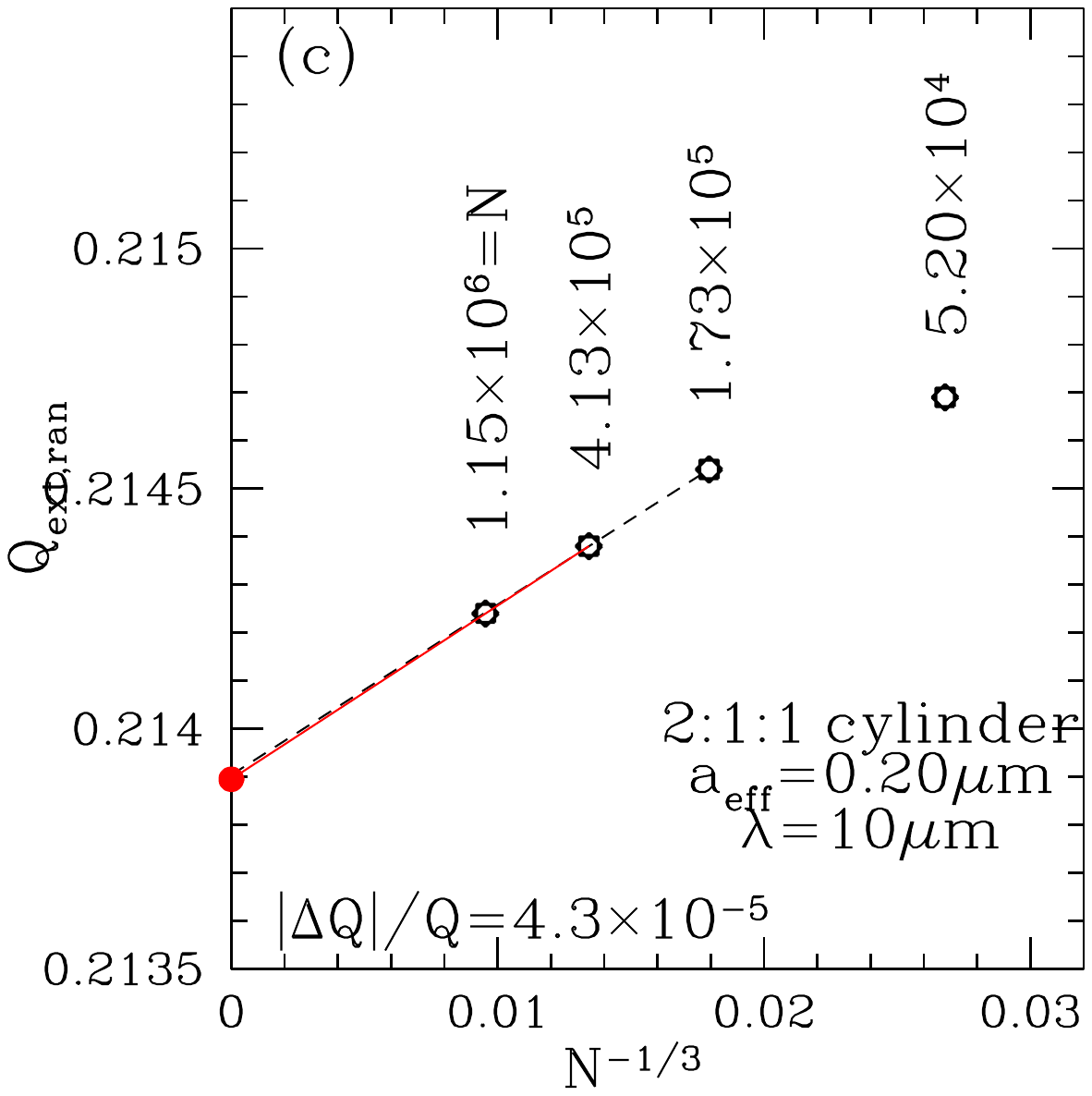}
\includegraphics[angle=0,width=4.4cm,
                 clip=true,trim=0.5cm 5.0cm 0.5cm 2.5cm]
{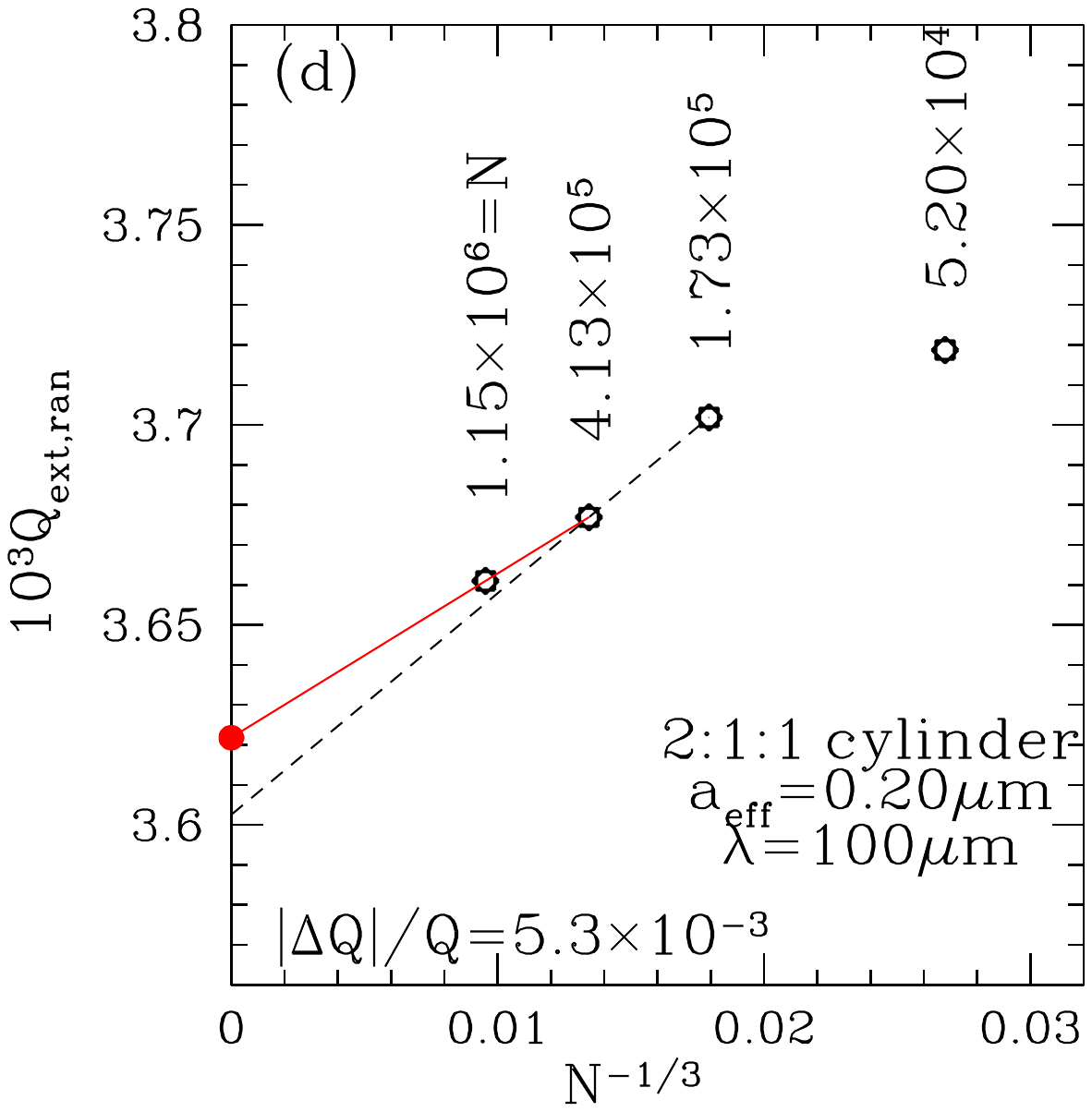}\\
\caption{\label{fig:qext_vs_N_cyl}\footnotesize Similar to Figure
  \ref{fig:qext_vs_N_sph}, but showing $\Qextran$ for
  randomly-oriented 2:1 cylinders.
  }
\end{center}
\end{figure}
\begin{figure}
\begin{center}
\renewcommand{\fwidth}{4.4cm}
\includegraphics[angle=0,width=\fwidth,
                 clip=true,trim=0.5cm 5.0cm 0.5cm 2.5cm]
{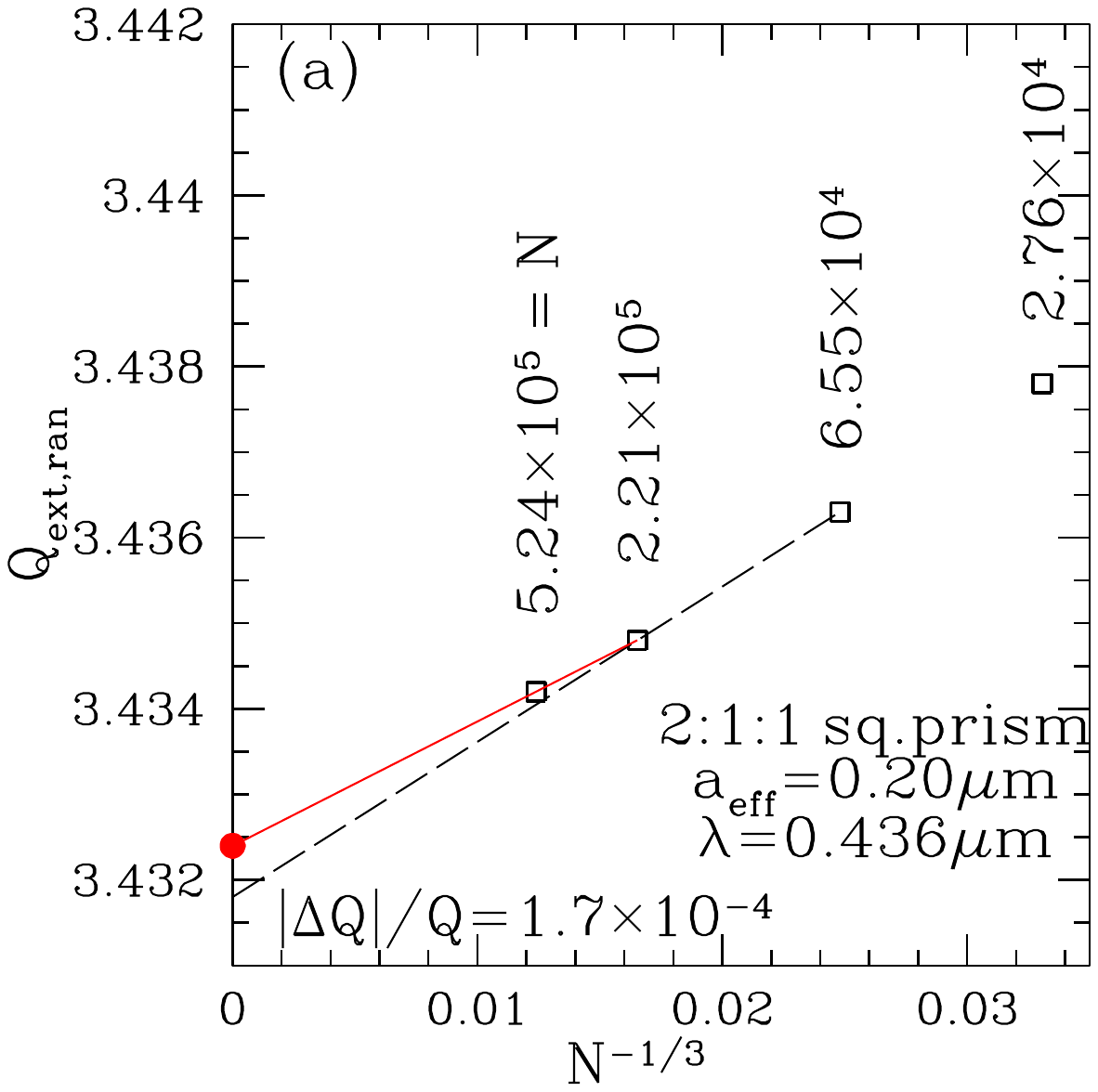}
\includegraphics[angle=0,width=\fwidth,
                 clip=true,trim=0.5cm 5.0cm 0.5cm 2.5cm]
{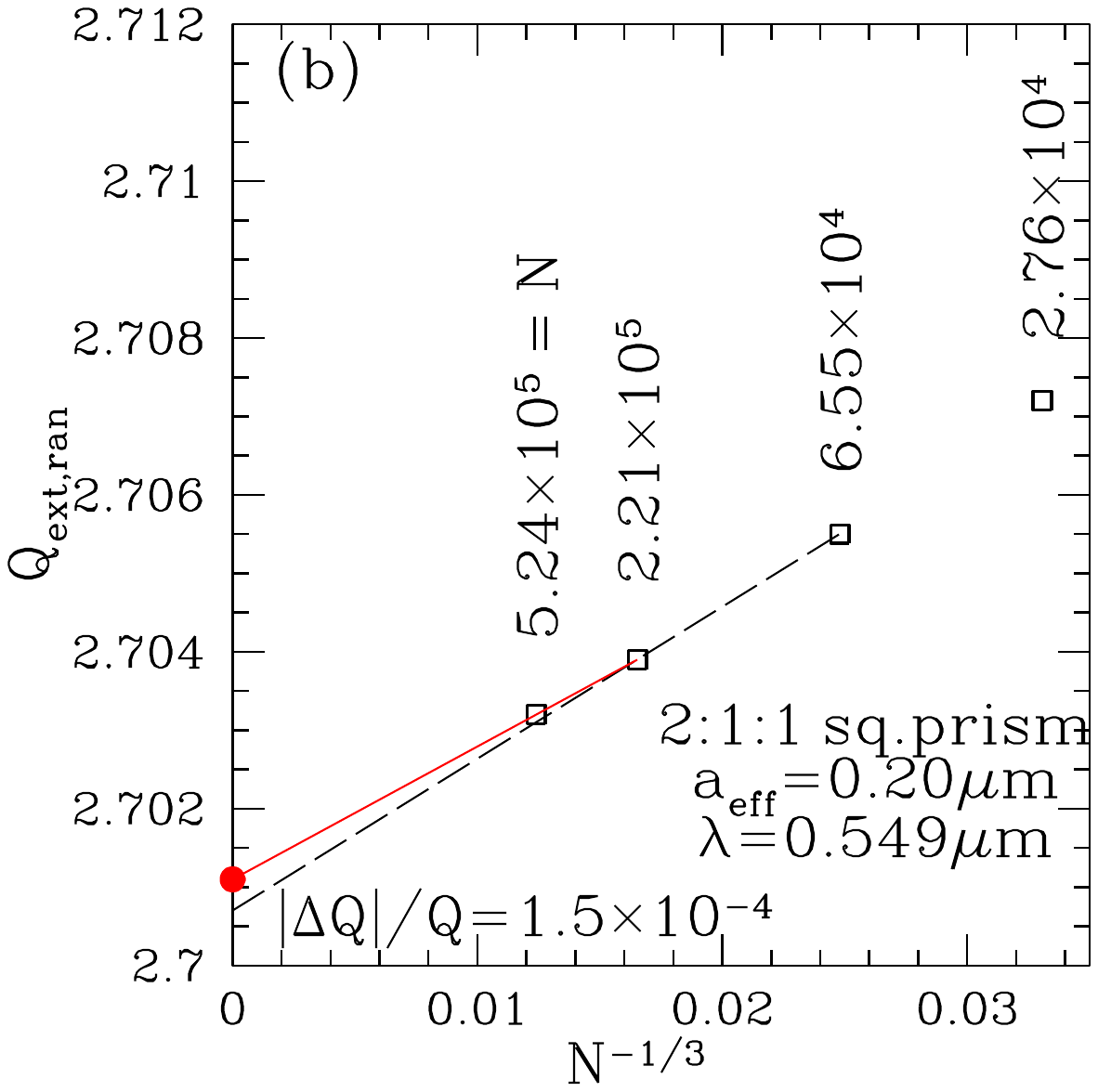}
\includegraphics[angle=0,width=\fwidth,
                 clip=true,trim=0.5cm 5.0cm 0.5cm 2.5cm]
{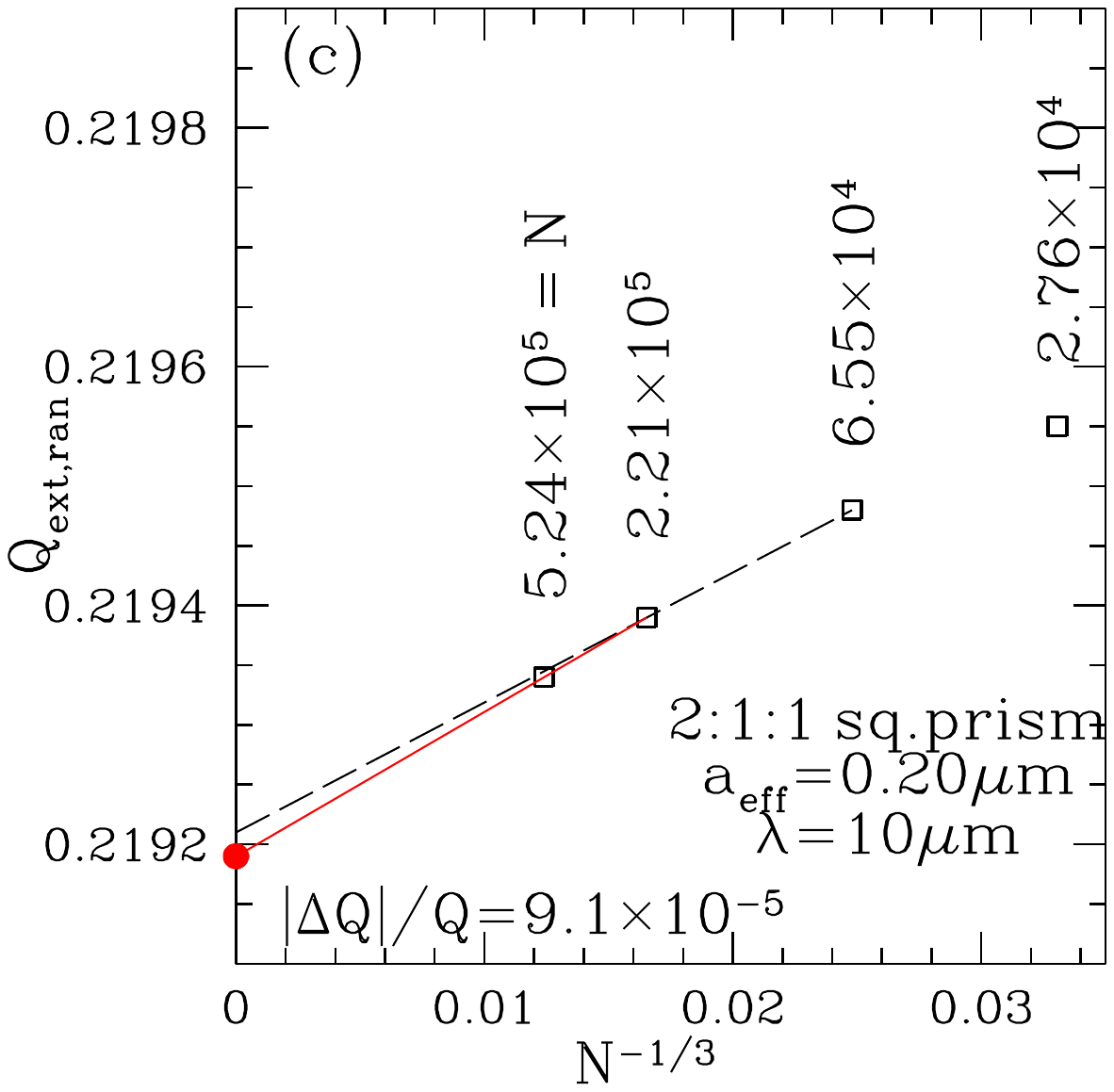}
\includegraphics[angle=0,width=\fwidth,
                 clip=true,trim=0.5cm 5.0cm 0.5cm 2.5cm]
{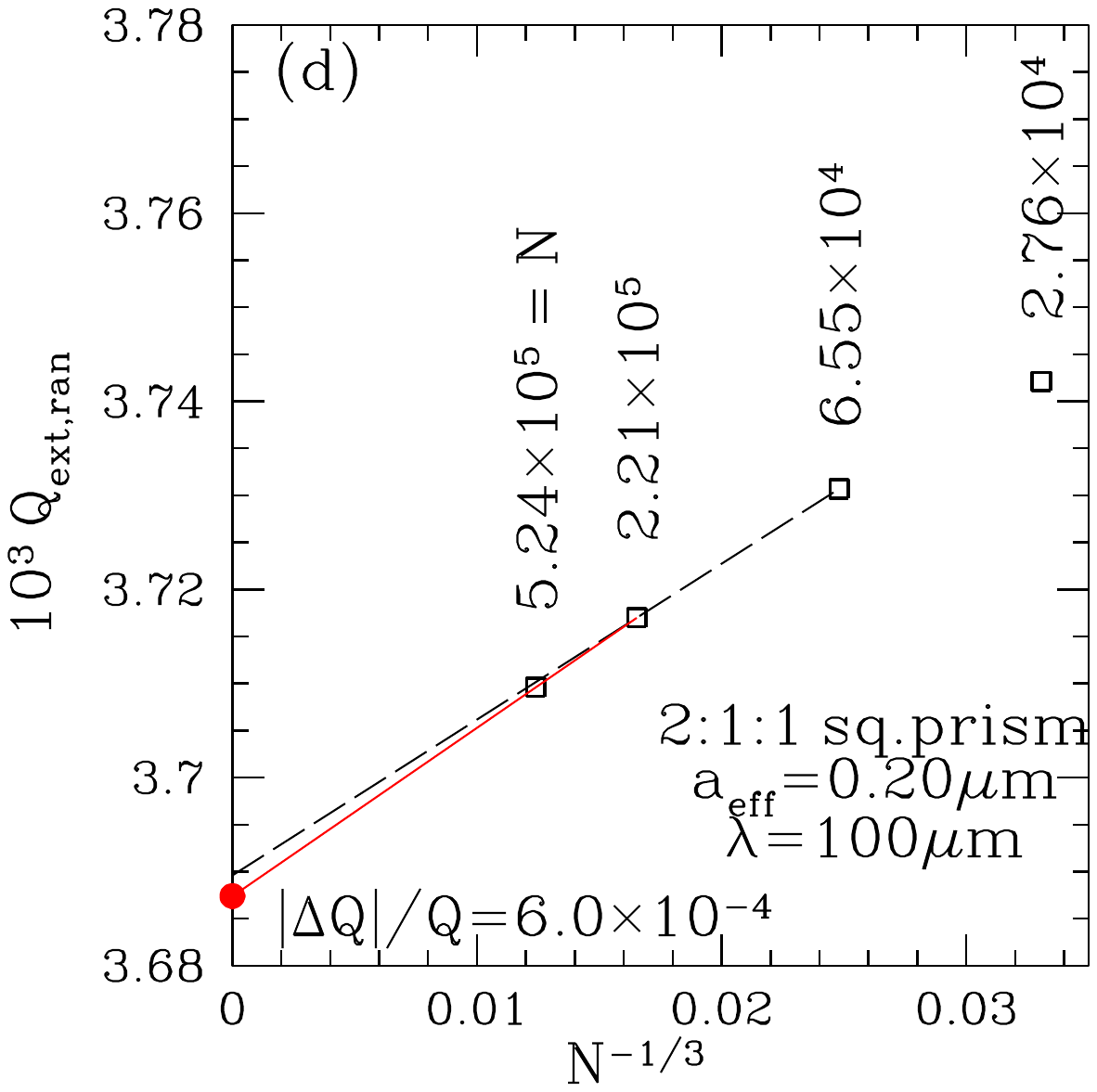}\\
\caption{\label{fig:qext_vs_N_bs}\footnotesize Similar to Figure
  \ref{fig:qext_vs_N_sph}, but showing $\Qextran$ for
  randomly-oriented 2:1:1 square prisms.
  }
\end{center}
\end{figure}

Figures \ref{fig:qext_vs_N_cyl} and \ref{fig:qext_vs_N_bs} show
$\Qextran$ vs.\ $N^{-1/3}$ for 2:1:1 cylinders and 2:1:1 square
prisms, both with $\aeff=0.2\micron$, at four selected wavelengths,
just as for 2:1:1 spheroids in Figure \ref{fig:qext_vs_N_sph}.  The
2:1:1 cylinder has a fractional uncertainty of $0.2\%$ in $\Qextran$
at $0.55\micron$ (Figure \ref{fig:qext_vs_N_cyl}b), and $0.5\%$ at
$100\micron$ (Figure \ref{fig:qext_vs_N_cyl}d).  these uncertainties
are not large enough to alter any of the conclusions in this paper.

\begin{figure}
\begin{center}
\includegraphics[angle=0,width=8.5cm,
                 clip=true,trim=0.5cm 0.5cm 0.5cm 0.5cm]
{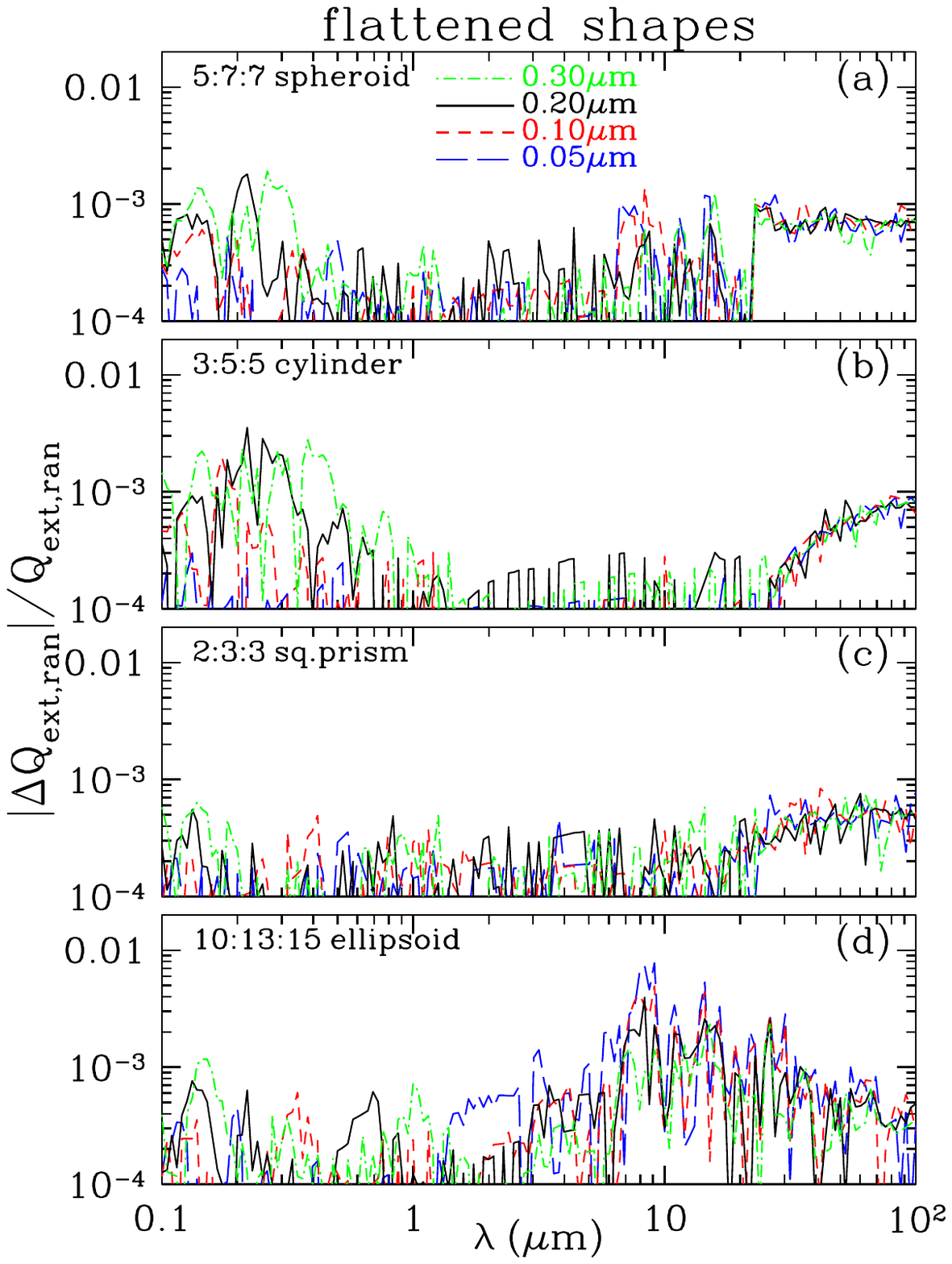}
\includegraphics[angle=0,width=8.5cm,
                 clip=true,trim=0.5cm 0.5cm 0.5cm 0.5cm]
{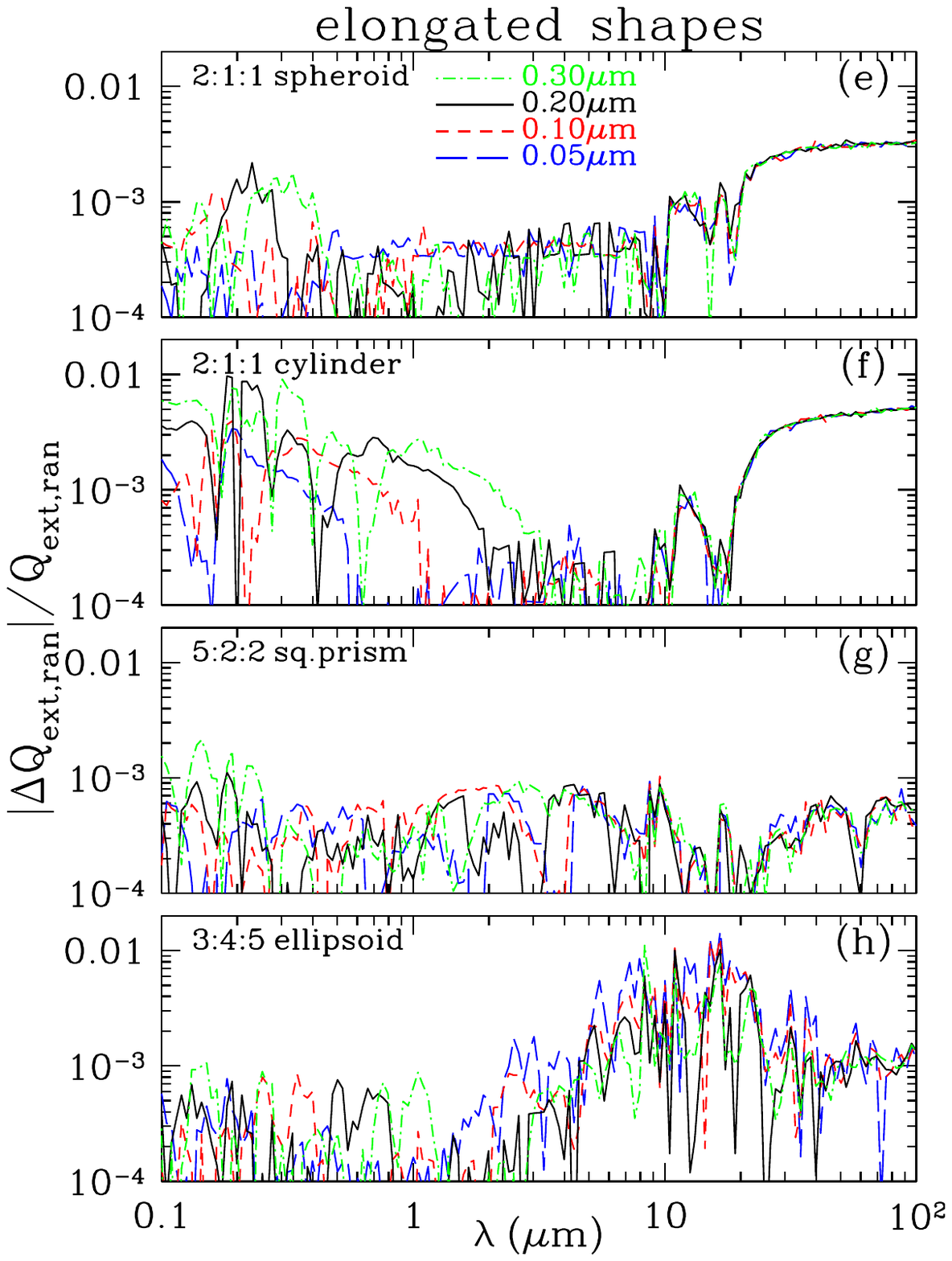}
\caption{\label{fig:ferrqext_vs_lambda}\footnotesize Fractional errors
  $|\Delta \Qextran|/\Qextran$ as a function of wavelength $\lambda$
  for the shapes in Figure \ref{fig:wave*Qext/a_vs_lambda}, for
  $\aeff=0.05, 0.10, 0.20, 0.30\micron$ and dielectric functions from
  Figure \ref{fig:diel}.  $\Qextran$ is extrapolated using Equation
  (\ref{eq:extrap}), and $\Delta\Qextran$ is obtained from Equation
  (\ref{eq:DeltaQ}) with the values of $N$ given in Table
  \ref{tab:N values}.
  }
\end{center}
\end{figure}

Figure \ref{fig:ferrqext_vs_lambda} shows the fractional uncertainties
in $\Qextran$ obtained from Equation (\ref{eq:extrap}) for 8 different
convex shapes, 4 values of $\aeff$, and the full range of wavelengths
$\lambda$ from $0.1\micron$ to $100\micron$.  For the values of $N$
used, the fractional errors in $\Qextran$ are generally quite small.
For these convex shapes, the DDA gives results for $\Qextran$ with
uncertainties well below $1\%$.

\begin{figure}
\begin{center}
\includegraphics[angle=0,width=8.5cm,
                 clip=true,trim=0.5cm 0.5cm 0.5cm 0.5cm]
{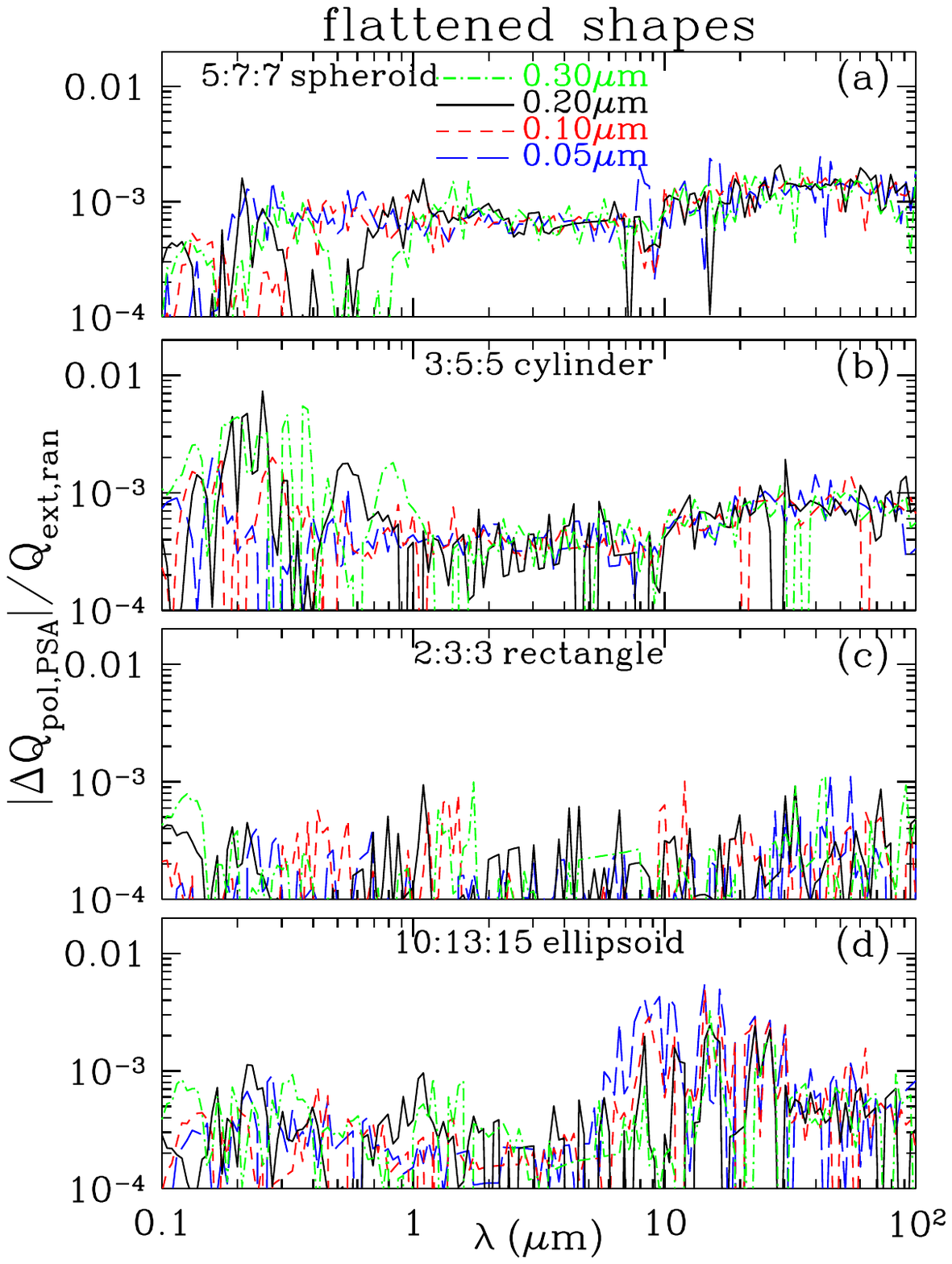}
\includegraphics[angle=0,width=8.5cm,
                 clip=true,trim=0.5cm 0.5cm 0.5cm 0.5cm]
{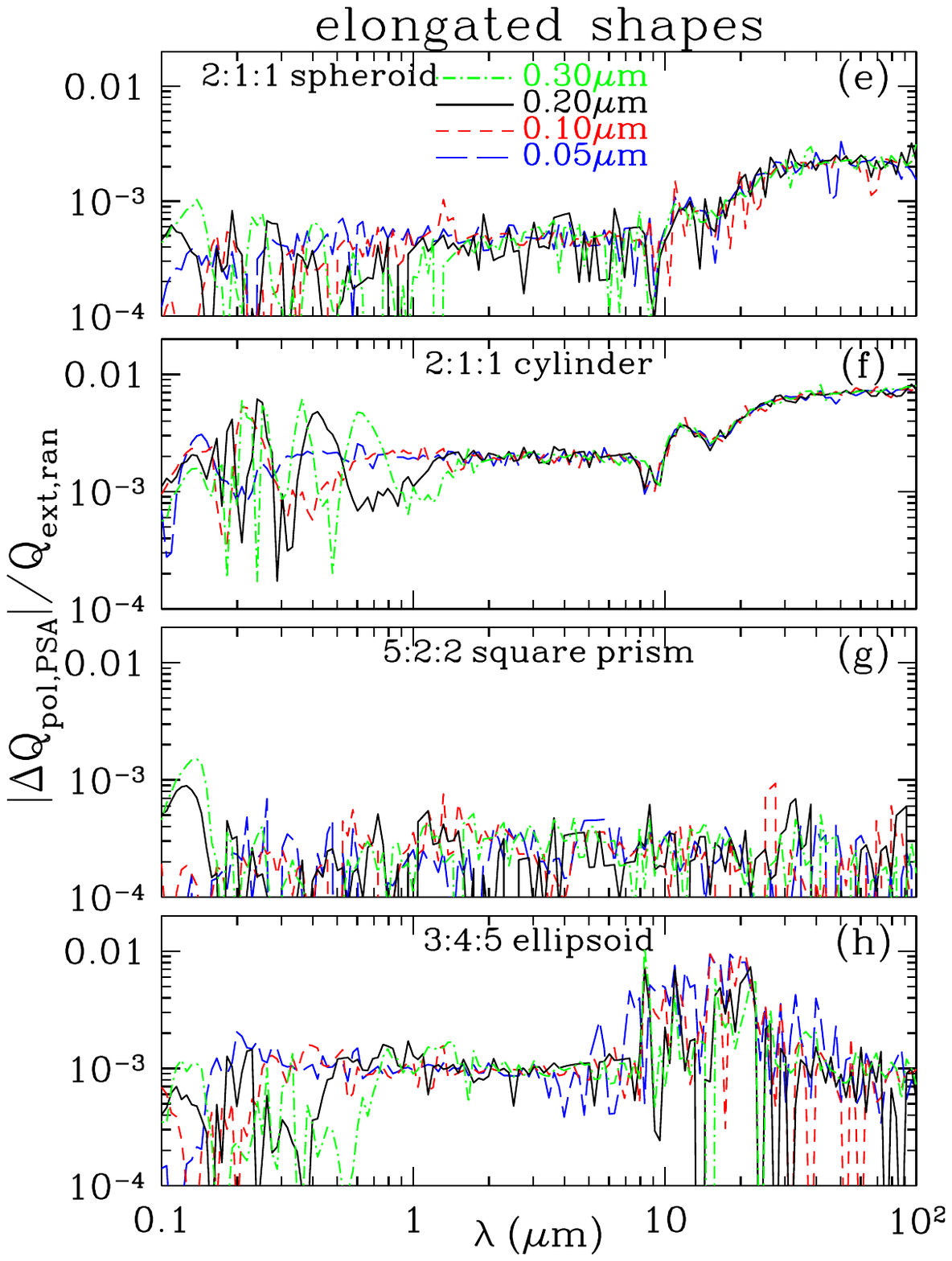}
\caption{\label{fig:ferrqpol_vs_lambda}\footnotesize
   $|\Delta\QpolPSA|/\Qextran$ as a function of wavelength $\lambda$
   for the shapes in Figure \ref{fig:wave*Qext/a_vs_lambda}, for   $\aeff=0.05, 0.10, 0.20, 0.30\micron$ and dielectric functions from
  Figure \ref{fig:diel}. $\QpolPSA$ is extrapolated using Equation
  (\ref{eq:extrap}), and $\Delta\QpolPSA$ is obtained from Equation
  (\ref{eq:DeltaQ}) with the values of $N$ given in Table
  \ref{tab:N values}. \btdnote{\bf OMIT THIS?}
  }
\end{center}
\end{figure}

Polarization cross sections depend on differences of cross
sections, and can be small or even negative at short wavelengths (see
Figure \ref{fig:wave*Qpol/a_vs_lambda}).  Thus the fractional error
$\Delta\QpolPSA/\QpolPSA$ can become large at short wavelengths,
even though the absolute error in $\QpolPSA$ may be small.  Therefore
we instead consider the ratio $|\Delta\QpolPSA|/\Qextran$.  Figure
\ref{fig:ferrqpol_vs_lambda} shows $|\Delta\QpolPSA|/\Qextran$ for the
eight shapes.  For the $N$ values used here, the uncertainties in
$\QpolPSA$ are generally well below $1\%$ of $\Qextran$.

The dependence of the extinction efficiency factor $\Qext(\Theta)$ and
the polarization efficiency factor $\Qpol(\Theta)$ on the orientation
angle $\Theta$ is shown in Figure \ref{fig:Qvsori} for
$\aeff=0.2\micron$ grains with the three different shapes.  For the
axisymmetric shapes considered here, the polarization efficiency
$\Qpol\rightarrow0$ for $\Theta\rightarrow0$, or
$\cos^2\Theta\rightarrow 1$.  For elongated grains in the Rayleigh
limit $\lambda\gg\aeff$, $\Qext(\Theta)$ has a minimum at $\Theta=0$,
but at optical wavelengths this need not be the case; for the examples
shown here, $\Qext$ is maximum for $\Theta=0$ for
$\lambda=0.44\micron$ and $\lambda=0.55\micron$.

\section{Effect of Shape on FIR Opacity}

\begin{figure}
\begin{center}
\includegraphics[angle=0,width=\fwidth,
                 clip=true,trim=0.5cm 0.5cm 0.5cm 0.5cm]
{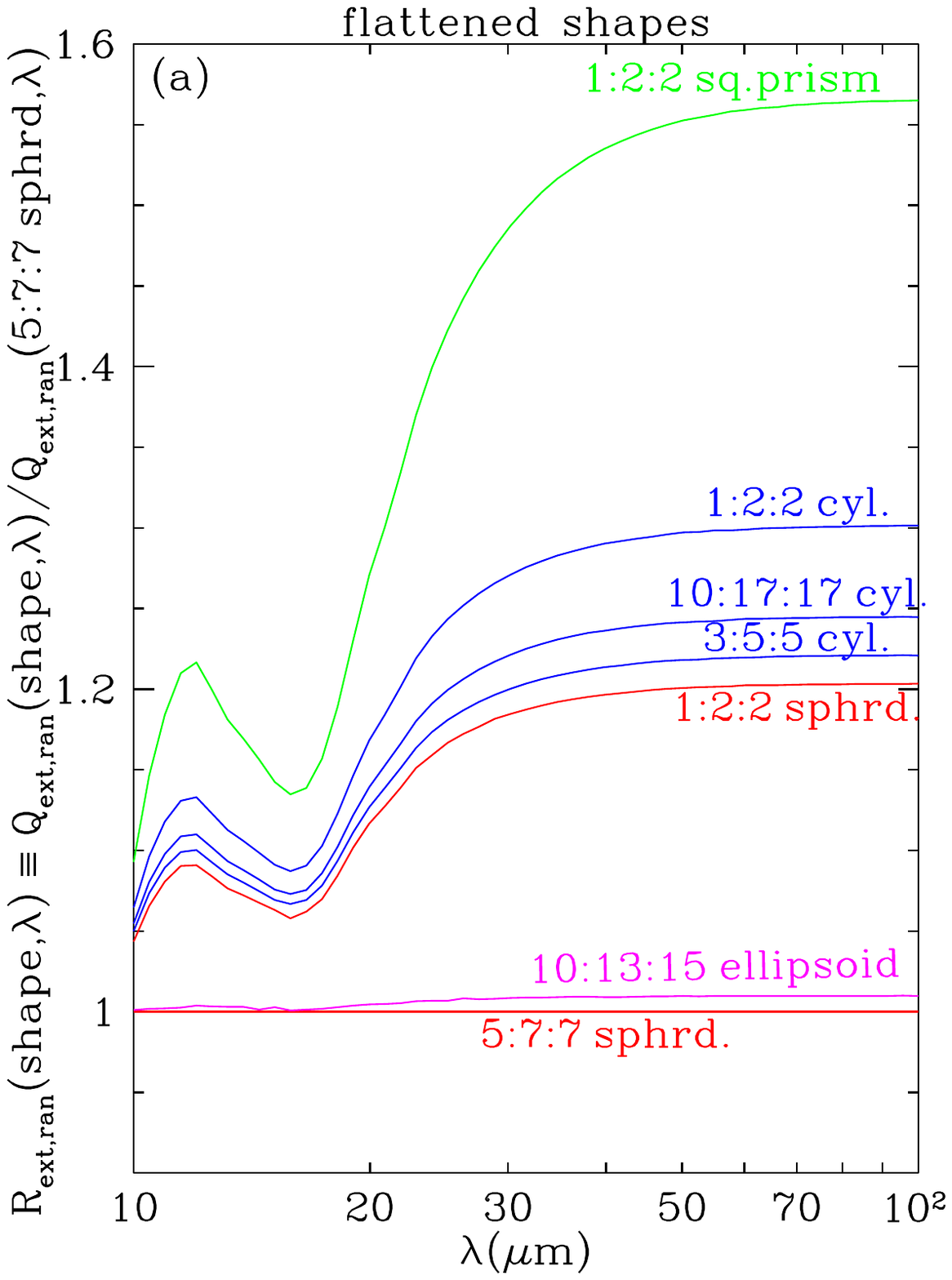}
\includegraphics[angle=0,width=\fwidth,
                 clip=true,trim=0.5cm 0.5cm 0.5cm 0.5cm]
{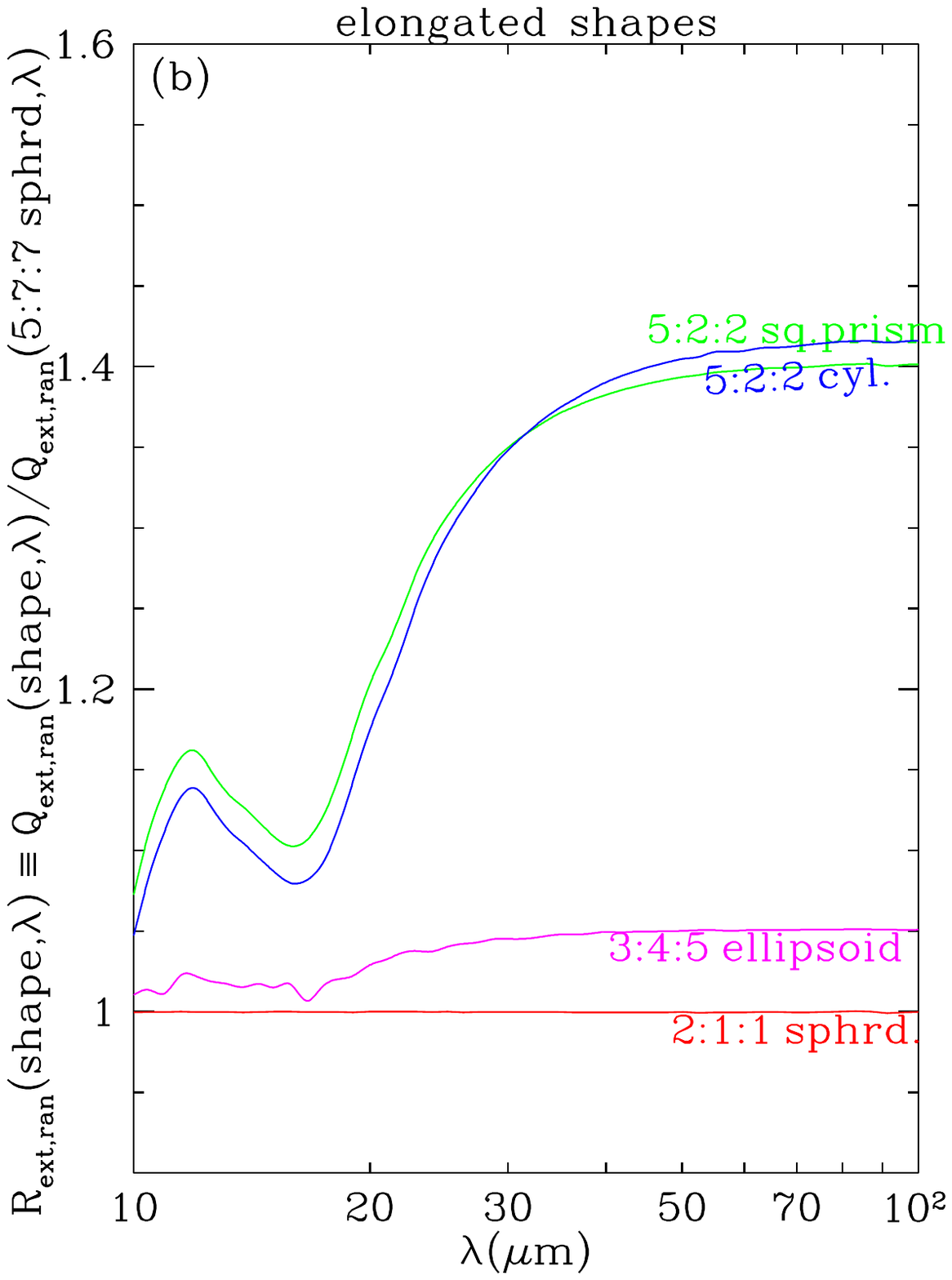}
\caption{\label{fig:figfir}\footnotesize Wavelength dependence of the
  ratio $R_{\rm ext,ran}$ of the infrared opacity, relative to the
  opacity of 5:7:7 oblate spheroids (see Equation \ref{eq:R}), for
  grain shapes with $\PhiPSA>0.7$ and $\sigmap<0.6$.  Cross sections
  were calculated using the ``oblate'' dielectric function
  $\epsilon(\lambda)$ from Figure \ref{fig:diel} for flattened shapes
  (left panel) and the ``prolate'' $\epsilon(\lambda)$ for elongated
  shapes (right panel).  All allowed shapes have $1 \leq R_{\rm
    ext,ran}(\lambda) \ltsim 1.6$ at $\lambda > 10\micron$.  The
  largest values of $R_{\rm ext,ran}$ are for the square prisms and
  cylinders; sharp edges and corners evidently enhance the FIR
  absorption relative to the smooth spheroids.
  }
\end{center}
\end{figure}

To see how shape influences the FIR absorption, Figure
\ref{fig:figfir} shows the ratio of the opacity of these different
shapes at wavelength $\lambda$ relative to the opacity for the 5:7:7
oblate spheroid:
\beq \label{eq:R}
R_{\rm ext,ran}({\rm shape},\lambda)
\equiv
\frac{Q_{\rm ext,ran}({\rm shape},\aeff,\lambda)}
     {Q_{\rm ext,ran}(5\!:\!7\!:\!7\,{\rm spheroid},\aeff,\lambda)}
~~~.
\eeq
The ``oblate'' dielectric function used for the flattened shapes was
derived by requiring that randomly-oriented 5:7:7 oblate spheroids
reproduce the observed FIR opacity of dust in the diffuse
ISM, while the ``prolate'' dielectric function used for the elongated
shapes was derived by requiring that randomly-oriented 2:1:1 spheroids
reproduce the observed opacity.  Thus it is unsurprising that Figure
\ref{fig:figfir}a has $R=1$ for 5:7:7 spheroids, and \ref{fig:figfir}b
has $R=1$ for 2:1:1 spheroids.  However, for other shapes, $R_{\rm
  ext,ran}(\lambda)$ remains of order unity, with $0.85 \leq R_{\rm
  ext,ran} < 1.6$ at $\lambda>10\micron$ for all shapes shown.  The
largest values of $R$ occur at the longest wavelengths, where
$\epsilon_1$ is large (see Figure \ref{fig:diel}) and the electric
field within the grains depends on the grain shape.  The sharp edges
and corners present in cylinders and square prisms lead to enhanced
absorption relative to spheroids in the FIR.

\section{\label{app:rotavg}
         Orientational Averaging}

\begin{figure}
\begin{center}
\includegraphics[angle=0,width=4.4cm,
                 clip=true,trim=0.5cm 0.5cm 0.5cm 0.5cm]
{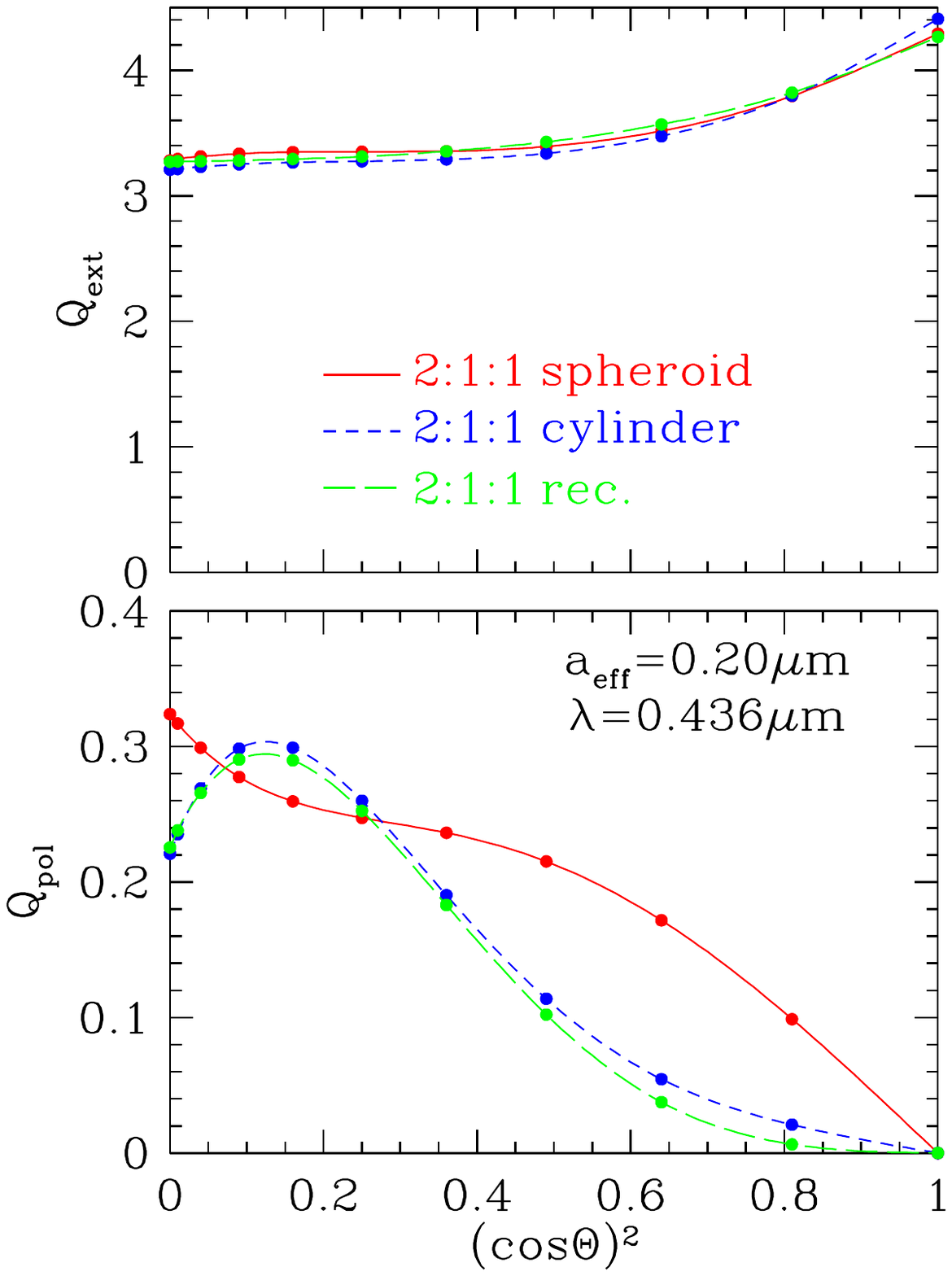}
\includegraphics[angle=0,width=4.4cm,
                 clip=true,trim=0.5cm 0.5cm 0.5cm 0.5cm]
{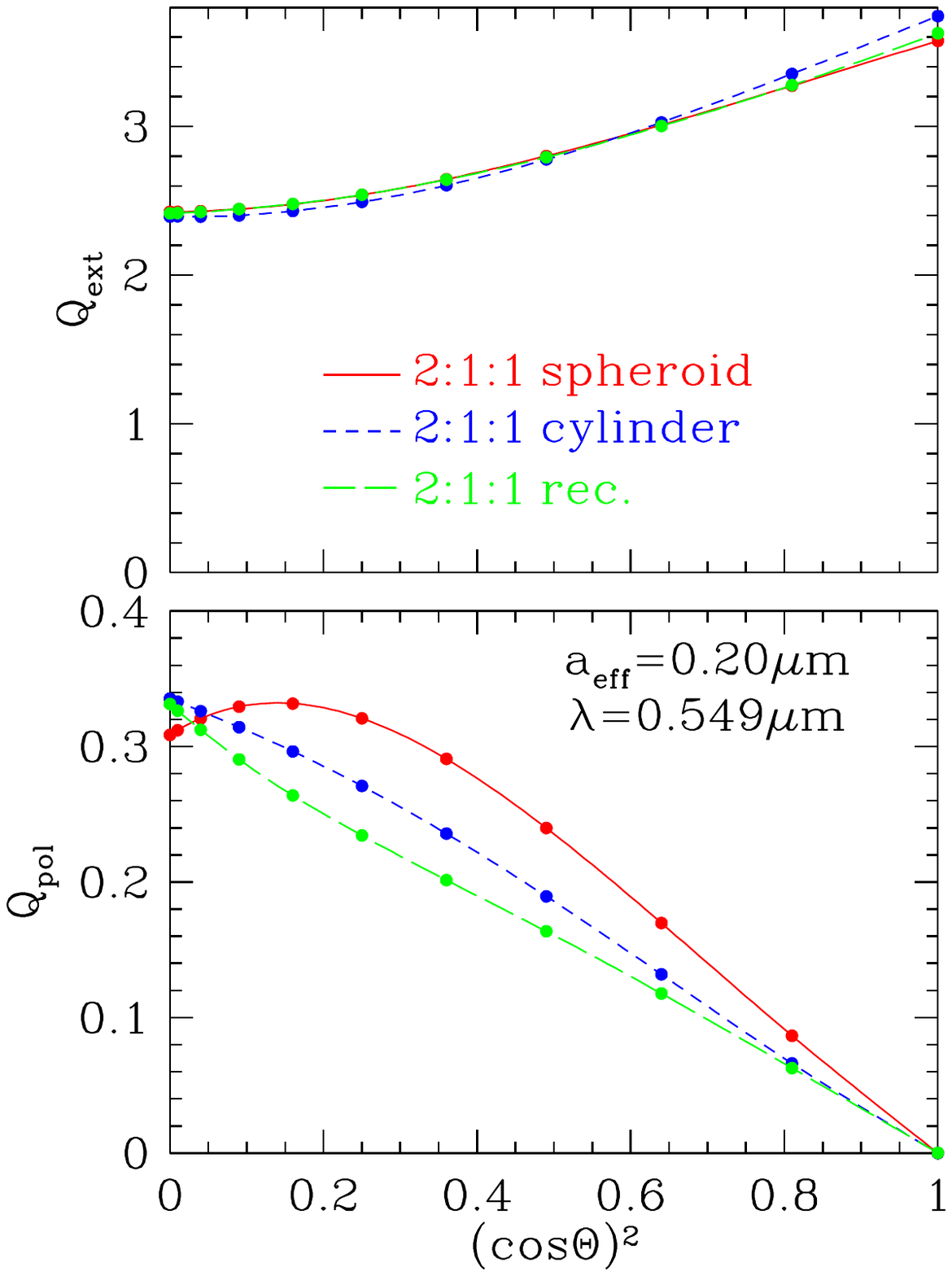}
\includegraphics[angle=0,width=4.4cm,
                 clip=true,trim=0.5cm 0.5cm 0.5cm 0.5cm]
{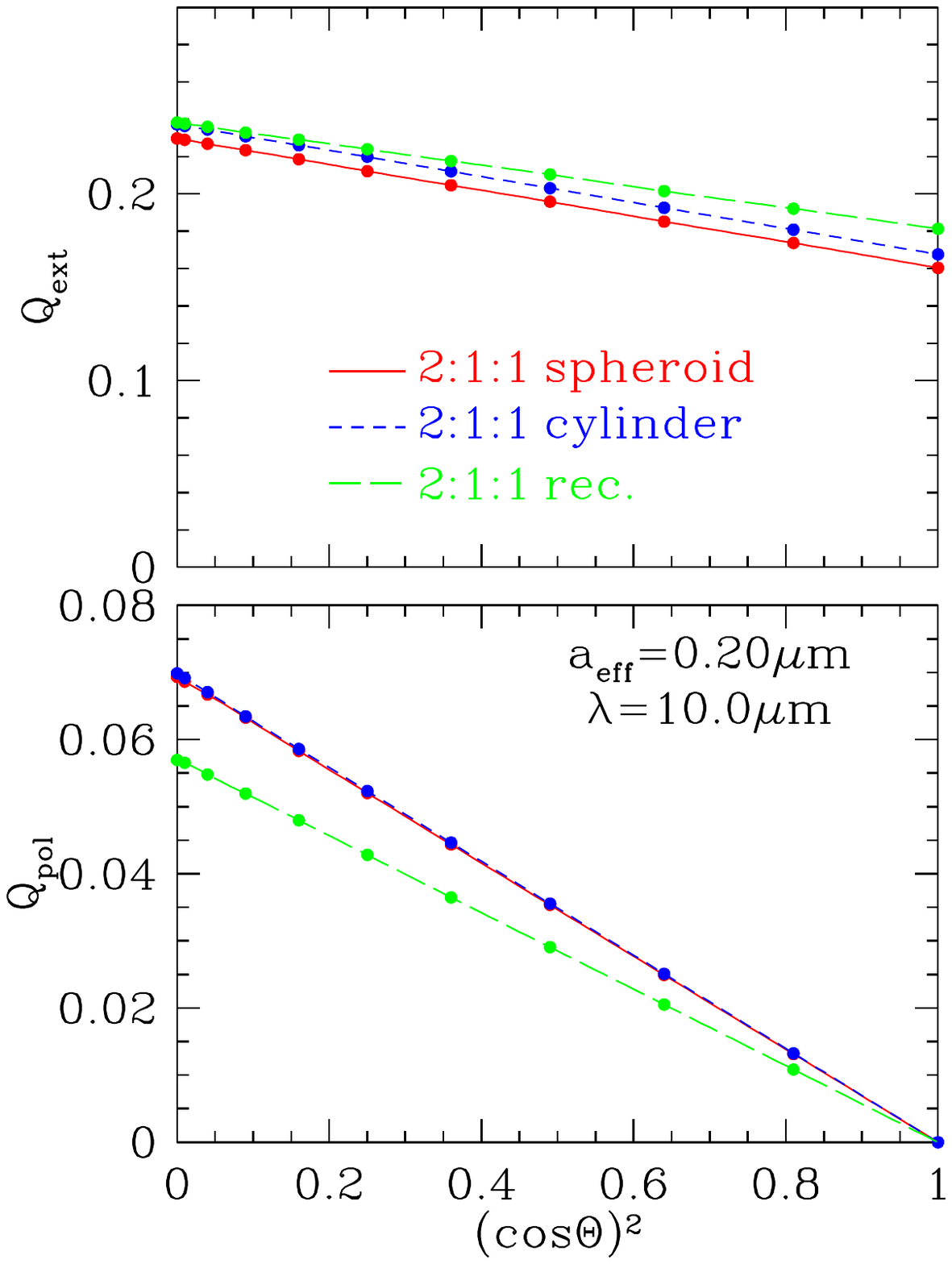}
\includegraphics[angle=0,width=4.4cm,
                 clip=true,trim=0.5cm 0.5cm 0.5cm 0.5cm]
{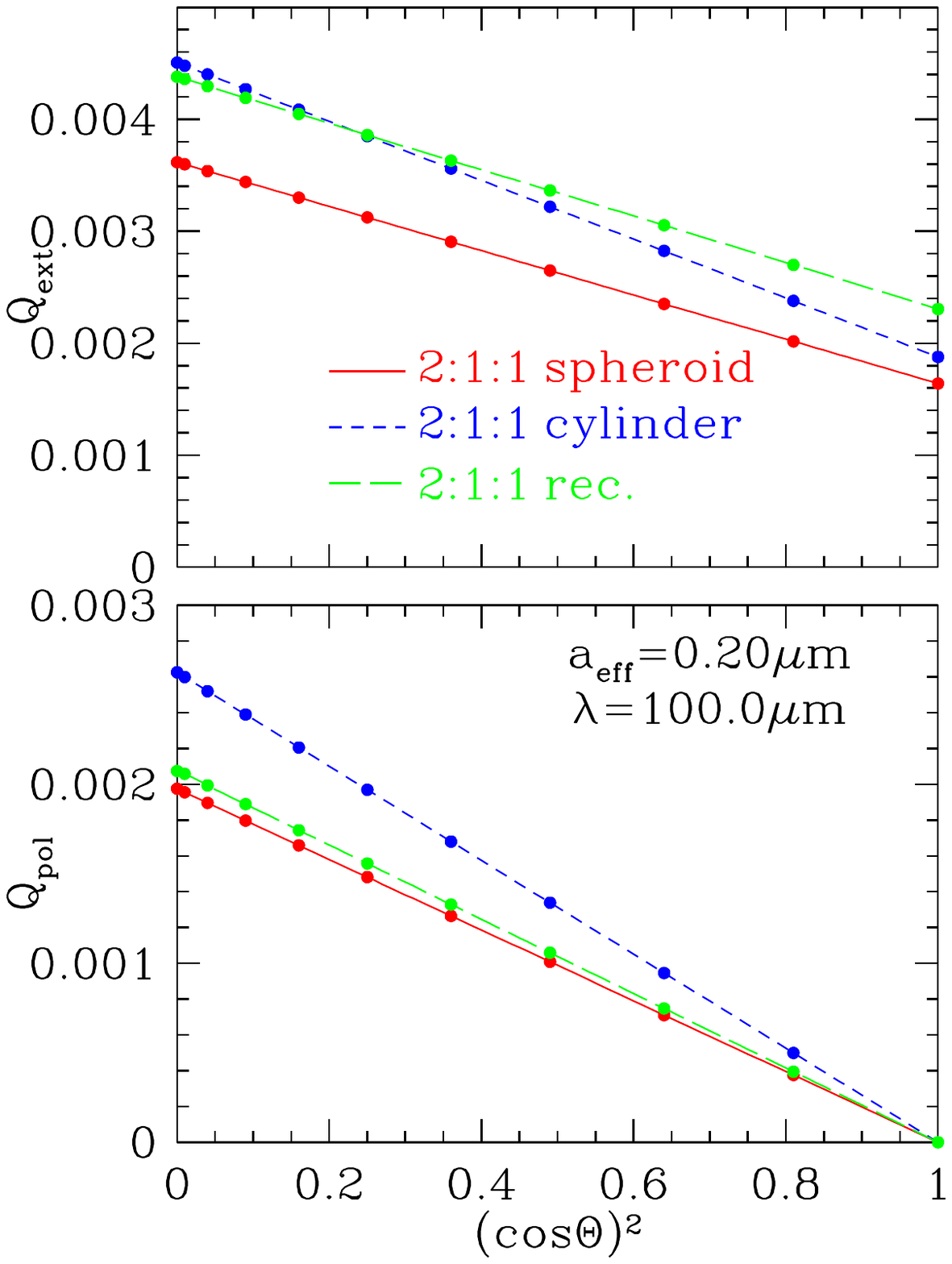}
\caption{\label{fig:Qvsori}\footnotesize Dependence of $\Qext(\Theta)$
  and $\Qpol(\Theta)$ on orientation angle $\Theta$ for four
  wavelengths $\lambda$ and three axisymmetric shapes, with size
  $\aeff=0.2\micron$.  For $\lambda\gg\aeff$ (panels c and d), the
  grain is in the Rayleigh limit, with $\Qext$ and $\Qpol$ both linear
  in $\cos^2\Theta$.  For $\lambda$ comparable to $\aeff$, the
  dependence on $\Theta$ becomes more complicated.  For
  axially-symmetric shapes, $\Qpol=0$ for $\cos^2\Theta=1$.
  }
\end{center}
\end{figure}

For polarization cross sections, we assume perfect spinning alignment
(PSA): the grain is spinning around the principal axis of largest
moment of inertia $\bahat_1$, with $\bahat_1\parallel\bB_0$.  We
consider the case where the radiation is propagating in the $\bzhat$
direction, and $\bB_0\parallel\bxhat$ (i.e., $\bB_0$ in the ``plane of
the sky'') (see Fig.\ \ref{fig:PSA}).

Suppose the grain has a symmetry axis $\bahat$.  Let
$\bahat_2\perp\bahat$ be a second axis fixed in the grain square
prism, let $\bahat_2$ be normal to one of the sides).  Let $\beta$
measure target rotations around $\bahat$, with $\beta=0$ when
$\bahat_2$ is in the $\bahat-\bkhat$ plane.  Let $\Theta$ be the angle
between $\bahat$ and the direction of propagation $\bkhat$.  $N$-fold
rotational symmetry implies
\beq \label{eq:sym1}
C(\Theta,\beta+2\pi/N)=C(\Theta,\beta)
\eeq
(e.g., $N=1$ for a general asymmetric grain, $N=4$ for a square
prism).

We require cross sections
over 
$\beta_{\rm min}<\beta<\beta_{\rm max}$ and $0<\Theta<\Theta_{\rm max}$.
Set $\beta_{\rm max}=\pi/N$. 
If the grain is symmetric under
reflection through the $\bahat_1-\bahat_2$ plane
\footnote{%
I.e.,
$C(\Theta,\beta)=C(\Theta,-\beta)$.}
(e.g., the square
prism), then $\beta_{\rm min}=0$; otherwise $\beta_{\rm min}=-\pi/N$.
If the grain has reflection symmetry through a plane perpendicular to
$\bahat$,\footnote{I.e.,
$C(\Theta,\beta)=C(\pi-\Theta,\beta)$.}  then
$\Theta_{\rm max}=\pi/2$; otherwise $\Theta_{\rm max}=\pi$.

Randomly-oriented grains have
\beq \label{eq:PSA1} 
\Qran =
\frac{1}{\pi\aeff^2}
\int_{\beta_{\rm min}}^{\beta_{\rm max}}
\frac{d\beta}{[\beta_{\rm max}-\beta_{\rm min}]}
\int_{0}^{\Theta_{\rm max}} 
\frac{Q_{\rm ext}(\Theta,\beta) \sin\Theta d\Theta}
{1-\cos(\Theta_{\rm max})}
\eeq

If the grain is spinning around axis $\bahat$ (e.g., flattened square
prism) then
\beq \label{eq:qpolPSAoblate}
\QpolPSA = 
\frac{1}{2\pi\aeff^2}
\int_{\beta_{\rm min}}^{\beta_{\rm max}}
\frac{
[C_{\rm ext,E}\left(\Theta=\frac{\pi}{2},\beta\right)-
C_{\rm ext,H}\left(\Theta=\frac{\pi}{2},\beta\right)
]\,d\beta}
{(\beta_{\rm max}-\beta_{\rm min})}
~~~.
\eeq
If the grain is spinning around an axis $\bahat_1\perp\bahat$ (e.g.,
elongated square prism) then
\beq \label{eq:qpolPSAprolate}
\QpolPSA = 
\frac{1}{2\pi\aeff^2}
\int_{\beta_{\rm min}}^{\beta_{\rm max}}
\frac{d\beta}
{(\beta_{\rm max}-\beta_{\rm min})}
\int_0^{\Theta_{\rm max}}
\left[
C_{\rm ext,E}\left(\Theta,\beta\right)-
C_{\rm ext,H}\left(\Theta,\beta\right)
\right]\frac{d\Theta}{\Theta_{\rm max}}
~~~.
\eeq
%

\end{document}